\documentclass[aps, prd, twocolumn, amsmath, floats,floatfix, superscriptaddress, nofootinbib]{revtex4}

\usepackage{amssymb}
\usepackage{amsmath}
\usepackage{verbatim}
\usepackage{mathrsfs}
\usepackage{amsfonts}
\usepackage{latexsym}
\usepackage{epsfig}
\usepackage{color}
\usepackage{graphicx,subfigure}
\usepackage{units}
\usepackage{slashbox}
\usepackage{natbib}
\usepackage[normalem]{ulem}

\begin{document}

%%%%%%%%%%%%%%%%%
%%%   TITLE   %%%
%%%%%%%%%%%%%%%%%
\title{%Mixed white-dwarf--boson stars  and possible observational consequences\\
Ultralight bosonic dark matter in white dwarfs and potential observational consequences%\\

%Possible dark matter signatures in the gravitational redshift of white dwarfs\\
%Possible boson star signatures in the gravitational redshift of white dwarfs\\
%Gravitational redshift of white dwarfs as a dark matter indicator
} 

\author{Nicolas Sanchis-Gual}
\affiliation{Departamento de
  Astronom\'{\i}a y Astrof\'{\i}sica, Universitat de Val\`encia,
  Dr. Moliner 50, 46100, Burjassot (Val\`encia), Spain}
\affiliation{Departamento  de  Matem\'{a}tica  da  Universidade  de  Aveiro  and  Centre  for  Research  and  Development in  Mathematics  and  Applications  (CIDMA),  Campus  de  Santiago,  3810-183  Aveiro,  Portugal}

\author{Paula Izquierdo}
\affiliation{
Instituto de Astrof\'isica de Canarias, 38205 La Laguna, Tenerife, Spain}
\affiliation{Department of Physics, University of Warwick, Coventry, CV4 7AL, UK}
%\affiliation{Departamento de Astrof\'isica, Universidad de La Laguna, 38206 La Laguna, Tenerife, Spain}

%%%%%%%%%%%%%%%%
%%%   DATE   %%%
%%%%%%%%%%%%%%%%

%\date{}

%%%%%%%%%%%%%%%%%%%%
%%%   ABSTRACT   %%%
%%%%%%%%%%%%%%%%%%%%

\begin{abstract}  
Fluid and ultralight bosonic dark matter can interact through gravity to form stable fermion-boson stars, which are static and regular mixed solutions of the Einstein-Euler-(complex, massive) Klein-Gordon system. In this work we study the dynamical formation via gravitational cooling of a spherical mixed white-dwarf--boson star, whose properties depend on the boson particle mass and the mass of the boson star. Due to the accretion of bosonic dark matter, the white dwarf migrates to a denser and more compact object with a boson star core, thus modifying its gravitational redshift and altering the electromagnetic radiation emitted from the photosphere. We discuss the implications of the changes in the gravitational redshift that in principle could be produced by any type of dark matter and that might lead to small discrepancies in the estimation of masses and radii derived from white dwarf observations. 
\end{abstract}

%%%%%%%%%%%%%%%%
%%%   PACS   %%%
%%%%%%%%%%%%%%%%

%\pacs{
%}

%%%%%%%%%%%%%%%%%%%%%%
%%%   MAKE TITLE   %%%
%%%%%%%%%%%%%%%%%%%%%%

\maketitle

%%%%%%%%%%%%%%%%%%%%%%%%%%%%%%%%%%%%%%%%%%%%%%%%%%%
\section{Introduction}\label{sec:introduction}
%%%%%%%%%%%%%%%%%%%%%%%%%%%%%%%%%%%%%%%%%%%%%%%%%%%

The progress of gravitational-wave astronomy~\cite{abbott2016observation,abbott2017gw170817,abbott2019gwtc,abbott2020gwtc,abbott2020gw190814,abbott2020gw190521} together with the imaging of the shadow of the supermassive black-hole M87*~\cite{event2019first,akiyama2019first} are stimulating and expanding the studies on dark compact objects. Even if the black hole hypothesis stands as the simplest, more robust explanation to most of the observational data, theoretical exotic models have been proposed as {\it black-hole mimickers} that could potentially lead to compelling degeneracies in the gravitational and electromagnetic phenomenologies~\cite{bustillo2021gw190521,olivares2020tell,herdeiro2021imitation}. These exotic solutions could be seen not as alternatives to black holes, but represent a new population of astrophysical compact objects co-existing with black holes in specific mass ranges and possibly forming new configurations~\cite{herdeiro2014kerr}.

In particular, one of the most studied models of exotic compact objects are scalar boson stars (BSs)~\cite{kaup1968klein,ruffini1969systems,jetzer1992boson,schunck2003general} (see~\cite{liebling2017dynamical} for a review). BSs are horizonless, localized, stationary solutions of the Einstein-(complex, massive) Klein-Gordon system, composed of ultralight fundamental bosonic fields that could account for (part of) dark matter. Such bosonic particles can have masses in the range of $10^{-22}$ to $10^{-10}$ eV~\cite{arvanitaki2010string,Freitas:2021cfi}, leading to configurations with astrophysical relevance, from stellar mass to dark matter halos in galaxies~\cite{matos2000scalar,hu2000fuzzy}. Vector boson stars, also known as Proca stars, which are solutions of the Einstein-(complex) Proca system, have also been proposed~\cite{brito2016proca}. Bosonic stars, both scalar and vector, have a well-known efficient dynamical formation mechanism called gravitational cooling~\cite{seidel1994formation,guzman2006gravitational,di2018dynamical} and their stability properties have been extensively investigated~\cite{gleiser1988stability,gleiser1989gravitational,lee1989stability,seidel1990dynamical,hawley2000boson,guzman2009three,Escorihuela-Tomas:2017uac,sanchis2017numerical}, even with angular momentum~\cite{sanchis2019nonlinear,DiGiovanni:2020ror,siemonsen2021stability}, making them dynamically robust and viable dark matter candidates.

BSs are only minimally coupled to Einstein's gravity and, therefore, are transparent to the electromagnetic radiation. In theory, they could only be detected through their associated gravitational phenomena: the gravitational-wave emission in binary mergers and gravitational collapse scenarios~\cite{palenzuela2017gravitational,bezares2017final,bezares2018gravitational,sanchis2019head,DiGiovanni:2020ror,bustillo2021gw190521,bezares2022gravitational}, or through their effective shadows~\cite{cunha2015shadows,cunha2017lensing,olivares2020tell,herdeiro2021imitation}. %Consequently, %obtaining the gravitational waveforms and shadows of these objects 
%computing these predictions is necessary to compare them with current and future observations.
On this matter, it was recently found that gravitational waves from Proca star collisions could fit the GW190521 event~\cite{abbott2020gw190521} just as well as black hole mergers~\cite{bustillo2021gw190521}, proving that gravitational radiation could indeed be a channel to discover these objects. 

Nonetheless, if ultralight bosonic fields exist in the Universe and can form massive compact stars, one may also wonder how these fundamental fields would interact with compact stars that emit electromagnetic waves and what observable consequences there might be. In this regard, several studies have addressed the stability and dynamical formation of fermion-boson stars~\cite{jetzer1990stability,henriques1990stability,valdez2013dynamical,valdez2020fermion,di2020dynamical,di2021dynamical}, where the fermion star is generally assumed to be akin to a neutron star~\cite{DiGiovanni:2021ejn}. These stars are mixed configurations that could be the result of scalar field accretion by the fermion star~\cite{brito2015accretion,brito2016interaction} or the incomplete gravitational collapse of a bosonic cloud around the star~\cite{di2020dynamical,di2021dynamical}. Mergers of mixed stars have also been performed in~\cite{bezares2019gravitational}.

Mixed solutions with a BS in an excited state (with one or more nodes in the scalar field radial profile), that are unstable when isolated~\cite{Lee:1988av,balakrishna1998dynamical, Sanchis-Gual:2021phr}, can become stable and form dynamically~\cite{di2020dynamical,di2021dynamical}. Thus, a compact star can play a fundamental role in the stability of BSs in excited states. This stabilization mechanism was already shown in~\cite{bernal2010multistate}, where a superposition of a BS in the fundamental state and an excited BS was considered, and in~\cite{Sanchis-Gual:2021edp} for rotating and dipolar stars. Mixed configurations can modify the stability and properties of isolated stars~\cite{alcubierre2019dynamical,guzman2020gravitational,jaramillo2020dynamical,Sanchis-Gual:2021edp}.

While neutron stars are one of the most compact objects in the Universe and could serve as seeds to the formation of mixed fermion-boson stars through accretion or dynamical collapse of bosonic clouds, in this paper we will consider another type of fermion star: white dwarfs. These objects are stellar core remnants of the evolution of main sequence stars with masses from about 0.07 to 10 $M_{\odot}$, which includes 97\% of the stars in the Milky Way~\cite{fontaine2001potential}. Compared to neutron stars, white dwarfs are less dense and compact objects, but could still accrete dark matter~\cite{leung2013dark,leung2019accretion,zha2019accretion}. Due to the weaker gravity and large radii in white dwarfs, a BS forming through accretion within the fermion star would likely be more dilute and less massive than in the neutron-boson star case.

We first study the time evolution of an initial cloud of scalar field with different total and scalar particle masses around a white dwarf. The fermion star is described by a polytropic equation of state as a toy model of astrophysical white dwarfs. We perform non-linear simulations of the Einstein-Euler-(complex, massive) Klein-Gordon system in spherical symmetry, showing that mixed white-dwarf--boson stars can be dynamically formed through the gravitational cooling mechanism. Then, we identify and build the corresponding static mixed model from the total mass of the final object, the BS and fermion masses, and central rest-mass density of the final object using the code described in~\cite{di2020dynamical,di2021dynamical}. Finally, we compare the fermion star with a white dwarf solution (without scalar field) with the same central rest-mass density. 

Our results show that the formation of a mixed star modifies the energy density radial distribution of the fermionic component. As the white dwarf collapses to a denser and more compact configuration due to the self-gravity and compactness of the BS, it approaches the radial density profile of an isolated white dwarf with the same value of the central density. %The radius is the same in both cases but the mass is different. 
Therefore, as a consequence of the changes endured by the fermion star due to the accretion of dark matter, %it is possible that some of basic parameters describing a white dwarf such as surface temperature and gravity and thus particularly 
its electromagnetic emission will also be modified. %of the new white dwarf in the mixed configuration 
%could coincide with those of an isolated and more massive white dwarf. 
It is then possible that a white dwarf with a BS core could imitate to some extend the electromagnetic phenomenology of a completely different white dwarf. However, the mass of the mixed star computed up to the white dwarf surface will not necessarily be equal to the mass of the mimicked isolated white dwarf. This implies that the measured gravitational redshift, which is a purely gravitational effect, at both photospheres would be different. The gravitational redshift can be used to obtain the mass of white dwarfs~\cite{parsons2017testing,joyce2018gravitational,romero2019white,chandra2020gravitational}, but it might not agree with other methods~\cite{bergeron1991spectroscopic,bergeron2011comprehensive,carrasco2014gaia,barstow2017sirius,joyce2018gravitational,pasquini2019masses}, if white dwarf models are considered instead of mixed solutions. So far, no large discrepancies in the estimated masses have been observed, but since the amount of (bosonic) dark matter in white dwarfs is expected to be small, this difference should also be small and probably within observational uncertainties.

This paper is organized as follows: in Section~\ref{sec:formalism} we define the basic equations, in Section~\ref{sec:numerics} we give the numerical details of the simulations, in Section~\ref{sec:num_results} we show our main results, and in Section~\ref{sec:conclusions} we discuss our findings. We use $c=G=\hbar=1$ units.

%%%%%%%%%%%%%%%%%%%%%%%%%%%%%%%%%%%%%%%%%%%%%%%%%%
\section{Basic equations}\label{sec:formalism}
%%%%%%%%%%%%%%%%%%%%%%%%%%%%%%%%%%%%%%%%%%%%%%%%%%

%%%%%%%%%%%%%%%%%
\subsection{Einstein's equations}
%%%%%%%%%%%%%%%%%

We solve numerically the coupled Einstein-Klein-Gordon system and the general relativistic hydrodynamics equations:
\begin{equation}
 R_{\alpha\beta}-\frac{1}{2}g_{\alpha\beta}R=8\pi T_{\alpha\beta} \ ,
\label{eq:Einstein}
\end{equation}
where $R_{\alpha\beta}$ is the Ricci tensor of the 4-dimensional
spacetime, $g_{\alpha\beta}$ is the spacetime metric,
$R$ is the Ricci scalar, and $T_{\alpha\beta}$ is the stress-energy tensor of the matter content. We assume that the perfect fluid and the bosonic field are minimally coupled to Einstein's gravity and only interact through gravity. The combined stress-energy tensor is given by
\begin{equation}
 T_{\alpha\beta} = T_{\alpha\beta}^{\text{F}}+ T_{\alpha\beta}^{\text{SF}},
\end{equation}
where 
\begin{equation}
T_{\alpha\beta}^{\text{F}}=\rho h u_{\alpha}u_{\beta} + p g_{\alpha\beta}\,,
\label{matter}
\end{equation}
is the stress-energy tensor of the fluid fermionic matter (F), and $T_{\alpha\beta}^{\text{SF}}$  is the scalar field (SF) counterpart, given by
\begin{equation}
T_{\alpha\beta}^{\text{SF}}=\partial_{\alpha}\Phi 
\partial_{\beta}\Phi-\frac{1}{2}g_{\alpha\beta}\left(\partial^{\sigma}\Phi\partial_{\sigma}\Phi+
\mu^2\Phi^2 \right)\ .
\label{eq:tmunu}
\end{equation} 

The equations of motion for the  scalar field are  given by the Klein-Gordon equation
\begin{equation}
 \Box \Phi-\mu^2\Phi=0 \ ,
\label{eq:KG}
\end{equation}
where we define the d'Alambertian operator $\Box:=
(1/\sqrt{-g})\partial_{\alpha}(\sqrt{-g}g^{\alpha\beta}\partial_{\beta})$. We
follow the convention that $\Phi$ is dimensionless and $\mu$ has
dimensions of (length)$^{-1}$. We refer the interested reader to previous papers~\cite{Montero:2012yr,Sanchis-Gual:2015,sanchis2015quasistationary,di2020dynamical} for further details on the evolution equations.  

The spacetime metric $g_{\alpha\beta}$ is given by
\begin{eqnarray} \label{metric}
ds^2 & = & g_{\alpha\beta} dx^\alpha dx^\beta \nonumber \\ 
& = & - \alpha^2 dt^2 + \gamma_{ij} (dx^i + \beta^i dt)(dx^j + \beta^j dt),
\end{eqnarray}
where $\alpha$ is the lapse function, $\beta^i$ the shift vector, and $\gamma_{ij}$ the spatial metric. Under the assumption of spherical symmetry the 1+1 line element can be reduced to
\begin{equation}
 ds^2 = e^{4\chi } (a(t,r)dt^2+ r^2\,b(t,r)  d\Omega^2) \ ,
\end{equation}
with $d\Omega^2 = \sin^2\theta d\varphi^2+d\theta^2$ being the solid angle element
and $a(t,r)$ and $b(t,r)$ two non-vanishing metric functions. Moreover, $\chi$ is related to
the conformal factor $e^\chi = (\gamma/\hat \gamma)^{1/12}$, with $\gamma$
and $\hat\gamma$ being the determinants of the physical and conformal 3-metrics, respectively. They are conformally related by $\gamma_{ij} = e^{4 \chi} \hat
\gamma_{ij}$.

We adopt the Baumgarte-Shapiro-Shibata-Nakamura (BSSN) formalism
of Einstein's equations~\cite{Baumgarte98,Shibata95} (see
\cite{Alcubierre:2010is,Montero:2012yr} for further details). The Hamiltonian and momentum constraints are given by the two following equations: 
\begin{eqnarray}
\mathcal{H}&\equiv& R-(A^{2}_{a}+2A_{b}^{2})+\frac{2}{3}K^{2}-16\pi 
\mathcal{E}=0,\label{hamiltonian}\\
\mathcal{M}_{r}&\equiv&\partial_{r}A_{a}-\frac{2}{3}\partial_{r}K+6A_{a}\partial_{r}\chi\nonumber\\
&+&(A_{a}-A_{b})\biggl(\frac{2}{r}+\frac{\partial_{r}b}{b}\biggl)-8\pi j_{r}=0,\label{momentum}
\end{eqnarray}
where $K$ is the trace of the extrinsic
curvature $K_{ij}$, and $A_{a}$ and $A_{b}$ are the contraction of the traceless part of the conformal extrinsic curvature, i.e. 
$A_{a}\equiv\hat A^{r}_{r},\, A_{b}\equiv\hat A^{\theta}_{\theta}$.
In addition to the evolution fields, there are two more variables to evolve: the lapse function $\alpha$, and the shift vector $\beta^{i}$. We 
use  the {``non-advective 1+log''} condition~\cite{Bona:1997prd} for the 
lapse, and a variation of the {``Gamma-driver''} condition for the shift vector 
\cite{Alcubierre:2003ab,Alcubierre:2010is}. 

The matter fields appearing on the right-hand-side of the evolution
equations for the gravitational fields (see~\cite{Alcubierre:2010is,Montero:2012yr}), $\mathcal{E}$,
$j_{i}$, and $S_{ij}$, corresponding to the energy density, momentum density, and spatial stress-energy tensor as measured by the Eulerian observers, include the contribution of both the fluid and the scalar field, 
i.e.~$\mathcal{E}=\mathcal{E}^{\rm F}+\mathcal{E}^{\rm{SF}}$, etc. The matter source terms for the 
fluid given by the stress-energy tensor of Eq.~\eqref{matter} read
\begin{eqnarray}
\mathcal{E}^{\rm F}&\equiv& n^{\alpha}n^{\beta}T^{\rm F}_{\alpha\beta},\label{eq17}\\
j_{r}^{\rm F}&\equiv&-\gamma^{\alpha}_{r}n^{\beta}T^{\rm F}_{\alpha\beta},\label{eq18}\\
S_{a}^{\rm F}&\equiv&(T^{\rm F})^{r}_{r},\label{eq19a}\\
S_{b}^{\rm F}&\equiv&(T^{\rm F})^{\theta}_{\theta}\label{eq19b},
\end{eqnarray}

and the matter source terms  for the scalar field are given by
\begin{eqnarray}
\mathcal{E}^{\rm{SF}}&\equiv&n^{\alpha}n^{\beta}T^{\rm{SF}}_{\alpha\beta}=\frac{1}{2}\biggl(\Pi^{2}
+\frac{\Psi^{2}}{ae^{4\chi}}\biggl)+\frac{1}{2}\mu^{2}\Phi^{2} \label{eq:rho}\ ,\\
j^{\rm{SF}}_{r}&\equiv&-\gamma^{\alpha}_{r}n^{\beta}T^{\rm{SF}}_{\alpha\beta}=-g_{rr}\Pi\Psi ,\\
S_{a}^{\rm{SF}}&\equiv&(T^{\rm{SF}})^{r}_{r}=\frac{1}{2}\biggl(\Pi^{2}+\frac{\Psi^{2}}{ae^{4\chi}}
\biggl)-\frac{1}{2}\mu^{2}\Phi^{2} \ ,\\
S_{b}^{\rm{SF}}&\equiv&(T^{\rm{SF}})^{\theta}_{\theta}=\frac{1}{2}\biggl(\Pi^{2}-\frac{\Psi^{2}}{ae^
{4\chi}}\biggl)-\frac{1}{2}\mu^{2}\Phi^{2} \ ,
\end{eqnarray}
where 
\begin{eqnarray}
\Psi &=& \partial_{r}\Phi,\\
\Pi&=& -\frac{1}{\alpha}(\partial_{t}\Phi-\beta^{r}\partial_{r}\Psi),
\end{eqnarray}

are two auxiliary first-order variables.

%%%%%%%%%%%%%
\subsection{Initial data}
%%%%%%%%%%%%%

We start with a scalar field described as a Gaussian distribution of the form

\begin{equation}\label{eq:pulse}
 \Phi(r)=\Phi_0\,e^{-(r-r_0)^2/\lambda^2} \ ,
\end{equation}
where $\Phi_0$ is the initial amplitude of the pulse, $r_0$ the center of the Gaussian, and
$\lambda$ its width. The auxiliary first-order quantities are
initialized as follows
\begin{eqnarray}\label{gaussianID}
 \Pi(t=0,r)&=&0 \ , \\ 
 \Psi(t=0,r) &=& -2\frac{(r-r_0)}{\lambda^2}\Phi_0e^{-(r-r_0)^2/\lambda^2} \ .
 \label{eq:iderivatives}
\end{eqnarray}

We choose a conformally flat metric with $a=b=1$ together with a time symmetry condition 
$K_{ij}=0$. These choices and $\Pi(t=0,r) = 0$ satisfy trivially the momentum and the 
Hamiltonian 
constraints, given in Eq.~\eqref{hamiltonian}, and yield the following equation for the conformal factor 
$\psi=e^{\chi}$,
\begin{equation}
 \partial_{rr}\psi + \frac{2}{r}\partial_{r}\psi +2\pi\psi^5 \mathcal{E}=0 \ .
\end{equation}
To solve this equation for our white dwarf, the conformal factor can be 
written as 
\begin{equation}
 \psi = \psi^{\text{TOV}}+ u(r) \ , 
\end{equation}
where $\psi^{\text{TOV}}$ corresponds to the part factor associated with the stellar 
Tolman-Oppenheimer-Volkoff (TOV) model. By substituting this ansatz in the Hamiltonian constraint we obtain 
\begin{eqnarray}
 \partial_{rr}u(r) &+&\frac{2}{r}\partial_{r}u(r)+2\pi\psi^5 \mathcal{E}^{\text{SF}}\nonumber\\ 
&+&2\pi(\psi^{5}-(\psi^{\text{TOV}})^{5})\mathcal{E}^{\text{TOV}}= 0 \ .
\label{eq:diff_eq_u}
\end{eqnarray}

Given a distribution of the scalar field density $\mathcal{E}^{\text{SF}}$, we solve the ordinary 
differential equation~(\ref{eq:diff_eq_u}) making use of a fourth-order Runge-Kutta integrator, assuming that 
$u\rightarrow 0$ when
$r\rightarrow \infty$ and regularity at the origin. 

The fermion star is a white dwarf modeled by a polytropic equation of state:
\begin{equation}\label{polytrope}
p=\kappa\rho^{\Gamma}
\end{equation}

\begin{table}
\caption{Stellar Tolman-Oppenheimer-Volkoff white dwarf model: $\Gamma=5/3$ polytrope with $\kappa=112$. From left to 
right the columns indicate  the central rest-mass density, $\rho_c$,  the radius of the star, $R_{\rm{WD}}$, the gravitational mass of the star, $M_{\rm{WD}}$, and 
the compactness, $M_{\rm{WD}}/R_{\rm{WD}}$. We note that all these quantities are in natural units.}\label{tab:table1}
\begin{ruledtabular}
\begin{tabular}{cccc}
$\rho_c$&$R_{\rm{WD}}$&$M_{\rm{WD}}$&$M_{\rm{WD}}/R_{\rm{WD}}$\\
\hline
$2.503\times10^{-8}$&318.2&0.562&0.0017\\
%\hline
%\hline
\end{tabular}
\end{ruledtabular}
\end{table}

where $p$ is the pressure, $\kappa$ the polytropic constant, $\rho$ the rest-mass density, with $\Gamma=1+1/N$, and $N$ the polytropic index. A similar initial setup was considered in~\cite{di2020dynamical,di2021dynamical} with neutron stars described by a polytrope with $N=1$ ($\Gamma=2$). The polytropic index for average low-mass white dwarfs is set to $N=1.5$ ($\Gamma=5/3$) and we take $\kappa=112$. With these parameters, we build the TOV model described in Table~\ref{tab:table1}. The mass of the white dwarf is $M_{\rm{WD}}=0.562$. If we assume that such mass is given in solar masses ($M_{\rm{WD}}=0.562\,M_{\odot}$), then the radius of our star in kilometers would be $R_{\rm{WD}}=468.35$ km, smaller than the $10^{4}-$km average radius of white dwarfs. Hence, in terms of compactness, our model is an intermediate step between white dwarfs and neutron stars. We take this value in order to reduce the dynamical timescale of the white dwarf evolution and the extent of our computational domain, since this study is a proof of concept. BSs (and mixed stars) scale with the particle mass $\mu$, therefore the whole system could be re-scaled as explained in~\cite{di2020dynamical}.

%%%%%%%%%%%%%%%%%%%%%%%%%%%%%%%%%%%%%%%%%%%%%%%%%%%%
\section{Numerics} 
\label{sec:numerics}
%%%%%%%%%%%%%%%%%%%%%%%%%%%%%%%%%%%%%%%%%%%%%%%%%%%%

The time evolution of the spacetime variables, the scalar field, and the fluid is performed with the techniques we have extensively 
tested and used in previous works (see in particular~\cite{Montero:2012yr,Sanchis-Gual:2015,sanchis2015quasistationary,di2020dynamical}). The evolution equations are integrated and updated using the second-order PIRK method developed by \cite{Isabel:2012arx,cordero2014partially}. The derivatives in the spacetime evolution are computed using a 
fourth-order centered finite difference approximation on the logarithmic grid presented and described in~\cite{sanchis2015quasistationary}, except for advection terms for which we adopt a fourth-order upwind scheme.  We also use fourth-order 
Kreiss-Oliger dissipation to avoid high frequency noise appearing near the outer boundary.
Correspondingly, the hydrodynamics equations are solved using the Harten-Lax-van Leer-Einfeldt approximate Riemann solver in tandem with the second-order monotonized central reconstruction scheme~\cite{Montero:2012yr}. The minimum resolution is set to $\Delta r = 0.1$, with a Courant factor $\Delta t = 0.3 \Delta r$. The outer boundary is located at $r_{\rm{max}}=52505$. We run the simulations up to a final time of $t_{\rm{final}}=84000$.

%%%%%%%%%%%%%%%%%%%%%%%%%%%%%%%%%%%%%%%%%%%%%%%%%%%%
\section{Results} 
\label{sec:num_results}
%%%%%%%%%%%%%%%%%%%%%%%%%%%%%%%%%%%%%%%%%%%%%%%%%%%

\subsection{Numerical simulations}
We have evolved three different spherically symmetric initial scalar distributions given by Eqs.~(\ref{eq:pulse})-(\ref{eq:iderivatives}) around a white dwarf described by the polytropic equation of state given in Eq.~(\ref{polytrope}) (see Table~\ref{tab:table1} for details). In particular we take Eq.~(\ref{eq:pulse}) with different amplitudes $\Phi_{0}$ (see Table~\ref{tab:table1b} for details). In all cases we keep $r_{0}=0$ and $\lambda=90$: Configurations 1, 2, and 3 have initial scalar field masses of $M_{\rm{SF}}=0.727\simeq1.3\,M_{\rm{WD}}$,  $0.183\simeq M_{\rm{WD}}/3$, and $0.051\simeq M_{\rm{WD}}/10$, respectively. %We also vary the particle mass by choosing $\mu=\lbrace0.1,0.5,1\rbrace$ while keeping the same initial masses, to explore its effect on the fermion star.
We denote $M_{\rm{SF}}$ as the initial scalar field cloud and $M_{\rm{BS}}$ as the final mass of the BS. 

Besides the three initial scalar field amplitudes, we have also chosen three different values of the scalar field particle mass $\mu=\lbrace1.0, 0.5, 0.1\rbrace$, to study its effect on the formation of mixed objects and on the white dwarf. The evolutions lead to the dynamical formation of two different types of mixed star: a white dwarf with a BS core or a white dwarf within a dilute BS. In either case, the scalar cloud suffers an incomplete gravitational collapse due to its self-gravity and thus condenses into a BS. The excess energy and scalar field are radiated through the gravitational cooling mechanism~\cite{seidel1994formation,di2018dynamical,di2020dynamical,di2021dynamical,Sanchis-Gual:2021phr}. 

\begin{figure*}[t!]
\begin{tabular}{ p{0.32\linewidth} p{0.32\linewidth}  p{0.32\linewidth} }
\centering Configuration 1, $\mu=1.0$&\centering Configuration 1, $\mu=0.5$ &\centering Configuration 1, $\mu=0.1$
\end{tabular}
\centering\\
\includegraphics[width=0.328\linewidth]{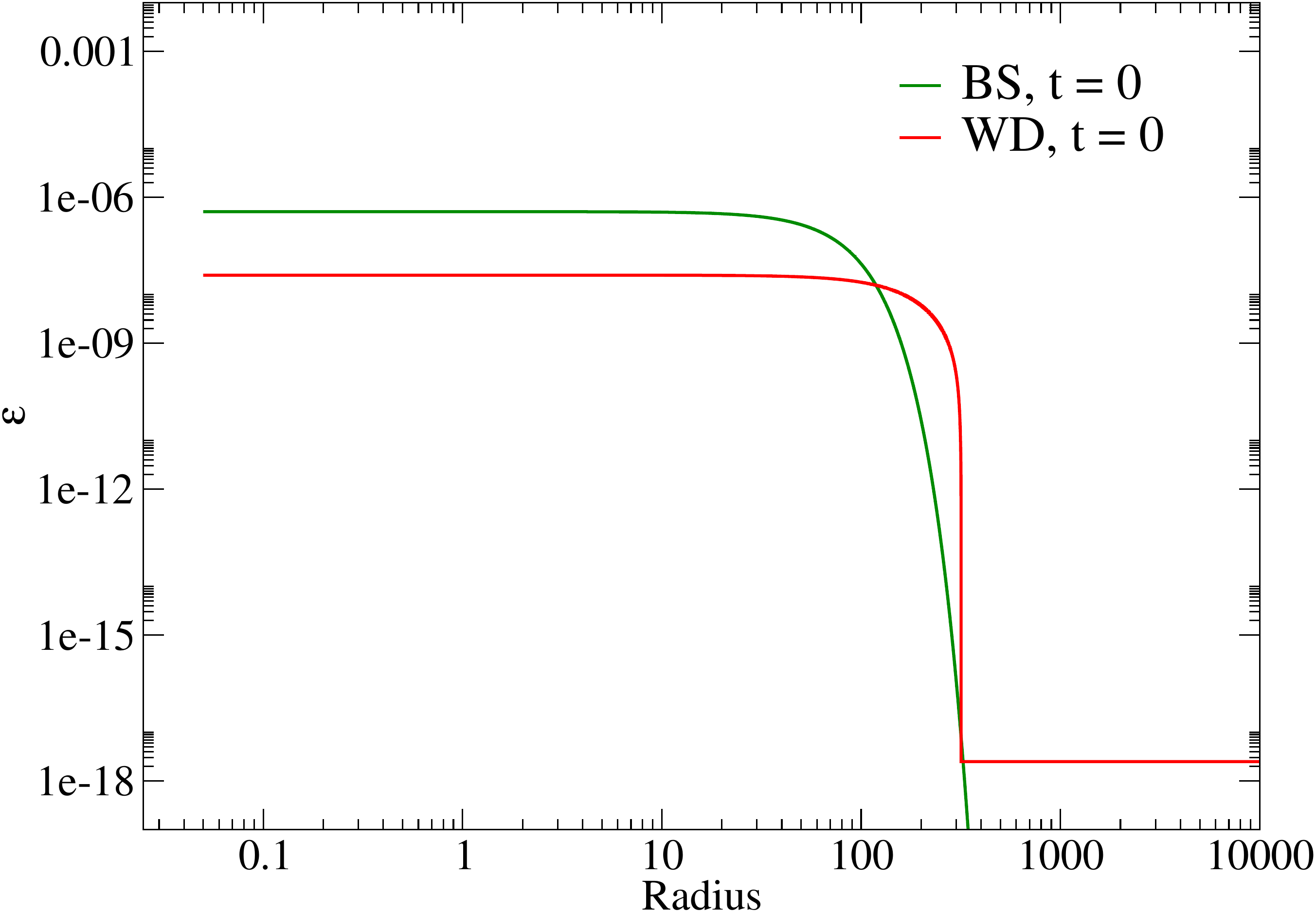}
\includegraphics[width=0.328\linewidth]{t0_Density_wd7.pdf}
\includegraphics[width=0.328\linewidth]{t0_Density_wd7.pdf}\\
\includegraphics[width=0.325\linewidth]{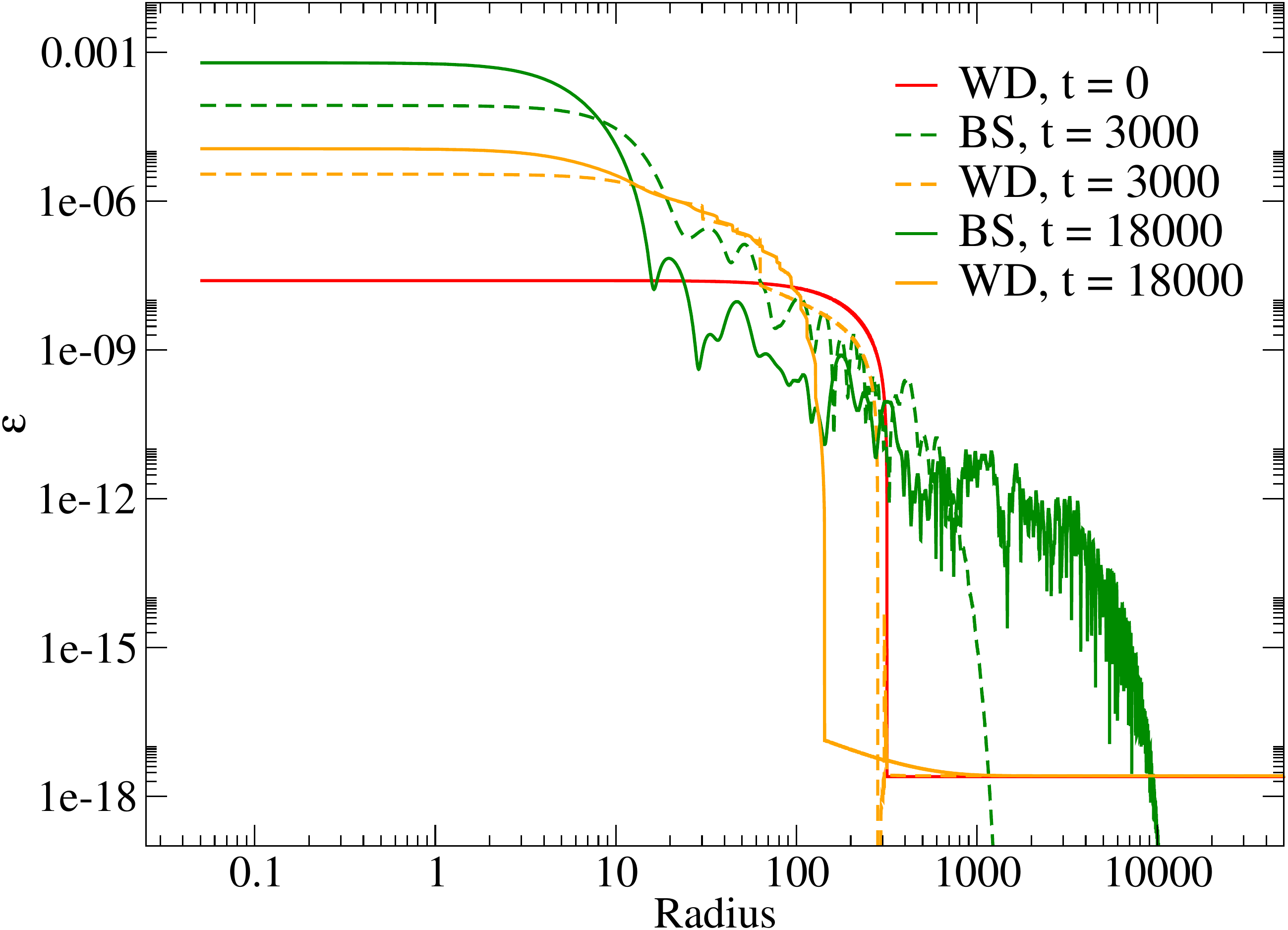}
\includegraphics[width=0.325\linewidth]{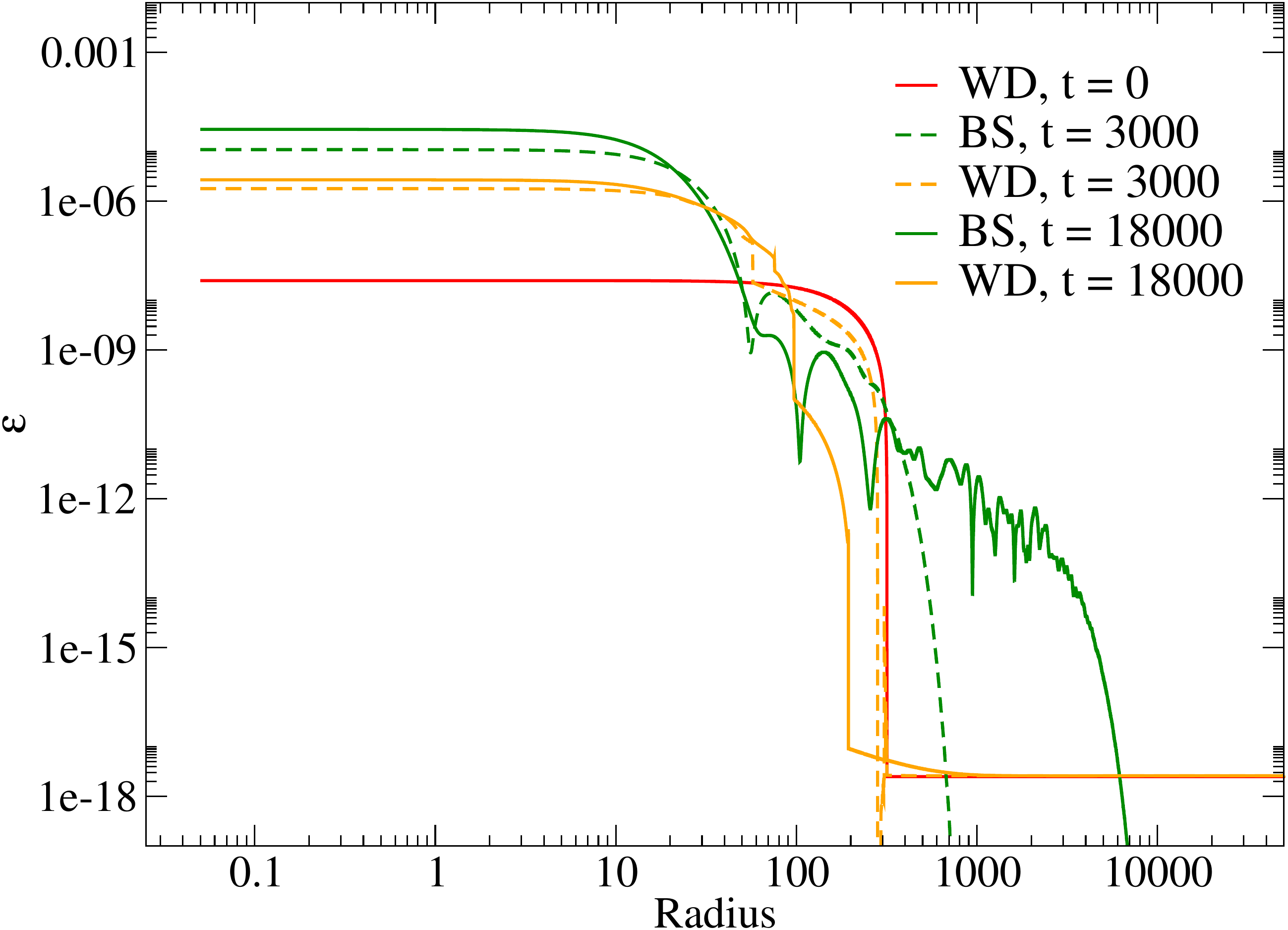}
\includegraphics[width=0.325\linewidth]{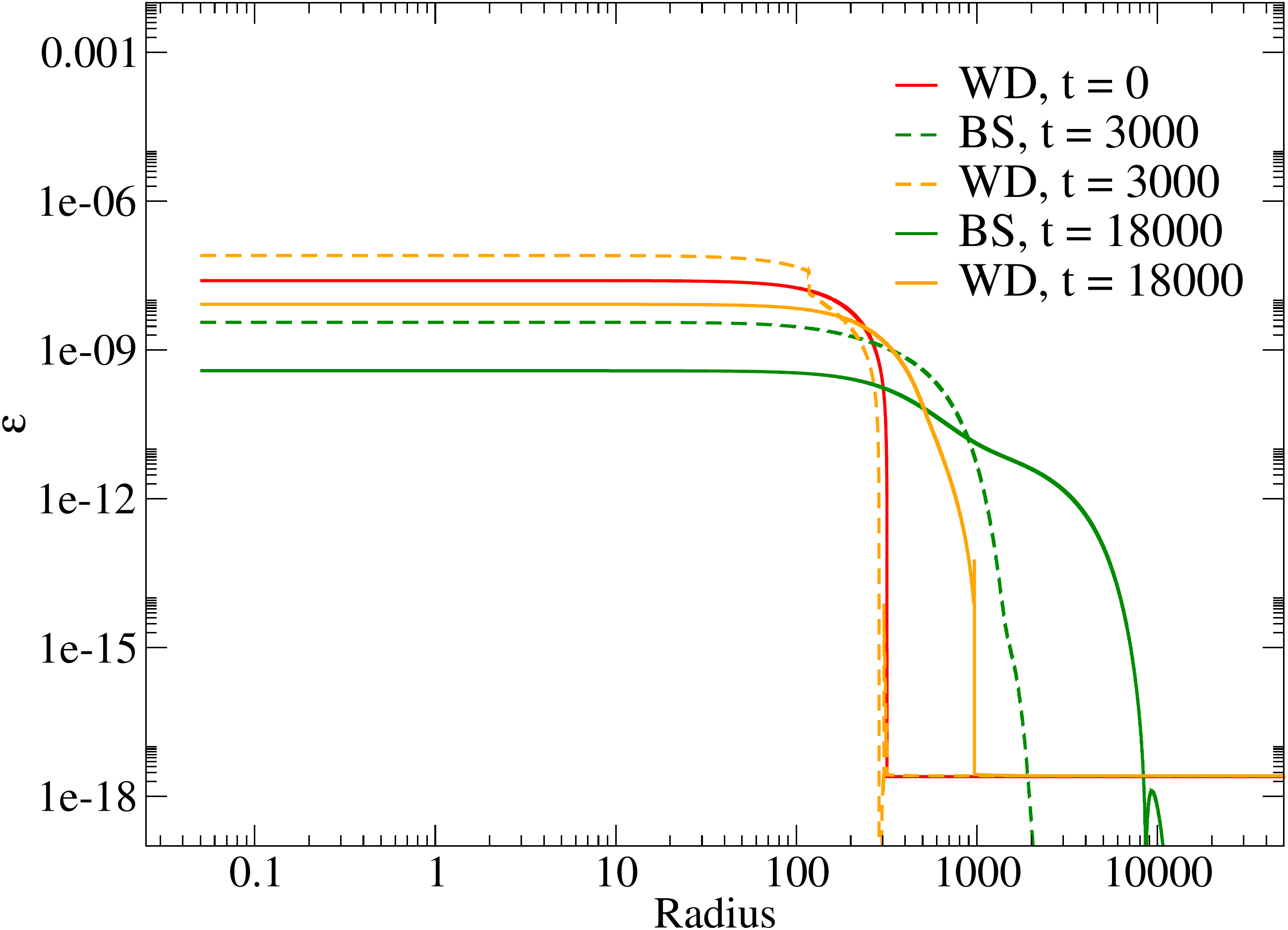}\\
\includegraphics[width=0.325\linewidth]{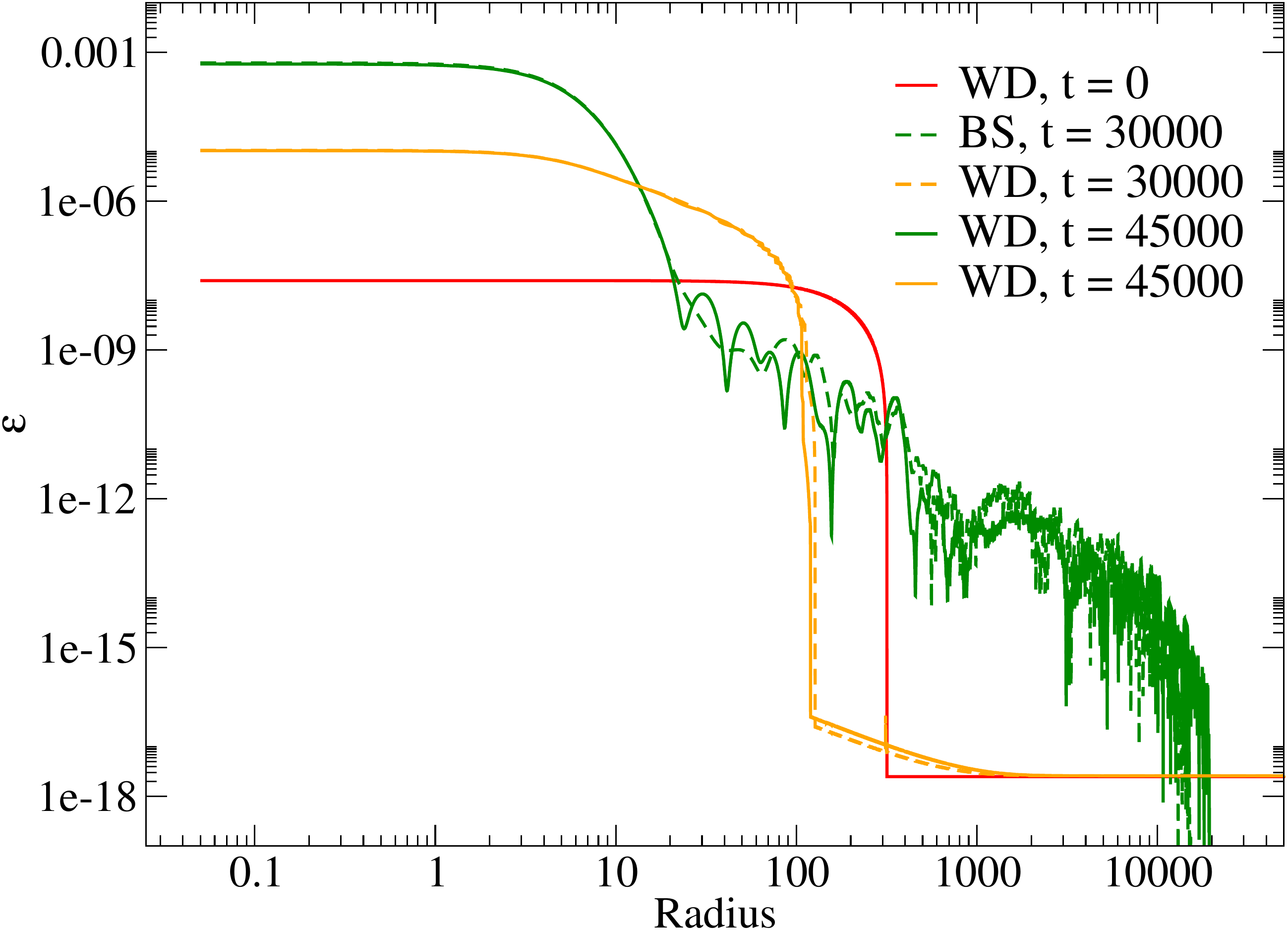}
\includegraphics[width=0.325\linewidth]{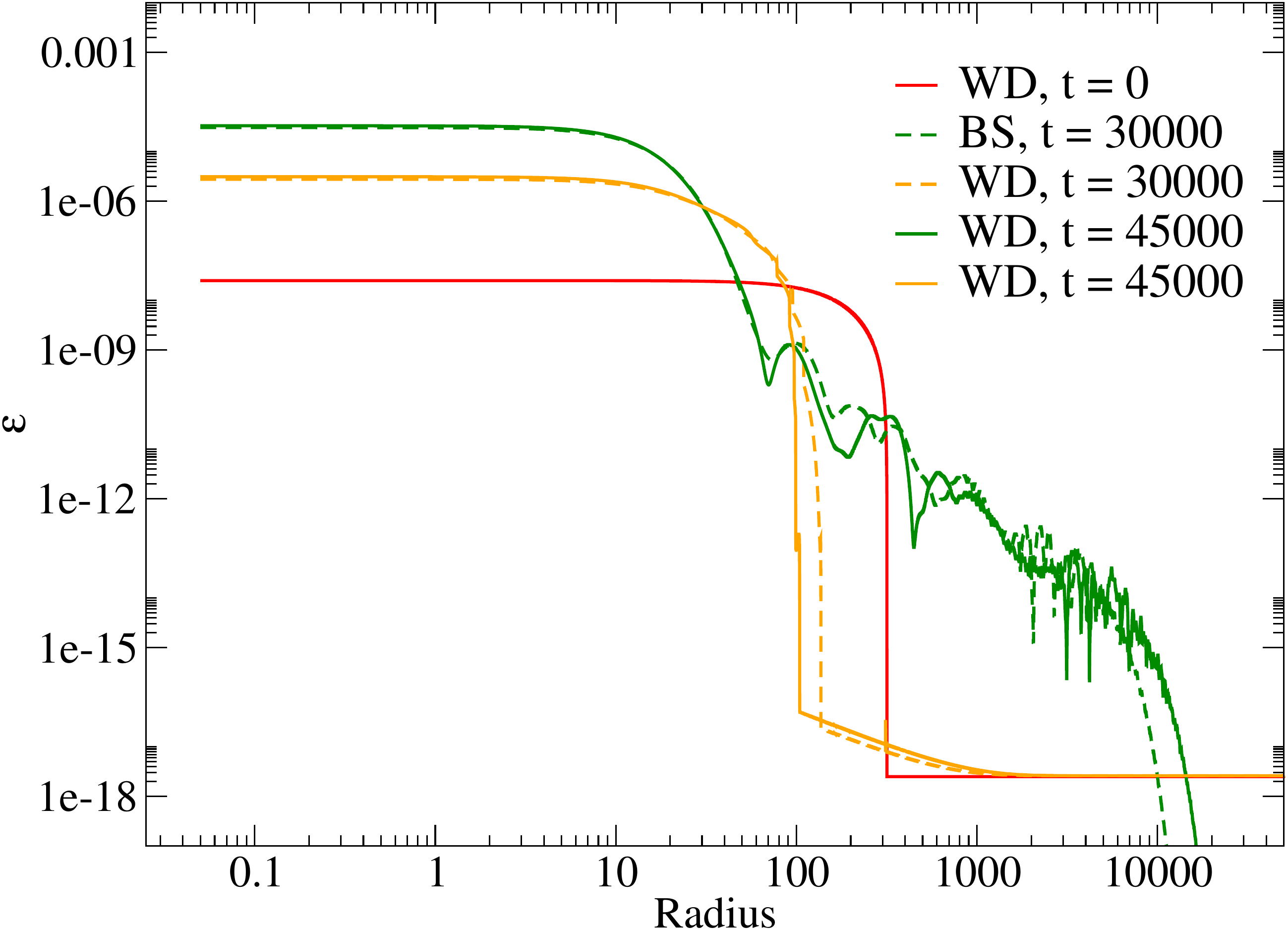}
\includegraphics[width=0.325\linewidth]{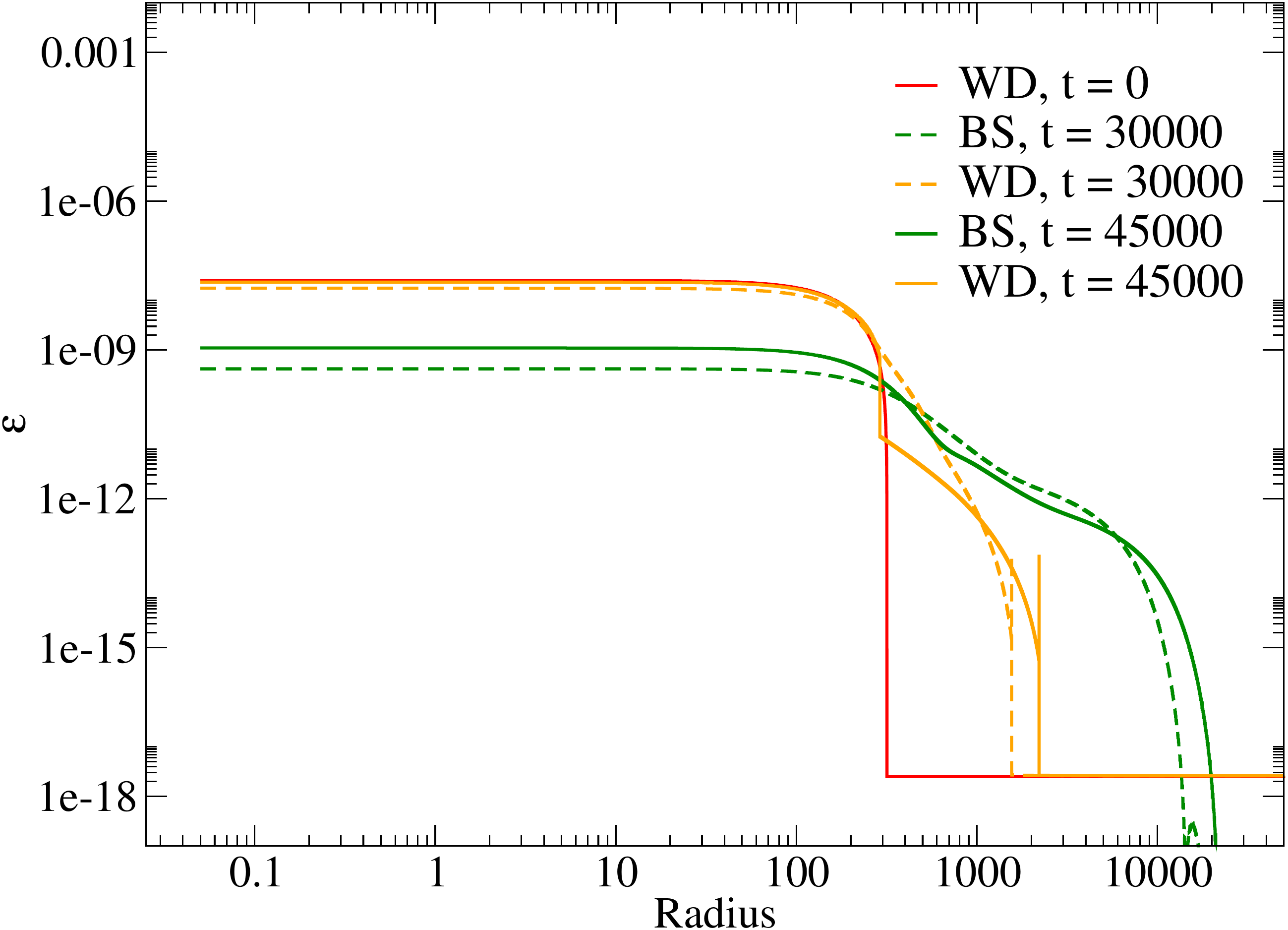}\\
\includegraphics[width=0.325\linewidth]{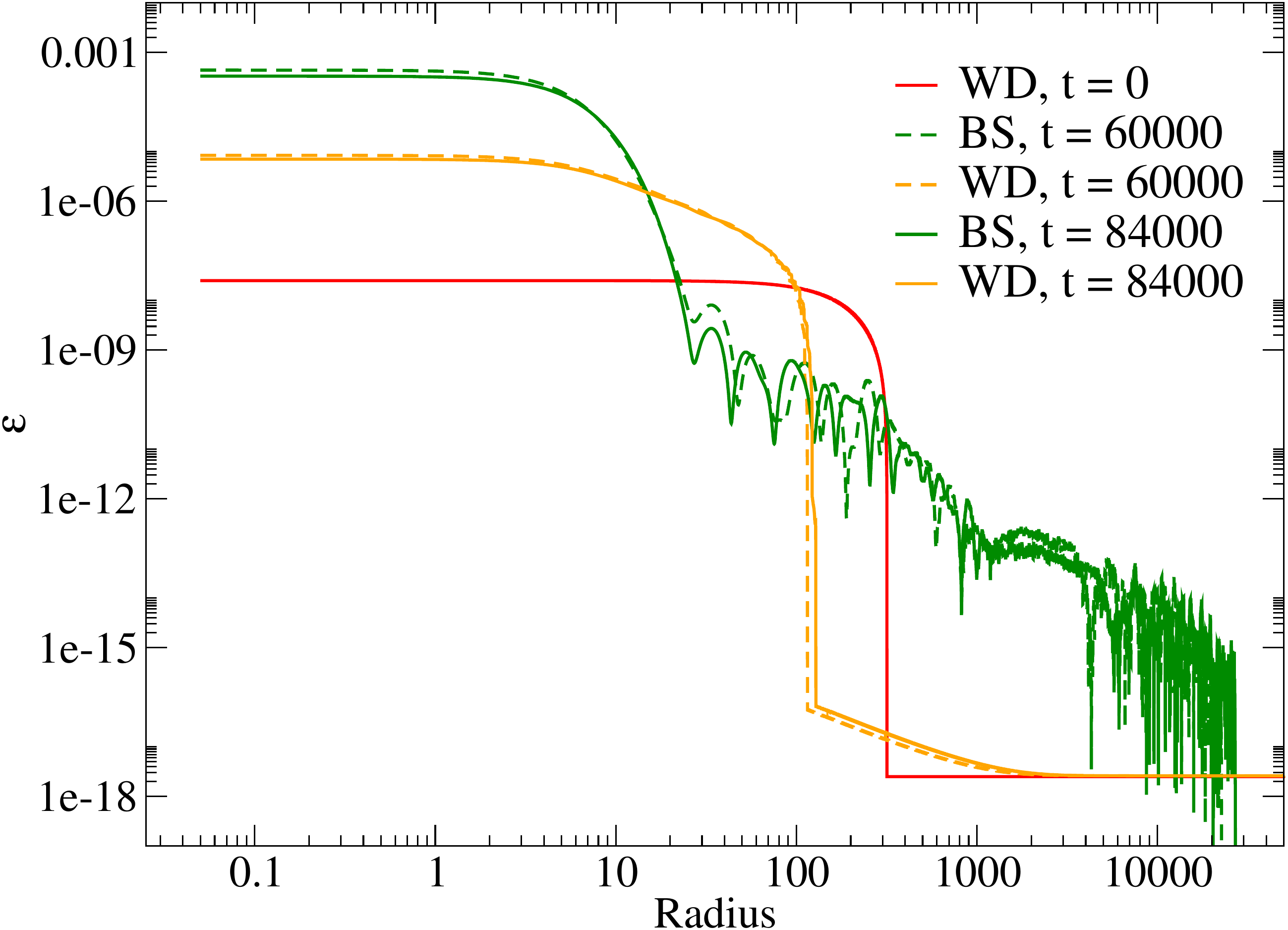}
\includegraphics[width=0.325\linewidth]{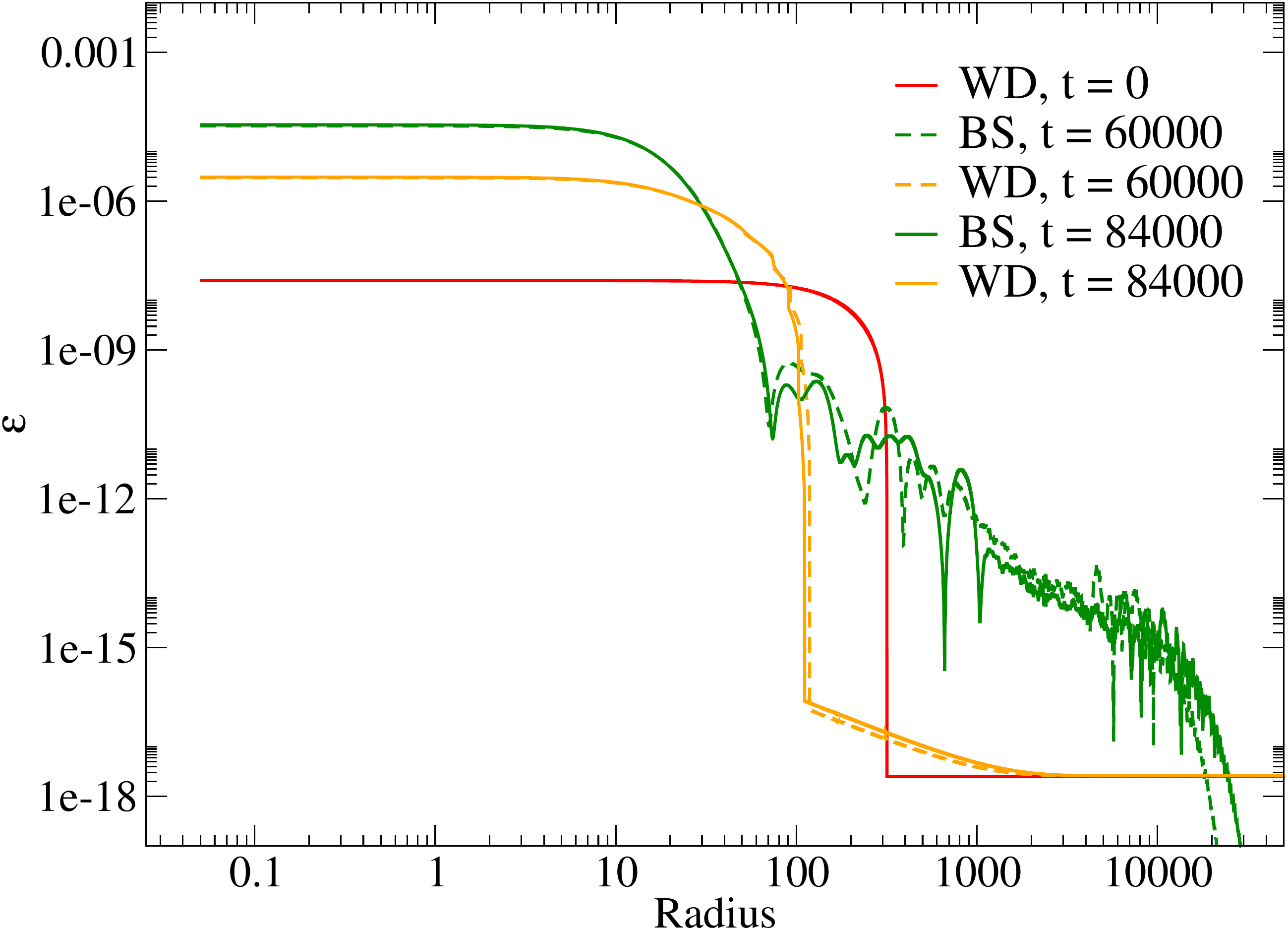}
\includegraphics[width=0.325\linewidth]{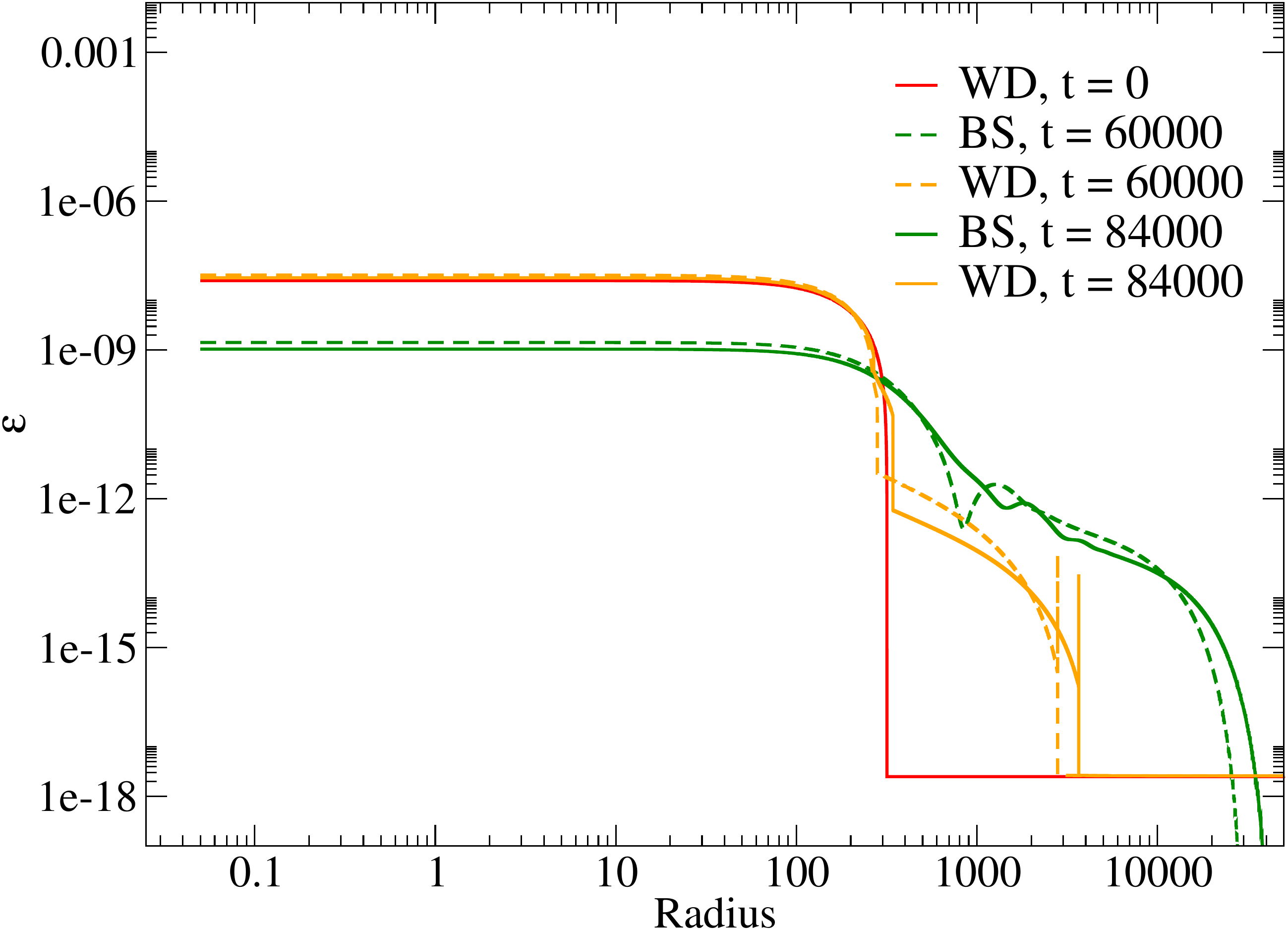}
\caption{Time evolution of the radial profiles of the scalar field (green lines) and white dwarf (orange lines) energy densities for Configuration 1 with $M_{\rm{SF}}=0.727$, and three different values of the particle mass $\mu=\lbrace1.0,0.5,0.1\rbrace$. The solid red line corresponds to the white dwarf energy density at $t=0$. Time runs top to bottom.}
\label{fig1}
\end{figure*}

In~\cite{di2020dynamical,di2021dynamical,valdez2013dynamical,valdez2020fermion}, the mixed fermion-boson star system scales with the scalar field particle mass, in the same way isolated BSs do. However, this means that fermion stars also scale with $\mu$ and could have non-astrophysical masses. Here we will assume that our fermion stars are analogous to astrophysical white dwarfs, with stellar masses up to the Chandrasekhar limit mass $M^{\rm{max}}_{\rm{WD}}\simeq1.4\,M_{\odot}$. This choice breaks the scaling and fixes the BS mass (and the boson particle mass $\mu$) relative to the white dwarf mass $M_{\rm{WD}}$. In this regard, let us recall that the maximum mass of isolated BSs without self-interactions, $M^{\rm{max}}_{\rm{BS}}$, is proportional to the inverse of the particle mass $\mu$~\cite{liebling2017dynamical}:
\begin{equation}
M^{\rm{max}}_{\rm{BS}}=0.633\,\frac{\displaystyle M^2_{\rm{Planck}}}{\displaystyle\mu}=0.633\,M_{\odot}\,\frac{1.34\times10^{-10}}{\mu\,[\rm{eV}]}.
\end{equation}
where $M_{\rm{Planck}}=\sqrt{\hbar c/G}$ is the Planck mass (which is one in our units).  

As mentioned above we consider that the fermion star is a white dwarf with mass $M_{\rm{WD}} =0.562\,M_{\odot}$. Taking $\mu=1.0$ in natural units ($c=G=\hbar=1$) the boson particle mass corresponds to $1.34\times10^{-10}$ eV and therefore, BSs have a maximum mass of $0.633\,M_{\odot}$. The other two choices of $\mu=\lbrace0.5, 0.1\rbrace$ are $6.70\times10^{-11}$ and $1.34\times10^{-11}$ eV, and correspond to $M^{\rm{max}}_{\rm{BS}}=1.266$ and 6.33 $M_{\odot}$, respectively. Since the scalar cloud has the same initial mass for each value of $\mu$, the setup leads to the formation of BSs with different masses, compactness, and radius. From now on we will use $M_{\odot}=1$. % Therefore, we have investigated the effect of the scalar field on the white dwarf depending on the compactness, scalar mass, and radius of the final BS.

\begin{table}
\caption{From left to 
right the columns indicate the configuration, the particle mass, $\mu$,  and the initial scalar field masses, $M_{\rm{SF}}$.}\label{tab:table1b}
\begin{ruledtabular}
\begin{tabular}{ccc}
Configuration&$\mu$&$M_{\rm{SF}}$\\
%\hline
%WD&-&0&0&318.2&-\\
\hline
1& (1, 0.5, 0.1) &0.727\\
%1&0.5&0.727\\
%1&0.1&0.727\\
\hline
2&(1, 0.5, 0.1)&0.183\\
%2&0.5&0.183\\
%2&0.1&0.183\\
\hline
3&(1, 0.5, 0.1) &0.051\\
%3&0.5&0.051\\
%3&0.1&0.051\\
%\hline
\end{tabular}
\end{ruledtabular}
\end{table}

\begin{figure*}[t!]
\begin{tabular}{ p{0.32\linewidth} p{0.32\linewidth}  p{0.32\linewidth} }
\centering Configuration 2, $\mu=1.0$&\centering Configuration 2, $\mu=0.5$ &\centering Configuration 2, $\mu=0.1$
\end{tabular}
\centering
\includegraphics[width=0.328\linewidth]{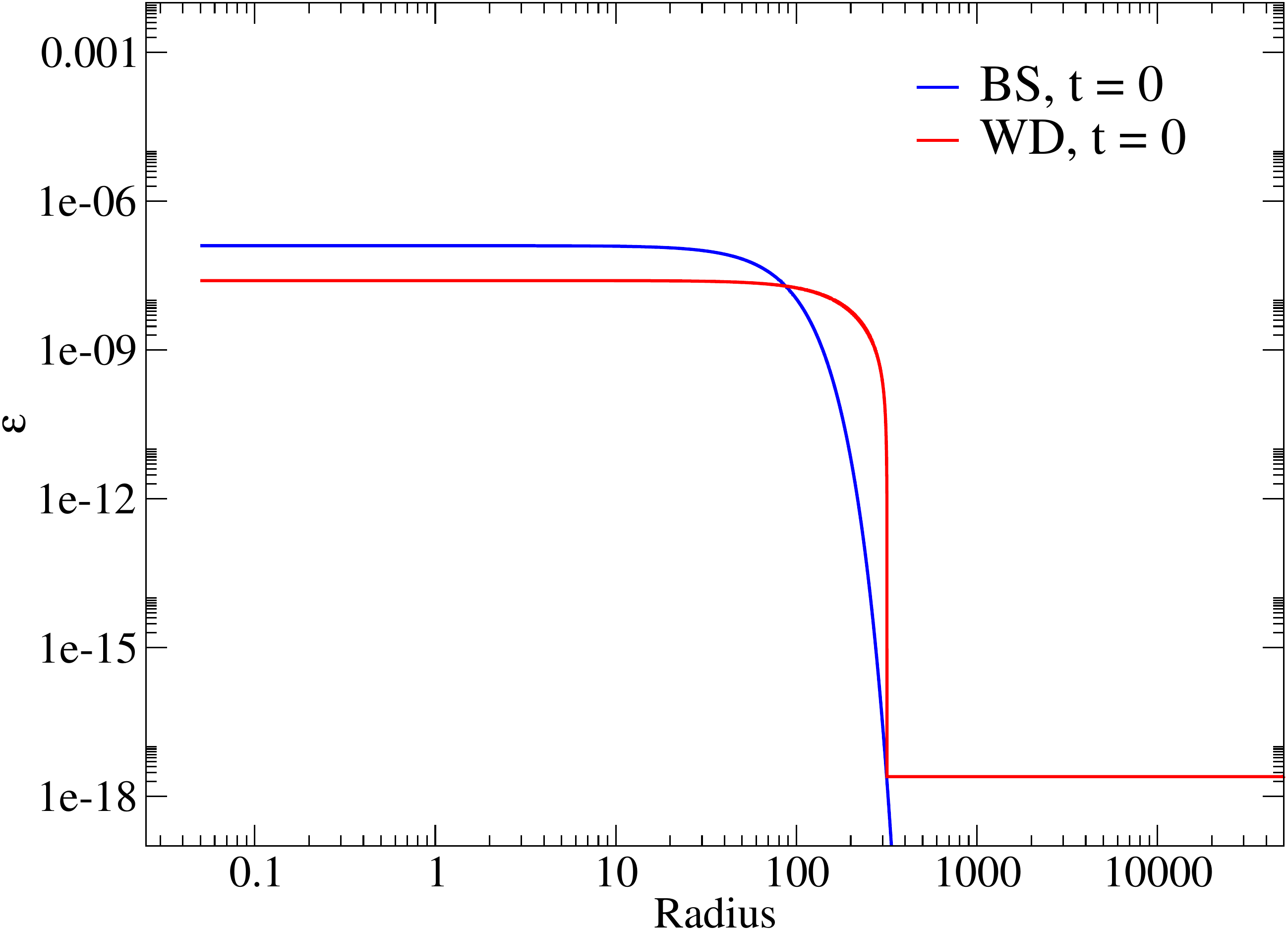}
\includegraphics[width=0.328\linewidth]{t0_Density_wd9.pdf}
\includegraphics[width=0.328\linewidth]{t0_Density_wd9.pdf}\\
\includegraphics[width=0.325\linewidth]{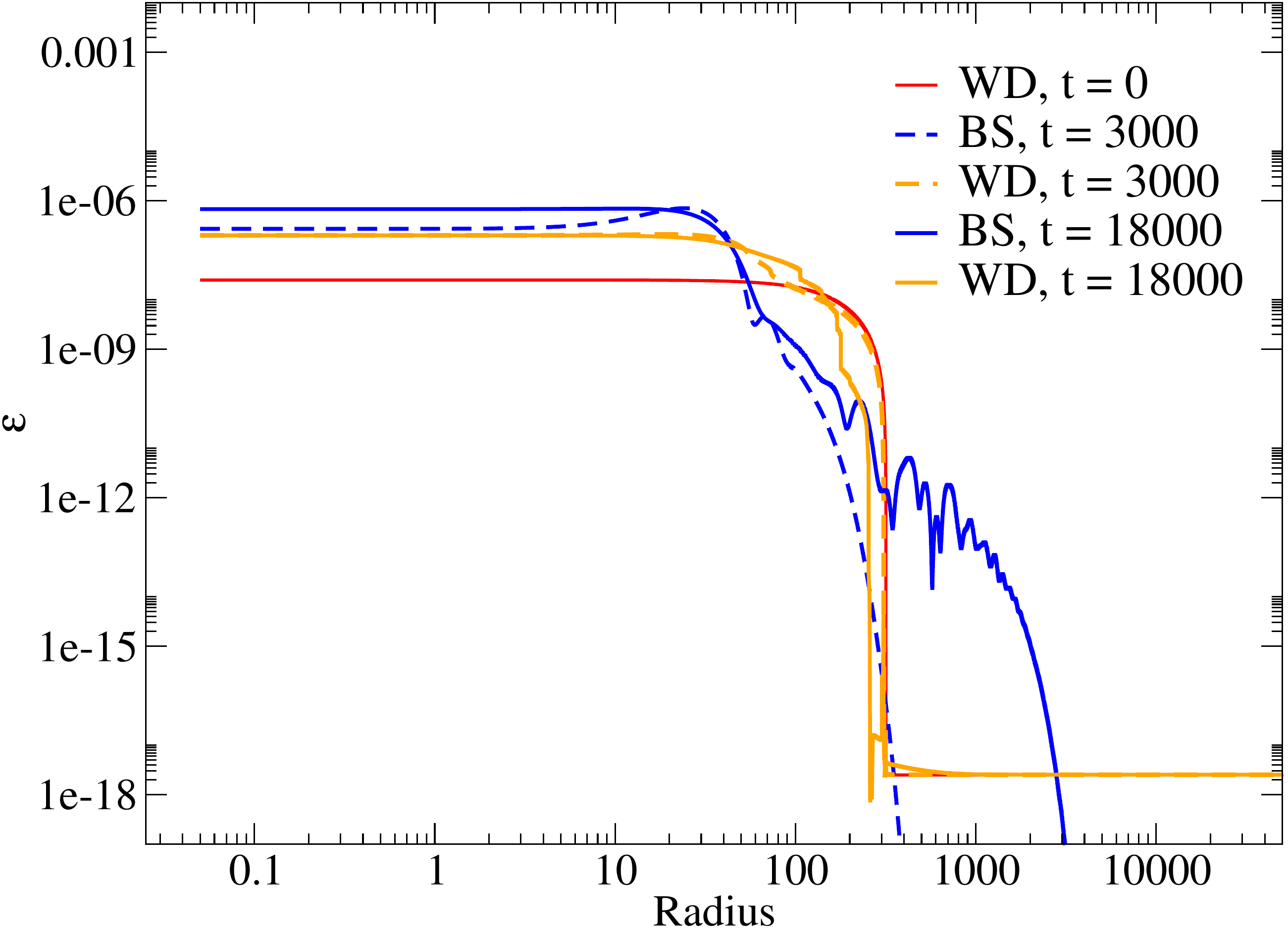}
\includegraphics[width=0.325\linewidth]{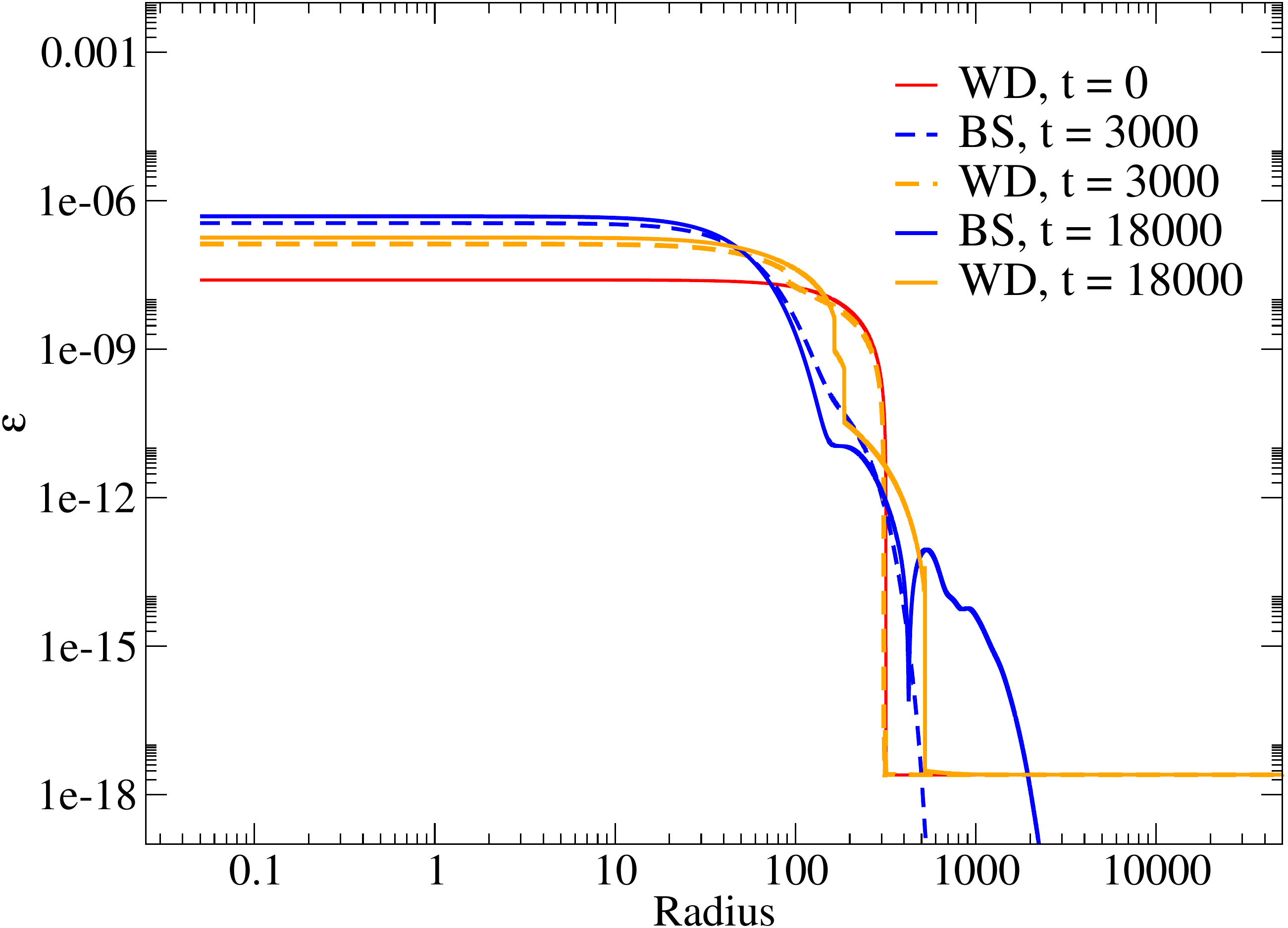}
\includegraphics[width=0.325\linewidth]{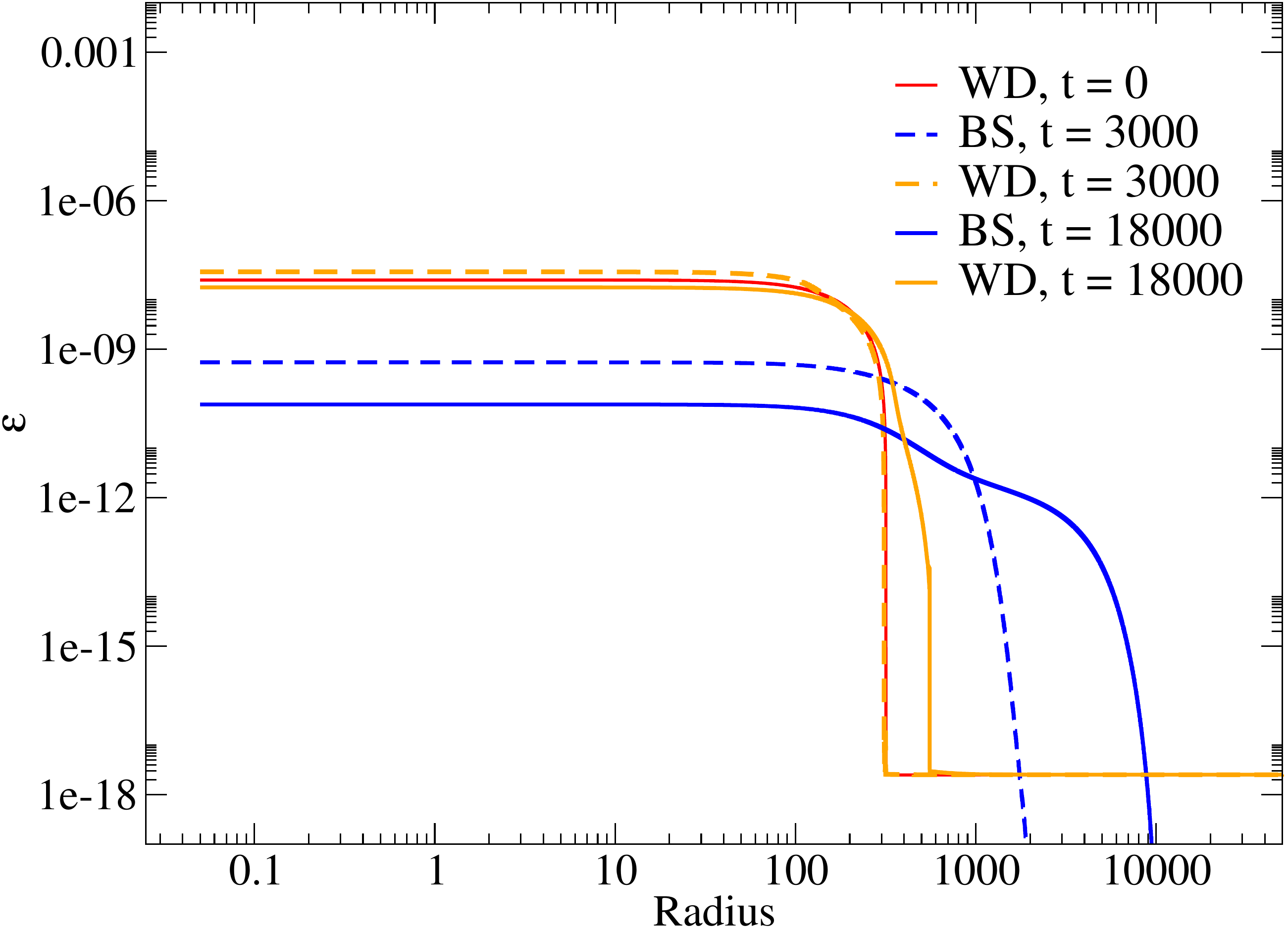}\\
\includegraphics[width=0.325\linewidth]{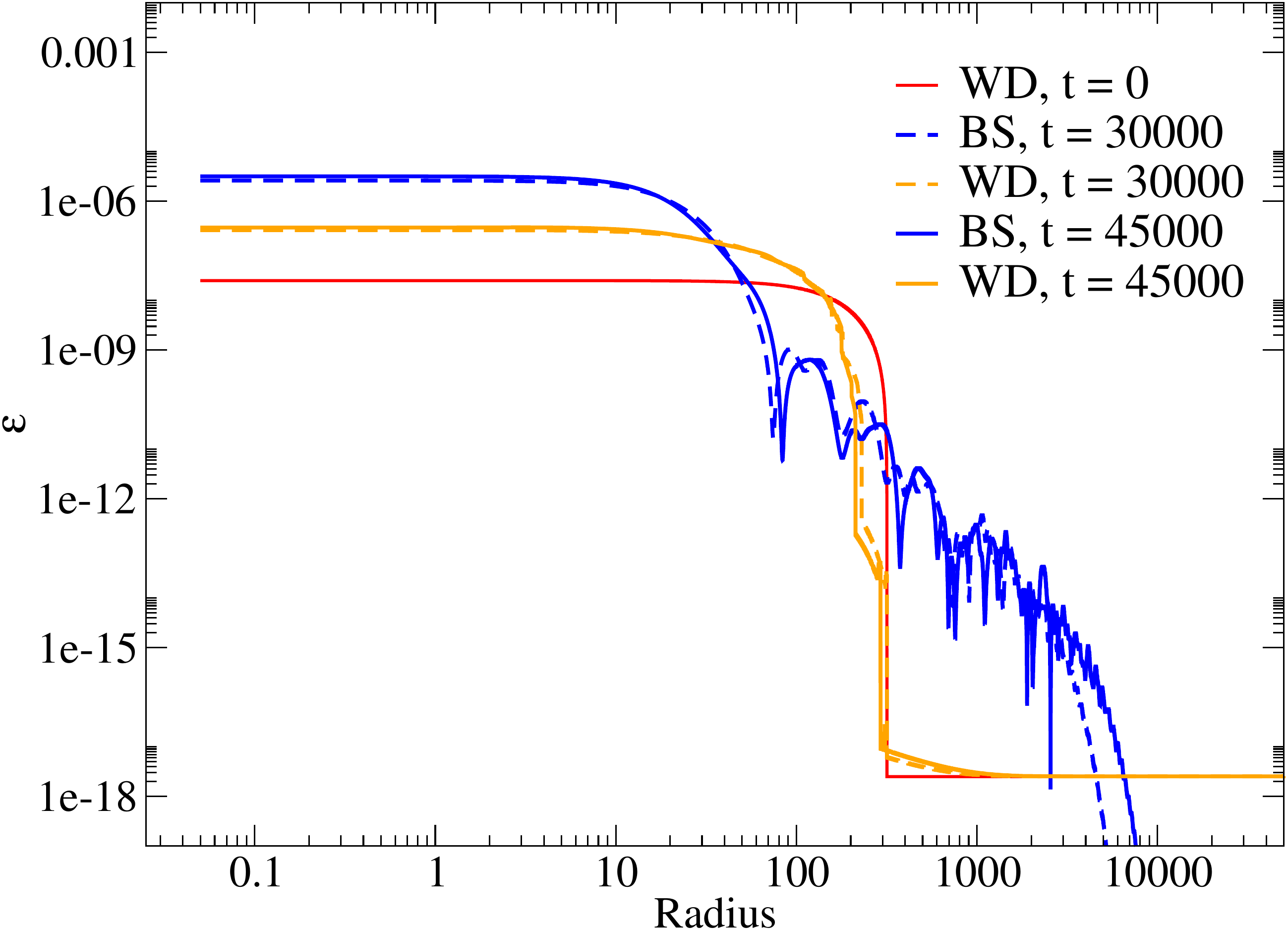}
\includegraphics[width=0.325\linewidth]{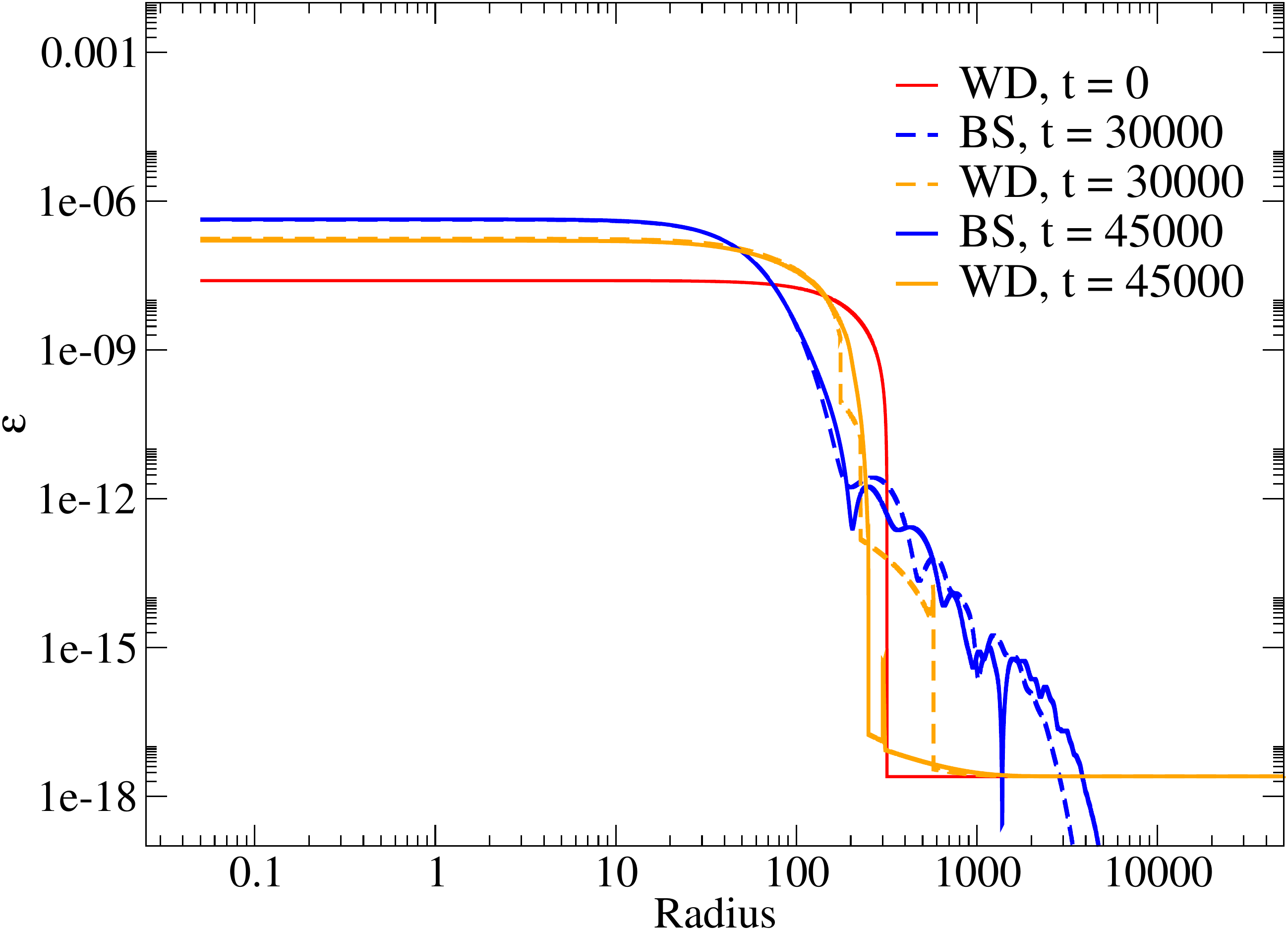}
\includegraphics[width=0.325\linewidth]{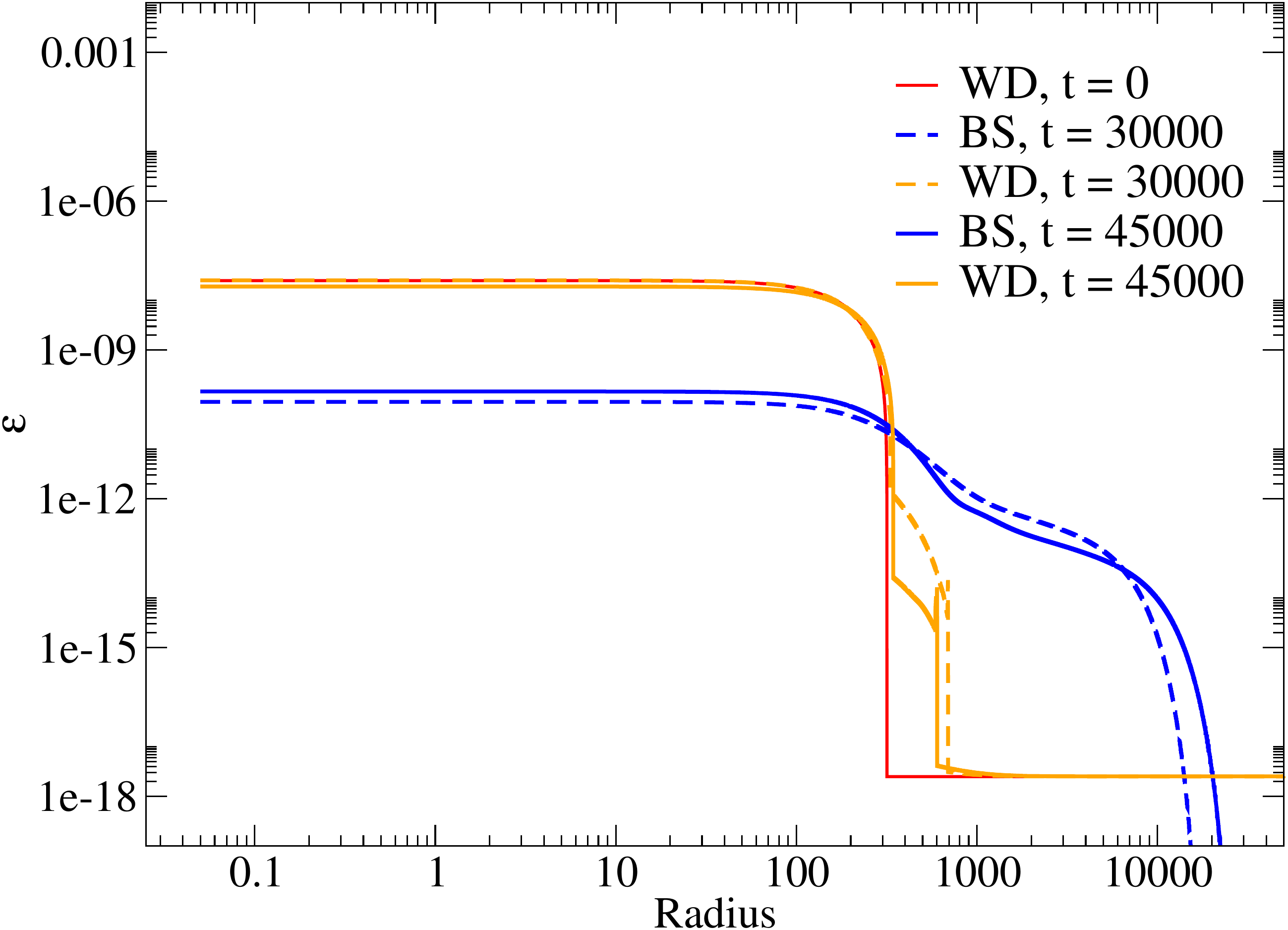}\\
\includegraphics[width=0.325\linewidth]{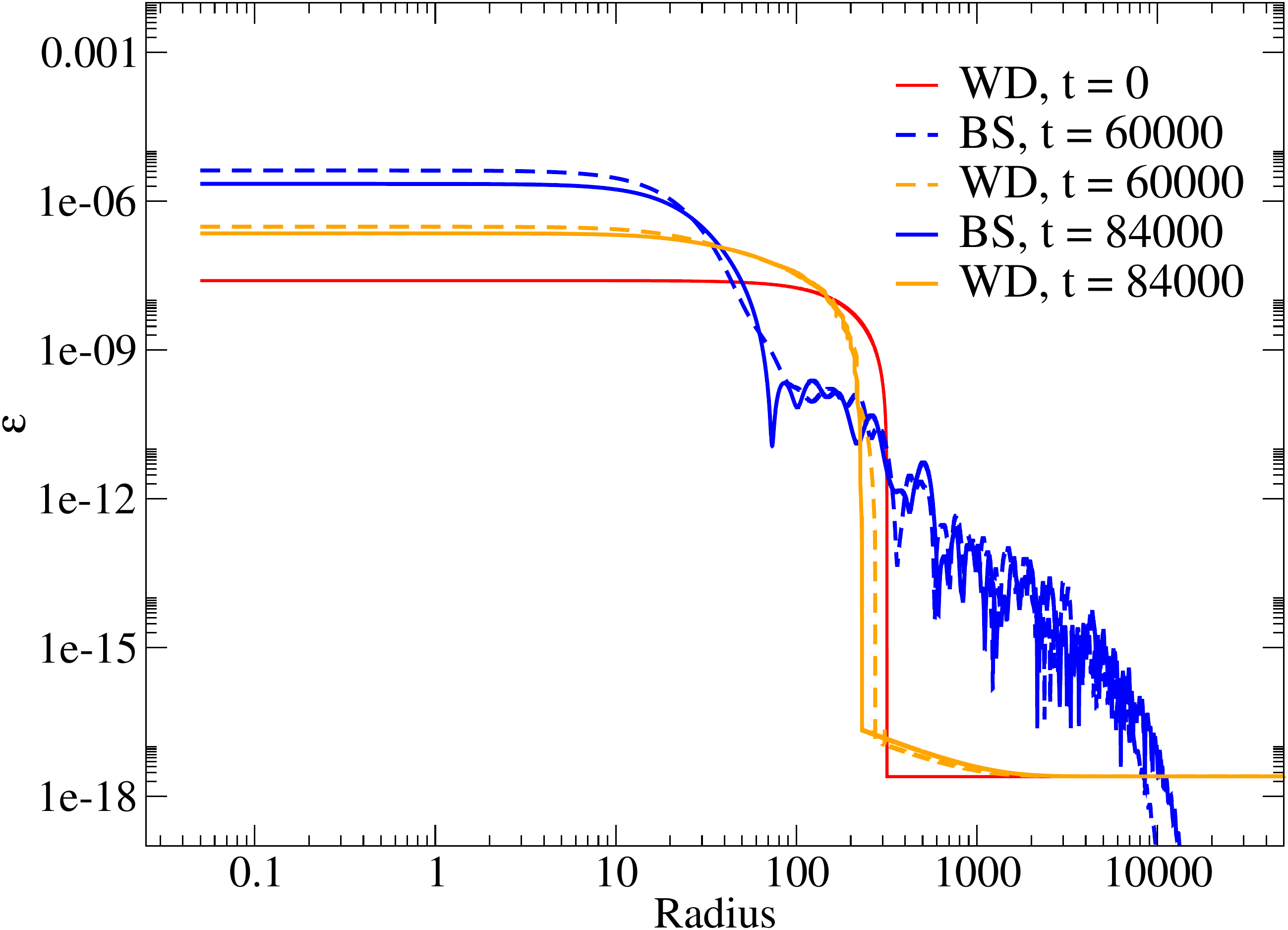}
\includegraphics[width=0.325\linewidth]{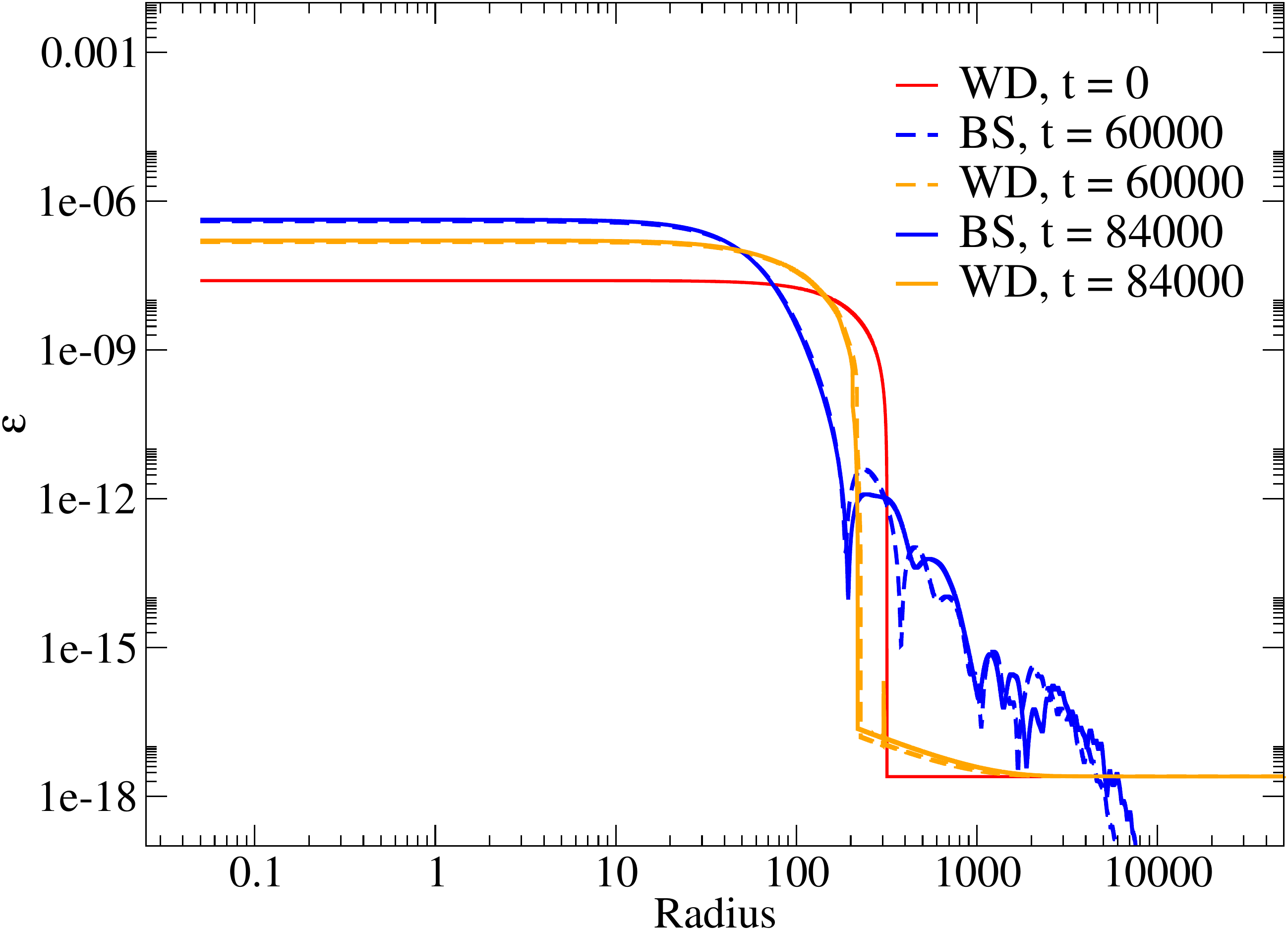}
\includegraphics[width=0.325\linewidth]{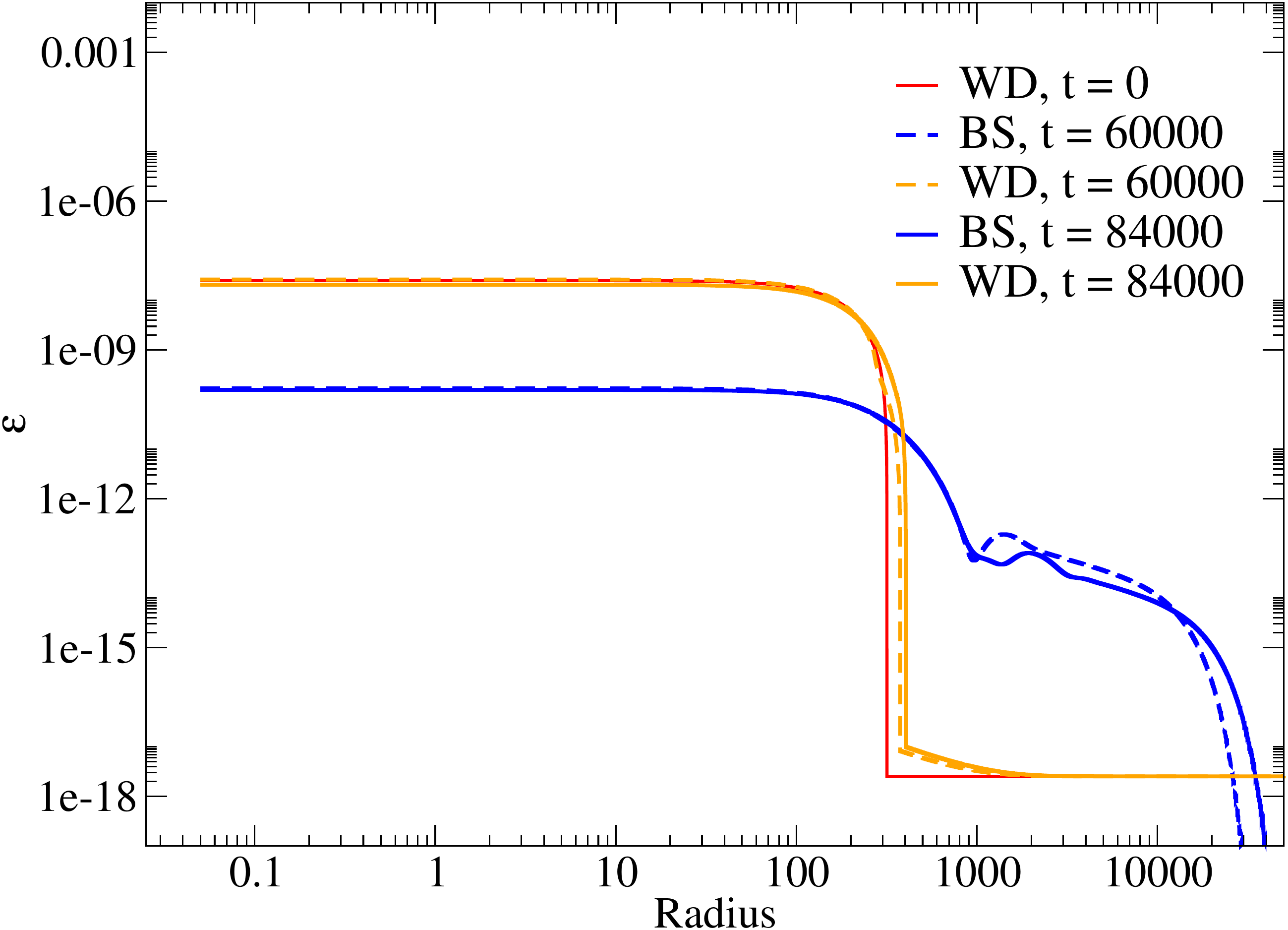}\\
\caption{Time evolution of the radial profiles of the scalar field (blue lines) and white dwarf (orange lines) energy densities for Configuration 2 with $M_{\rm{SF}}=0.183$, and three different values of the particle mass $\mu=\lbrace1,0.5,0.1\rbrace$. The solid red line corresponds to the white dwarf energy density at $t=0$. Time runs top to bottom.}
\label{fig2}
\end{figure*}

\begin{figure*}[t!]
\begin{tabular}{ p{0.32\linewidth} p{0.32\linewidth}  p{0.32\linewidth} }
\centering Configuration 3, $\mu=1.0$&\centering Configuration 3, $\mu=0.5$ &\centering Configuration 3, $\mu=0.1$
\end{tabular}
\centering
\includegraphics[width=0.328\linewidth]{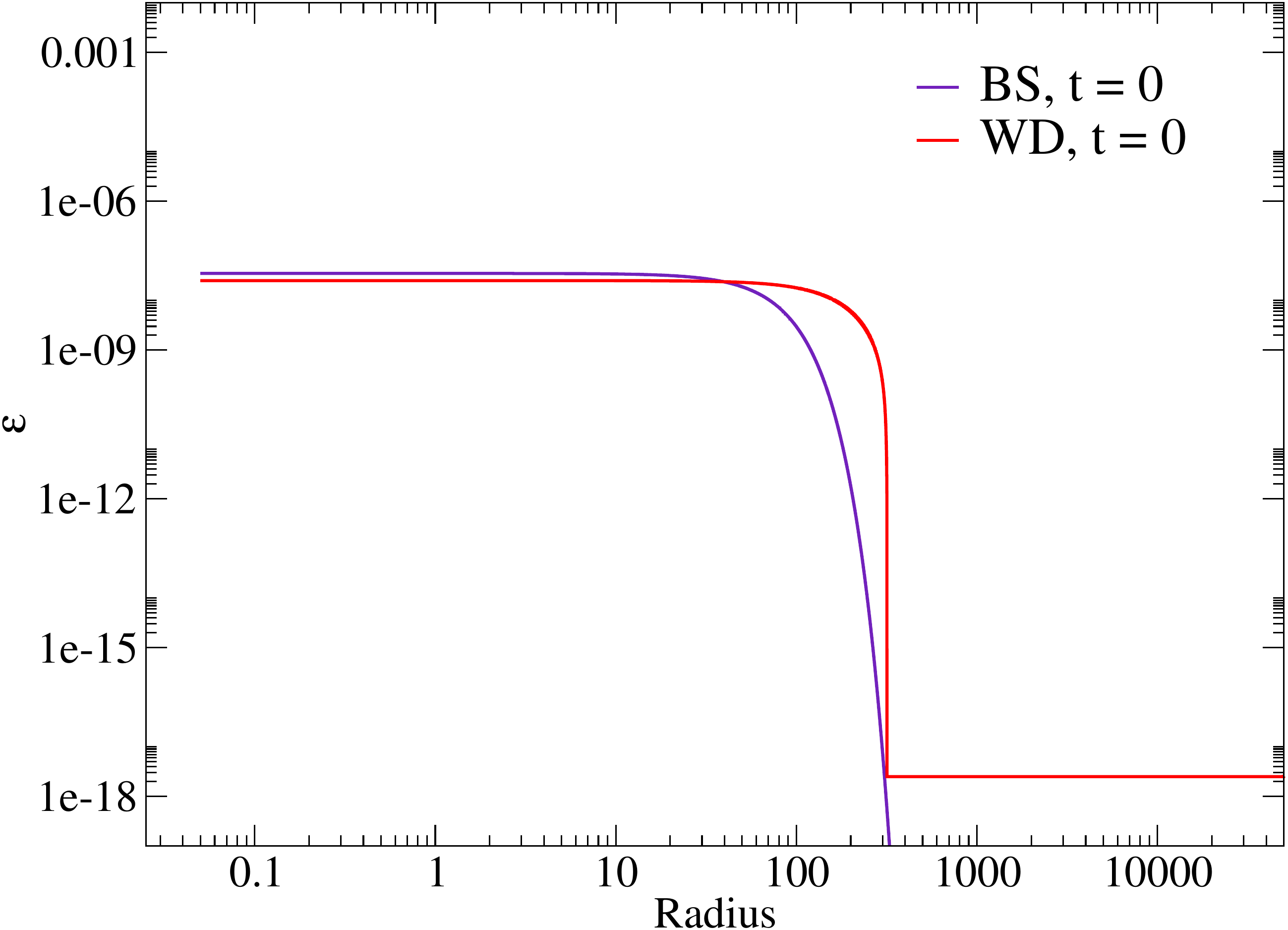}
\includegraphics[width=0.328\linewidth]{t0_Density_wd12.pdf}
\includegraphics[width=0.328\linewidth]{t0_Density_wd12.pdf}\\
\includegraphics[width=0.325\linewidth]{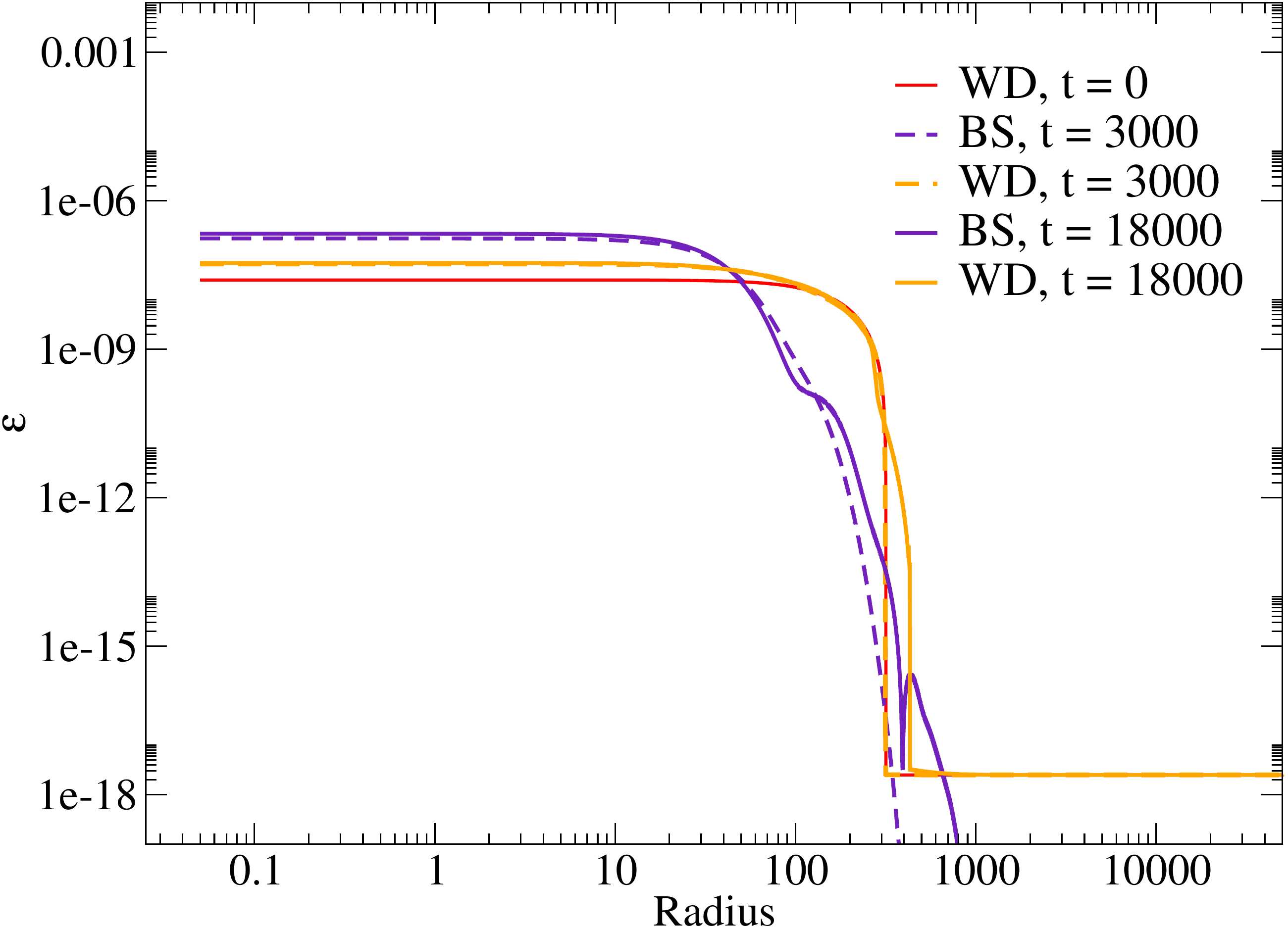}
\includegraphics[width=0.325\linewidth]{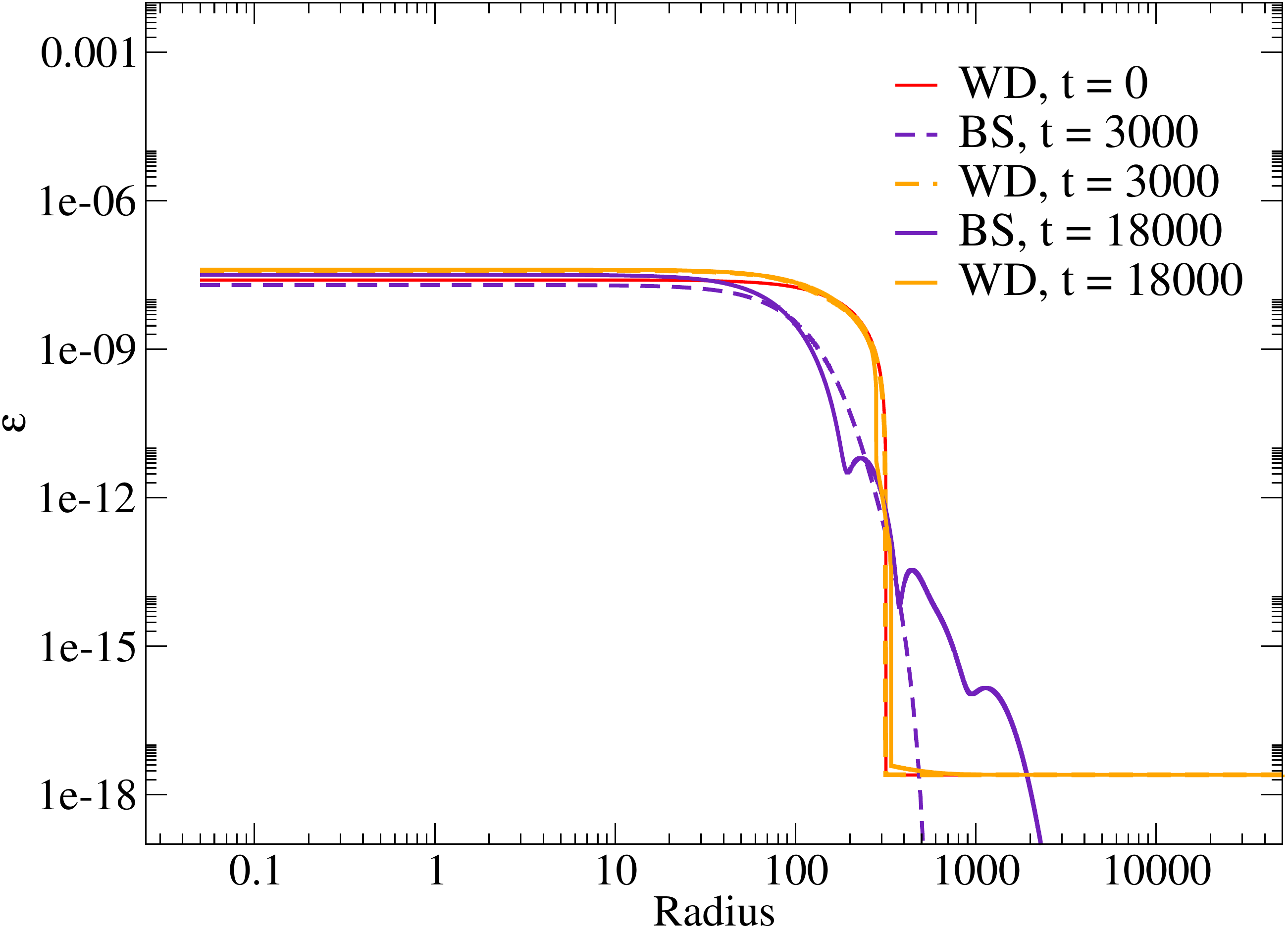}
\includegraphics[width=0.325\linewidth]{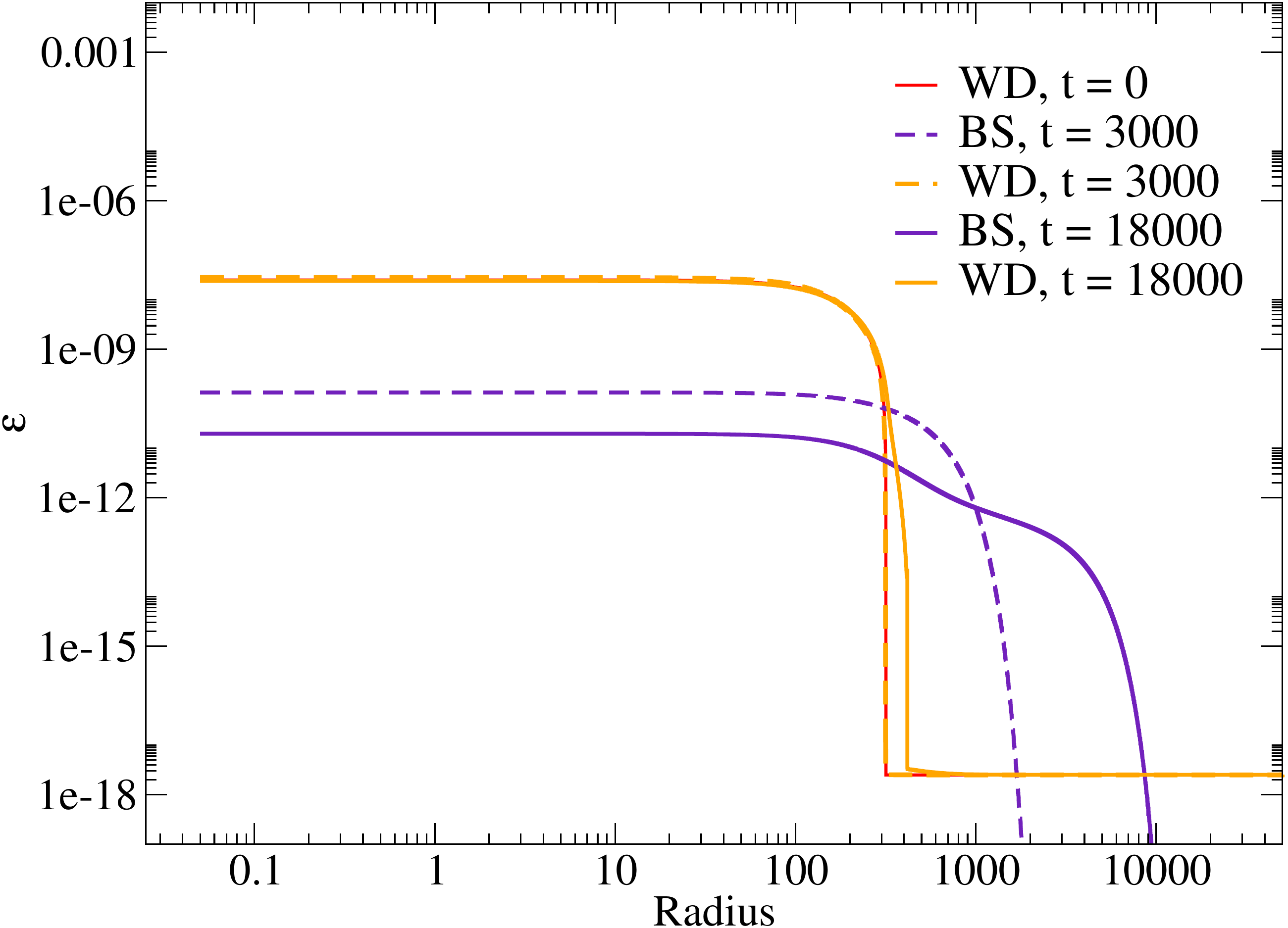}\\
\includegraphics[width=0.325\linewidth]{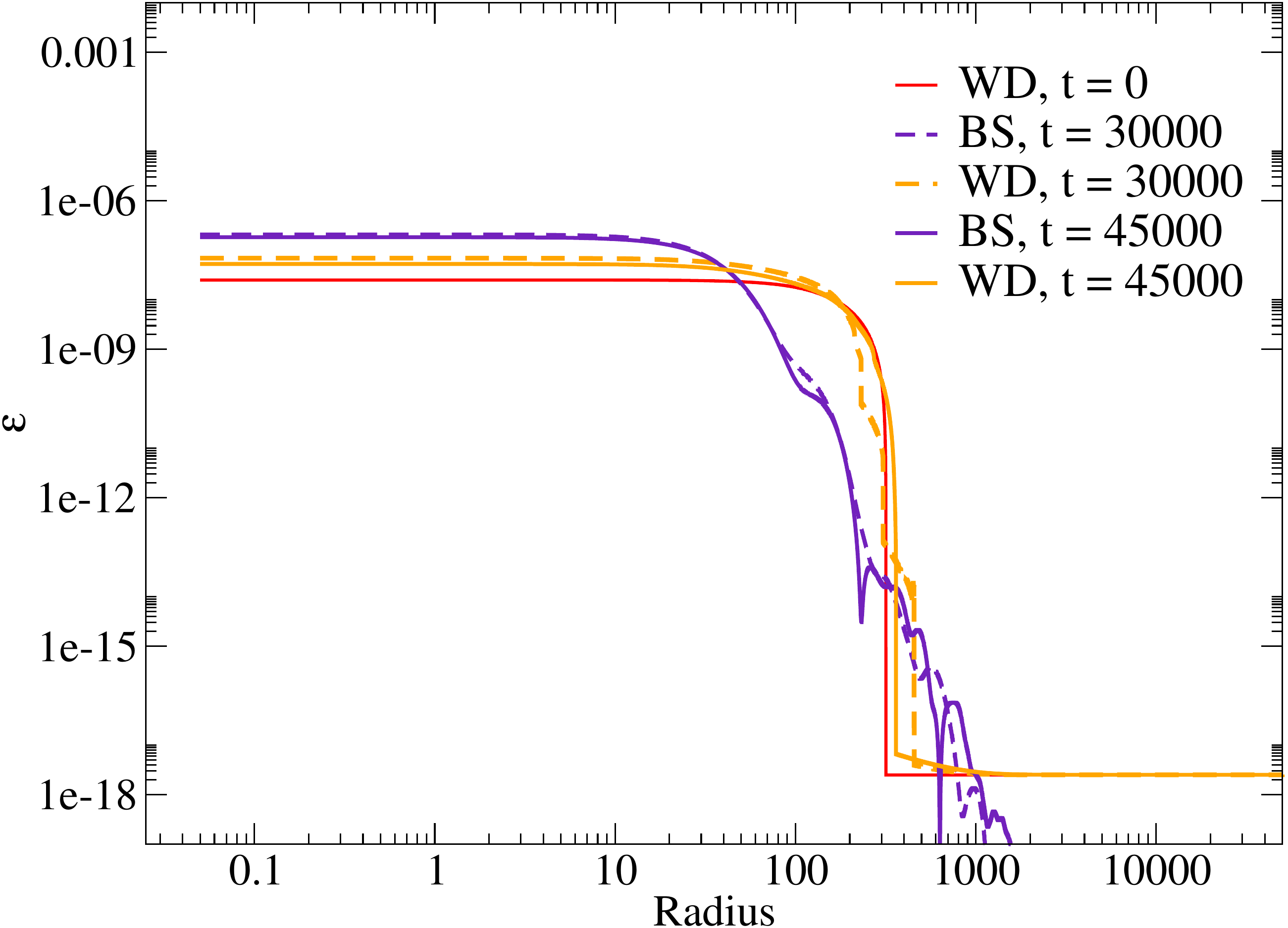}
\includegraphics[width=0.325\linewidth]{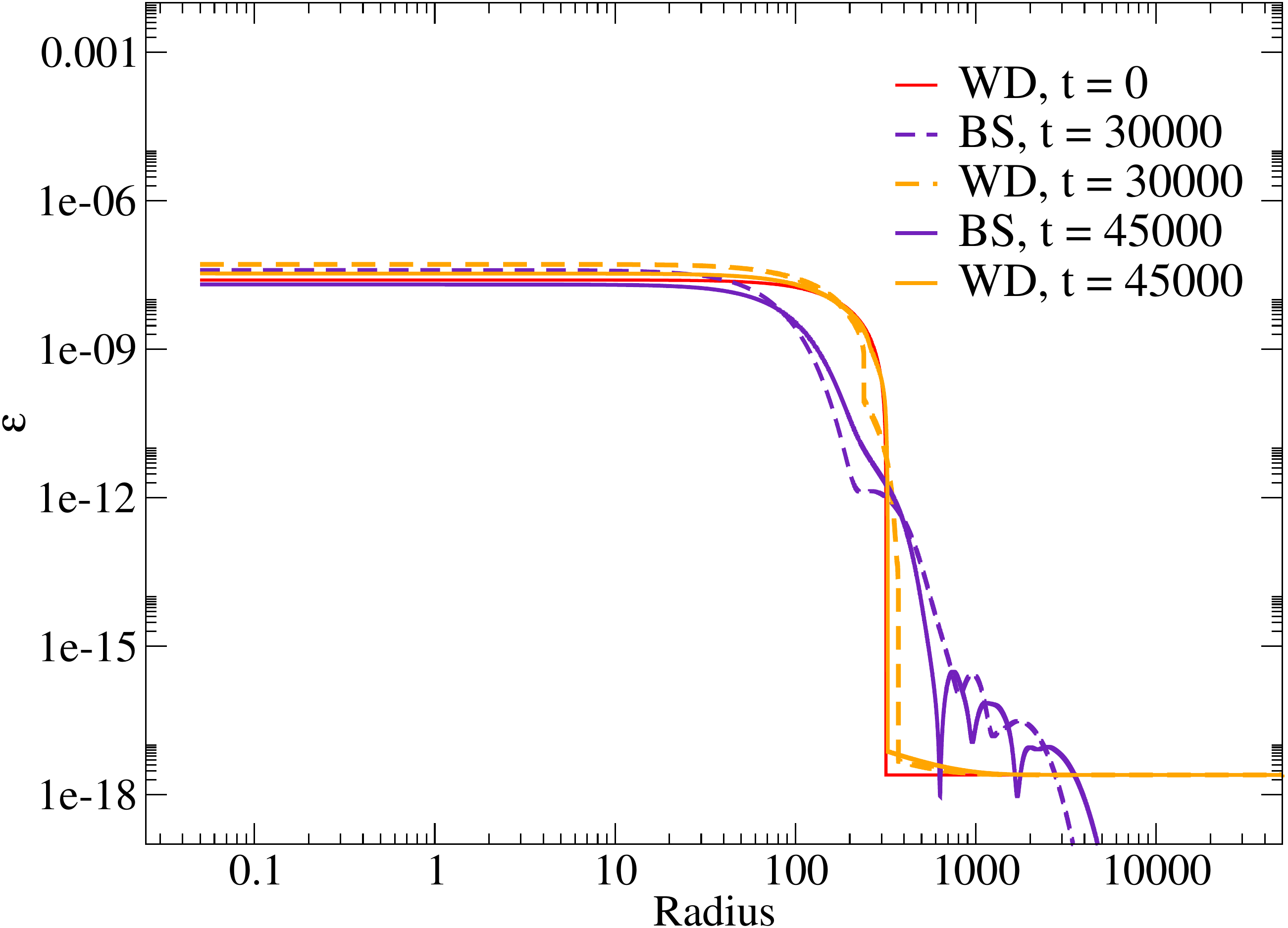}
\includegraphics[width=0.325\linewidth]{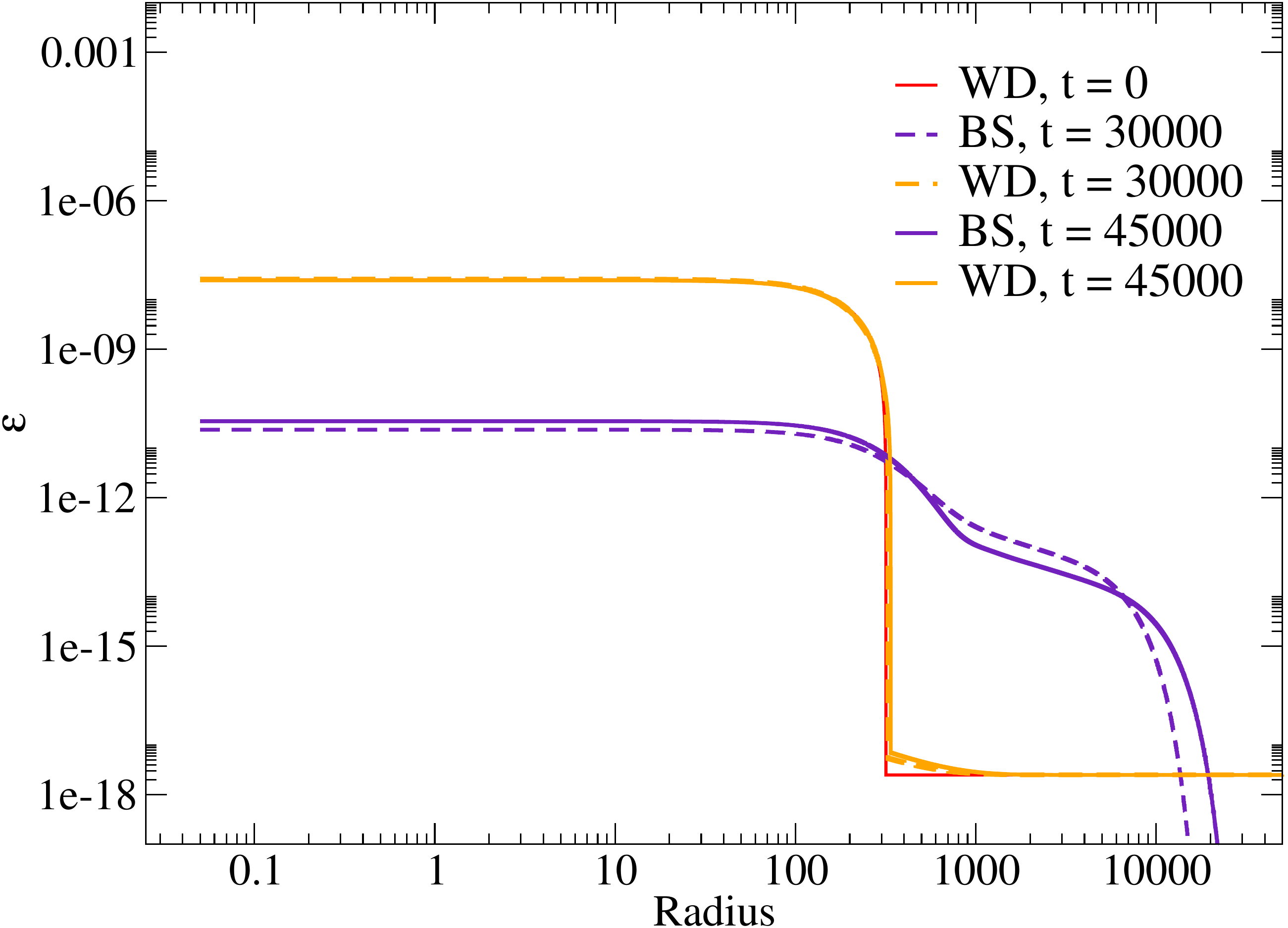}\\
\includegraphics[width=0.325\linewidth]{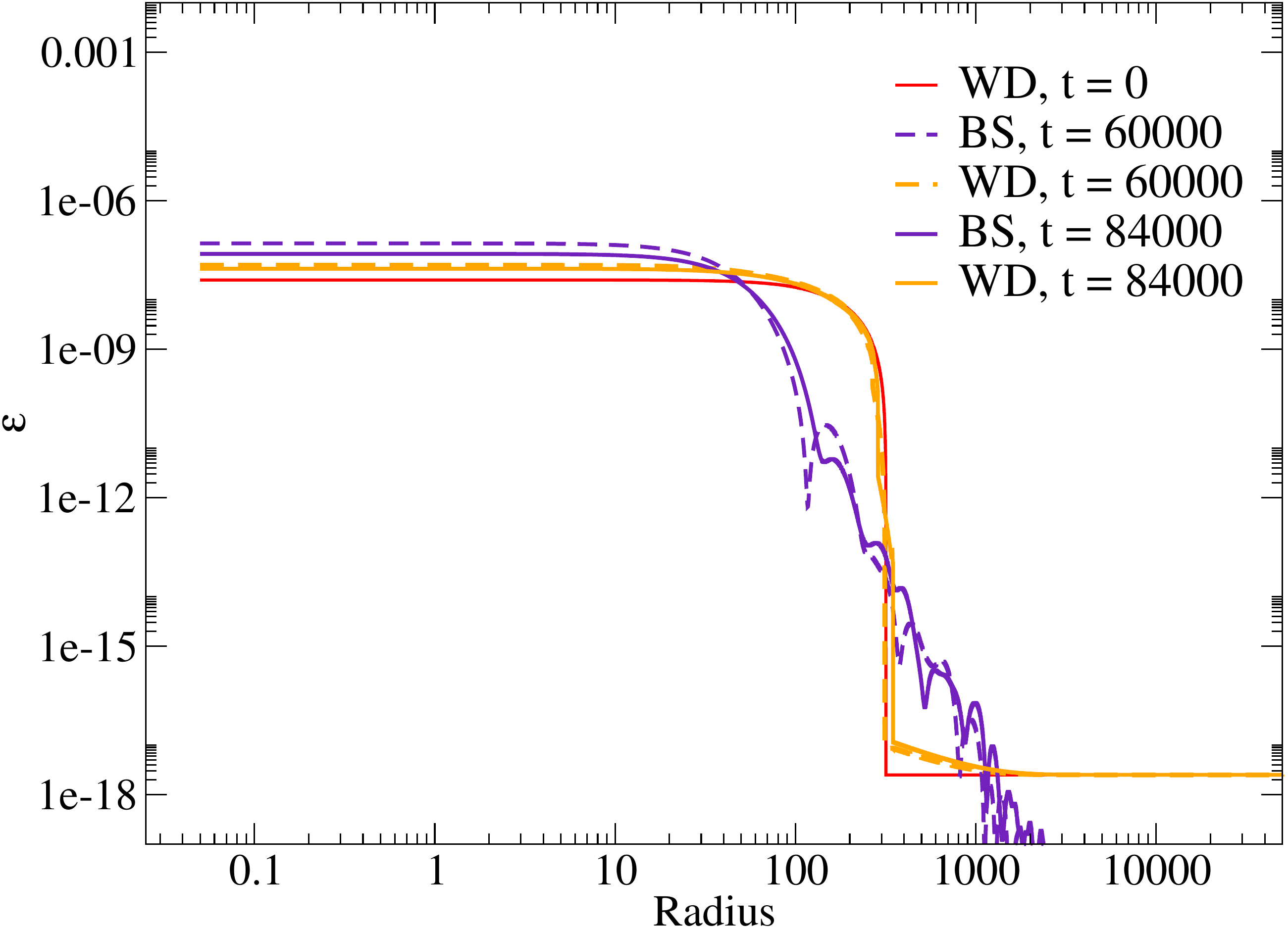}
\includegraphics[width=0.325\linewidth]{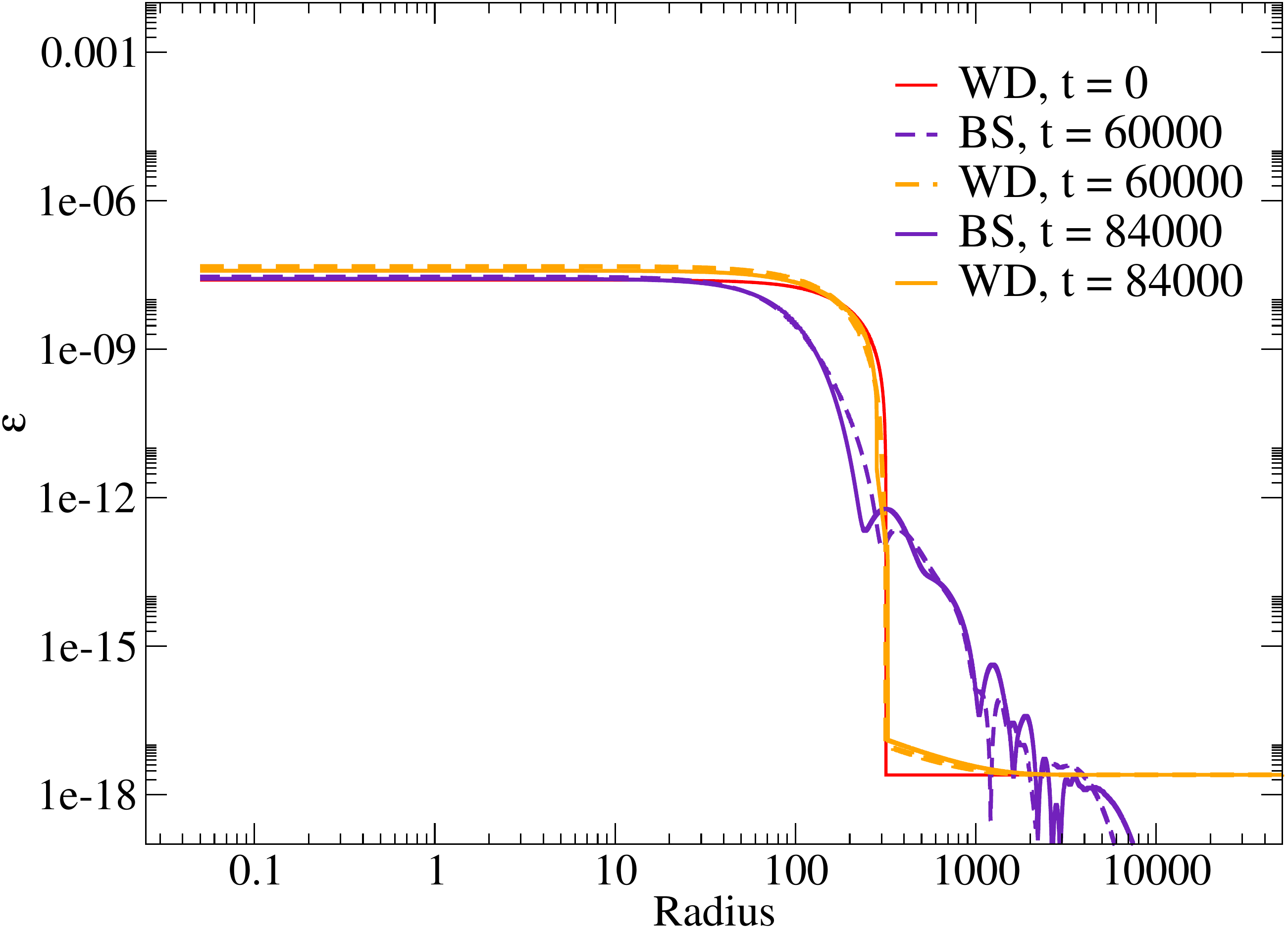}
\includegraphics[width=0.325\linewidth]{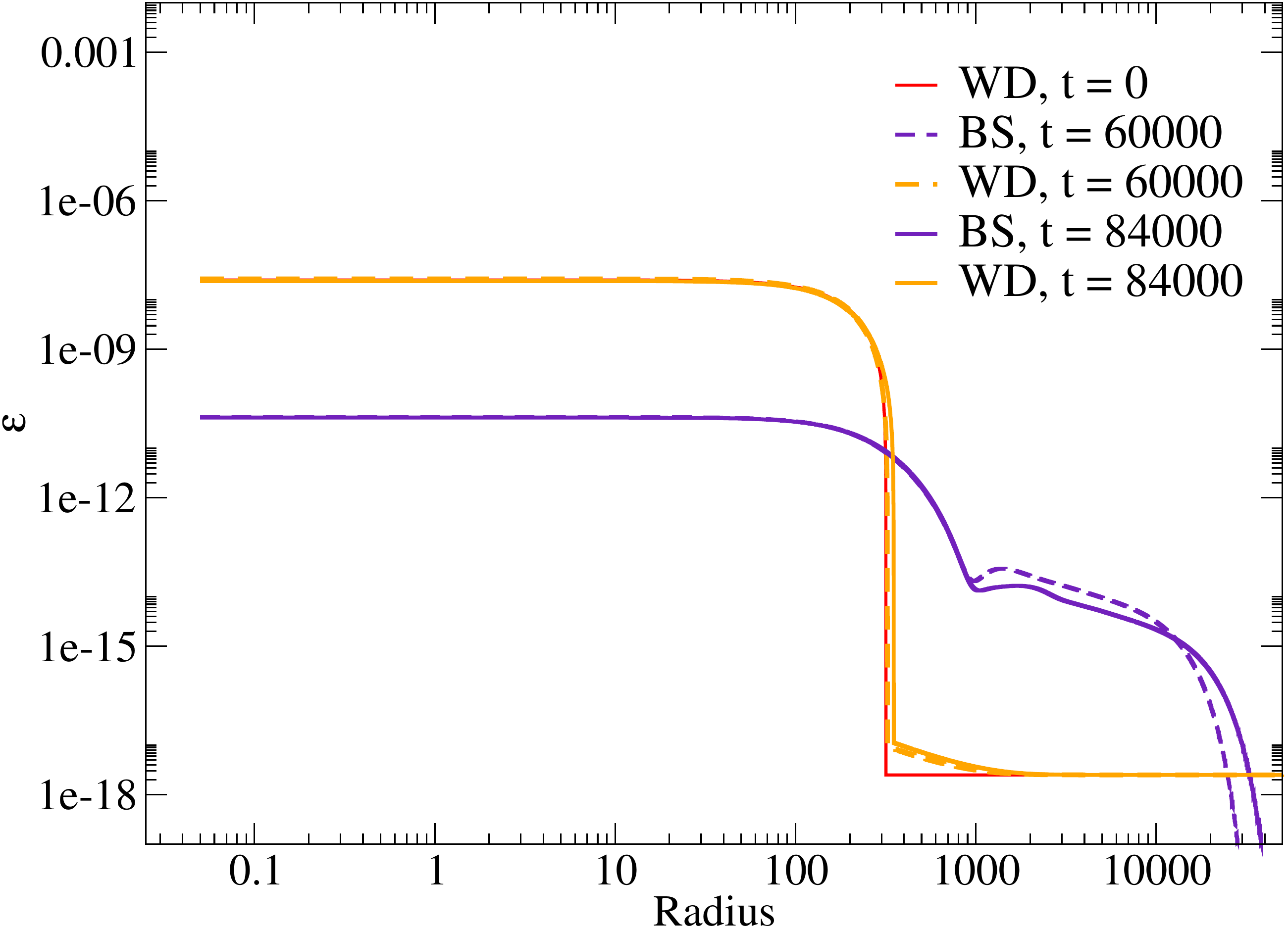}\\
\caption{Time evolution of the radial profiles of the scalar field (blue lines) and white dwarf (orange lines) energy densities for Configuration 3 with $M_{\rm{SF}}=0.051$, and three different values of the particle mass $\mu=\lbrace1,0.5,0.1\rbrace$. The solid red line corresponds to the white dwarf energy density at $t=0$. Time runs top to bottom.}
\label{fig3}
\end{figure*}

In Figs.~\ref{fig1}, \ref{fig2}, and \ref{fig3} we show the radial profiles of the scalar field density $\mathcal{E}^{\rm{SF}}$ for the three different configurations (green, blue, and purple lines, respectively) and of the fermionic energy density $\mathcal{E}^{\rm{F}}$ (orange lines for the three figures) at different times from $t=0$ up to $t_{\rm{final}}=84000$. The red line is shown for reference and marks the initial energy density of the isolated white dwarf model at $t=0$.

In Configuration 1 (Fig.~\ref{fig1}), the initial cloud is $M_{\rm{SF}}=0.727$ for the three boson particle masses. However, this mass is above the maximum allowed for a BS with $\mu=1.0$, but the scalar field does not collapse to a black hole since the excess mass is dissipated through gravitational cooling, reaching a final BS mass of $M^{\mu=1.0}_{\rm{BS}}=0.51$, i.e. approximately the same mass as the white dwarf. For $\mu=0.5$ and $0.1$, $M_{\rm{SF}}$ is well below their respective maximum masses. In the case of $\mu=0.5$, the final BS is able to retain a larger part of the initial scalar field mass, ending up with $M^{\mu=0.5}_{\rm{BS}}=0.68$, while for $\mu=0.1$ only a small fraction forms the BS with $M^{\mu= 0.1}_{\rm{BS}}=0.09$, probably due to the lower final compactness compared with that of the initial cloud which leads to a larger dispersion of the field.

\begin{figure}[t!]
\begin{tabular}{ p{0.5\linewidth}  p{0.5\linewidth} }
\end{tabular}
\\
\includegraphics[width=0.9\linewidth]{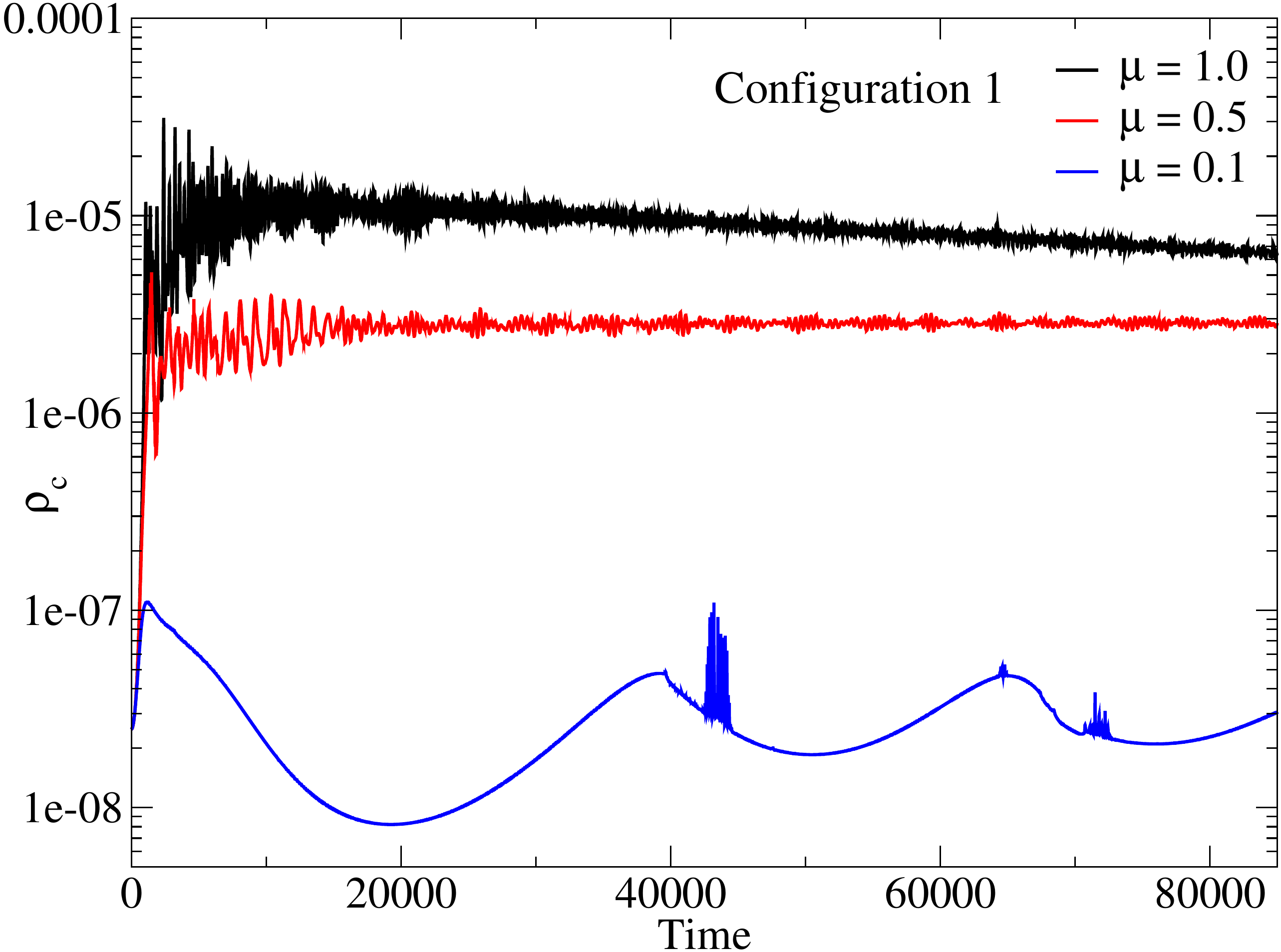}
\includegraphics[width=0.9\linewidth]{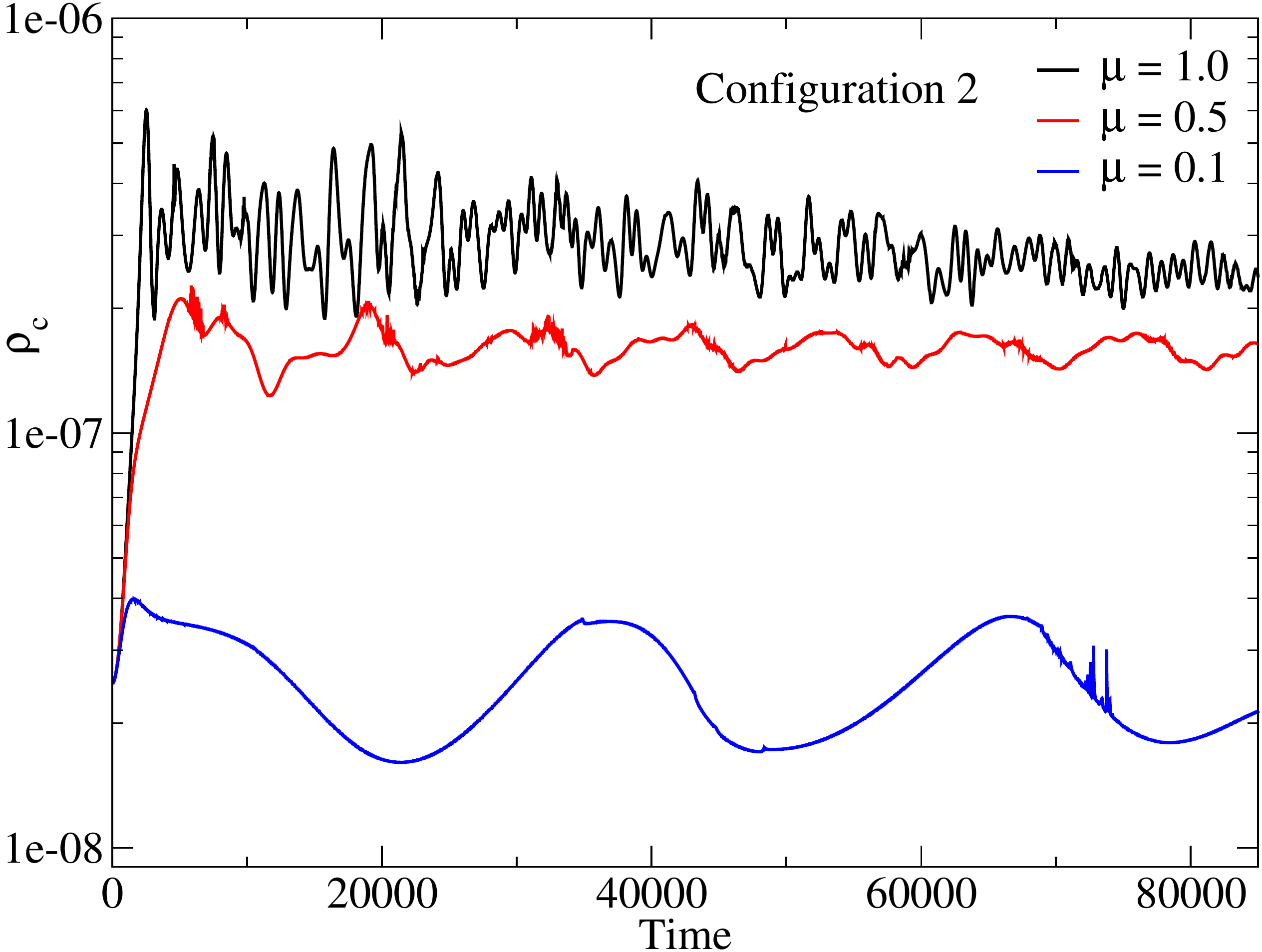}
\includegraphics[width=0.9\linewidth]{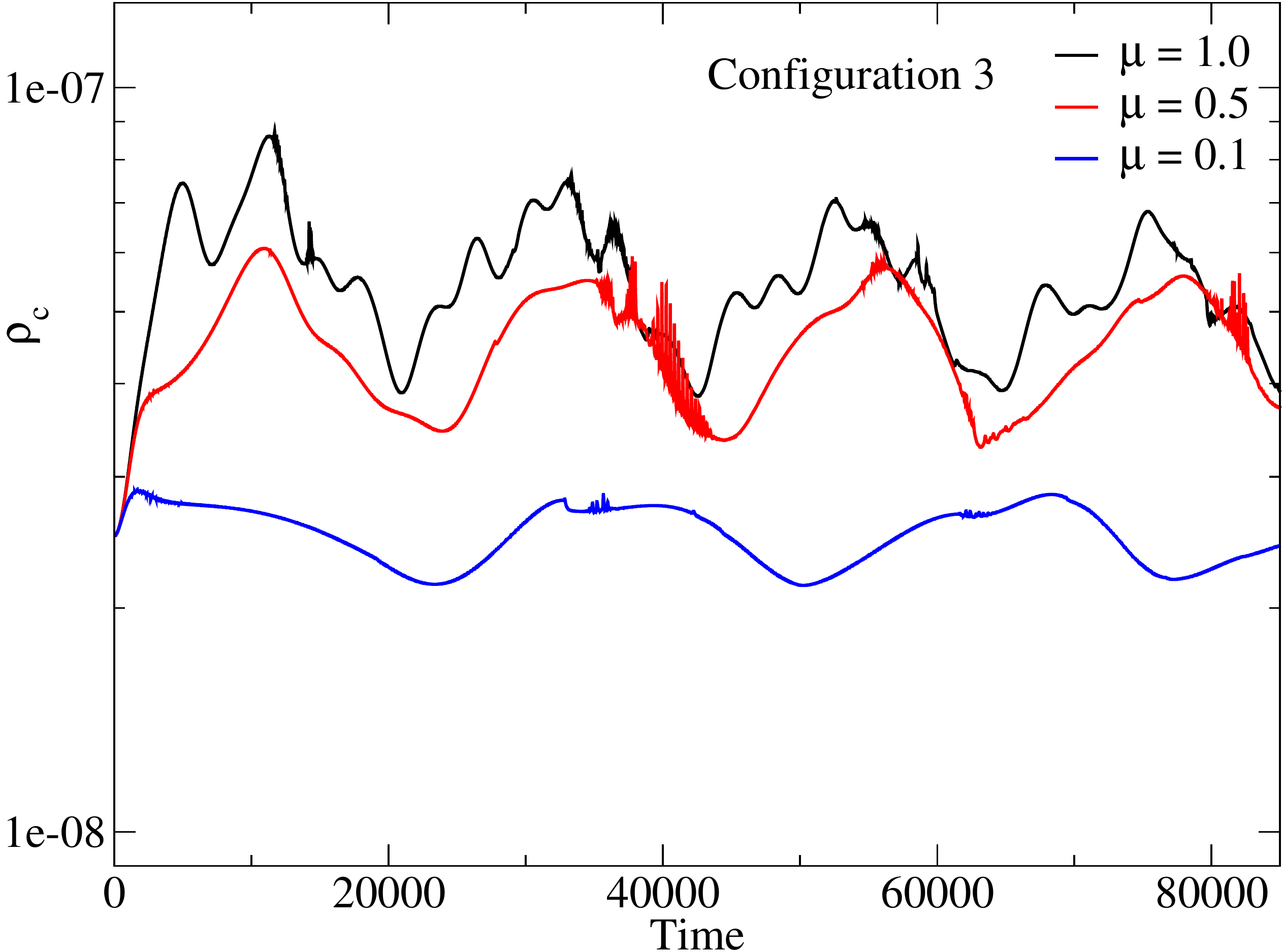}
\caption{Time evolution of the central values of the rest-mass density $\rho_c$ for Configuration 1, 2 and 3 (top, middle and bottom panels, respectively) for three values of the scalar particle mass.
}
\label{fig4}
\end{figure}

Despite the similar BS and white dwarf masses for the $\mu=1.0$ and $0.5$ cases, the BSs are far more compact than the initial white dwarf, with approximate radii three-to-five times smaller. As already mentioned, the final system is a mixed fermion-boson star. During the evolution and as a consequence of the mixing process, the white dwarf decreases its radius to $R^{\rm{final}}_{\rm{WD}}\sim130$, roughly three times smaller than its initial value and now comparable with the BS. To illustrate this, we show in the top panel of Fig.~\ref{fig4} the time evolution of the rest-mass density central value $\rho_c$ of the fermionic part of the mixed star. It increases between two and three orders of magnitude, demonstrating that the formation of a BS induces the collapse of the white dwarf into a denser fermion star (an object that could be much closer to a neutron star~\cite{leung2019accretion,zha2019accretion}).

The dynamics and outcome change for $\mu=0.1$ (right column of Fig.~\ref{fig1}). The BS is less massive and compact than in the other cases, with a radius several times larger than the initial radius of the white dwarf. The time evolution shows that during the simulation a very dilute ``atmosphere'' made of white dwarf material lingers around the mixed star due the gravitational dynamics of the BS. The white dwarf becomes slightly more compact due to the additional scalar mass and the central value of its rest-mass density undergoes a small increase (around a 10\%) with respect to its initial value (blue line in the top panel of Fig.~\ref{fig4}).

\begin{figure}[t!]
\begin{tabular}{ p{0.5\linewidth}  p{0.5\linewidth} }
\end{tabular}
\\
\includegraphics[width=0.95\linewidth]{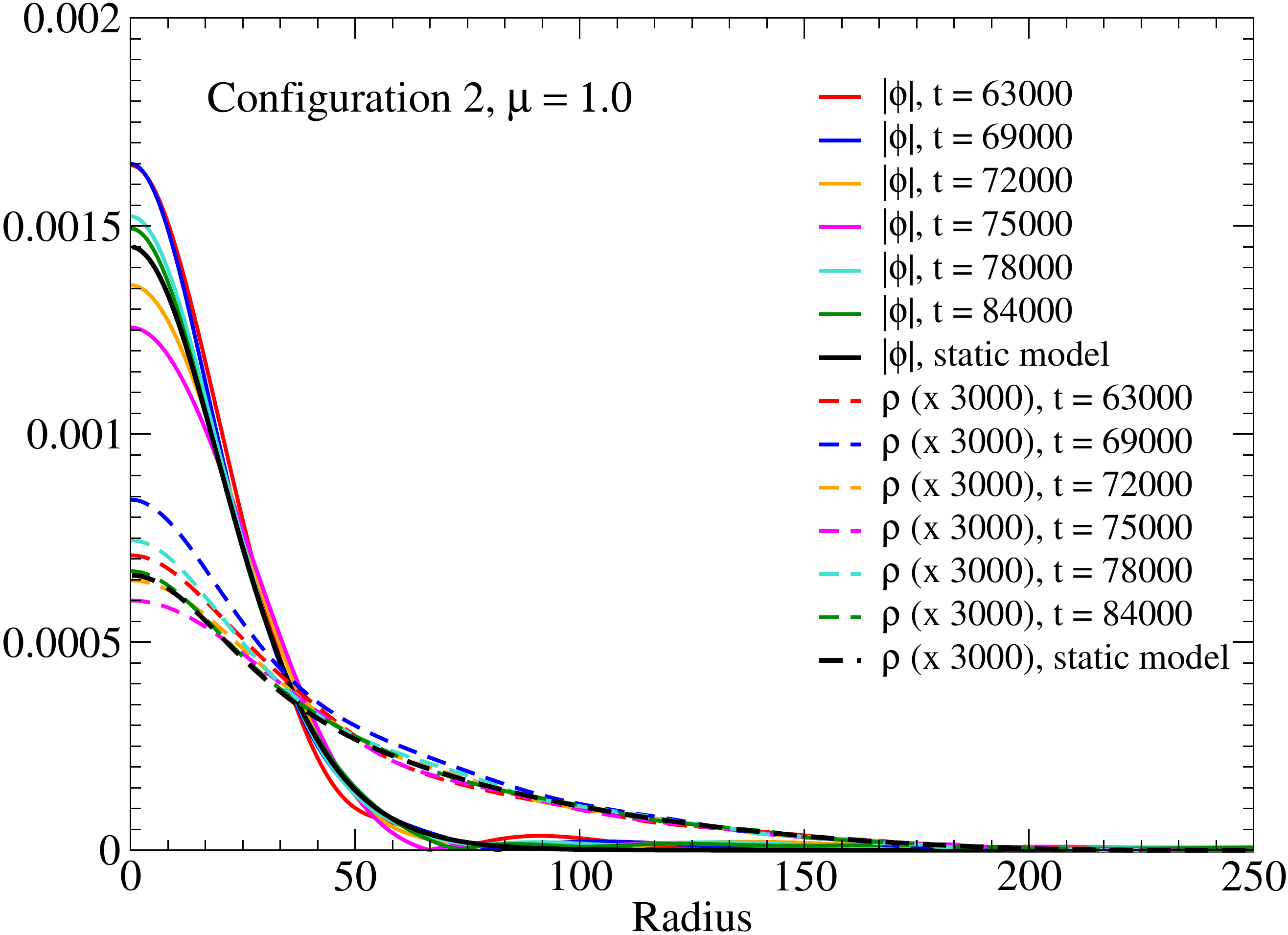}
\includegraphics[width=0.95\linewidth]{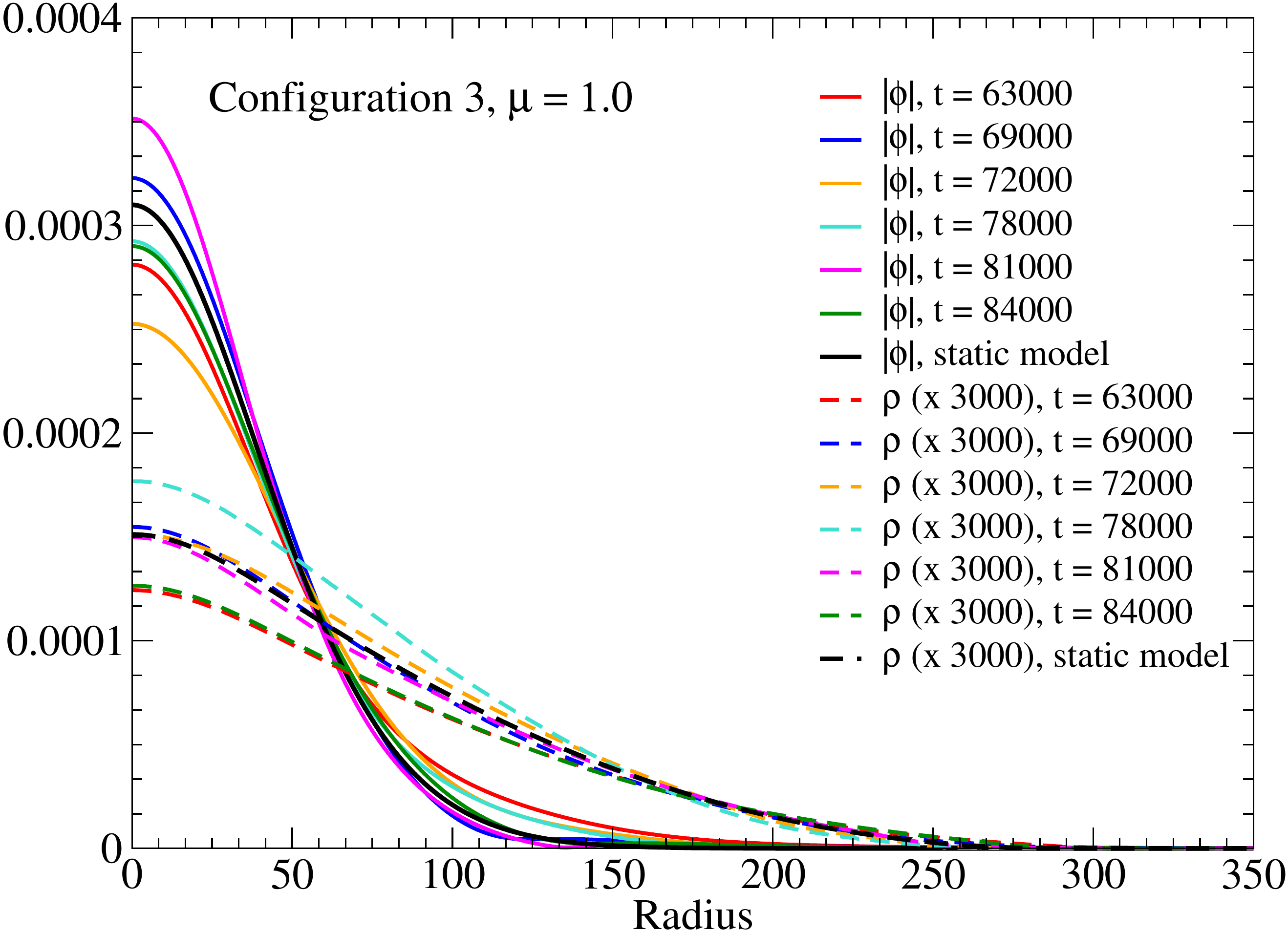}
\caption{Radial profiles of the magnitude of the scalar field $|\phi(r)|$ and the rest-mass density $\rho$ at different times, for Configuration 2 and 3 (top and bottom panels, respectively) for a particle mass $\mu=1.0$. Black solid and dashed lines correspond to the static model fitted to the final BS and white dwarf masses.}
\label{fig5}
\end{figure}

\begin{figure}[t!]
\begin{tabular}{ p{0.5\linewidth}  p{0.5\linewidth} }
\end{tabular}
\\
\includegraphics[width=0.95\linewidth]{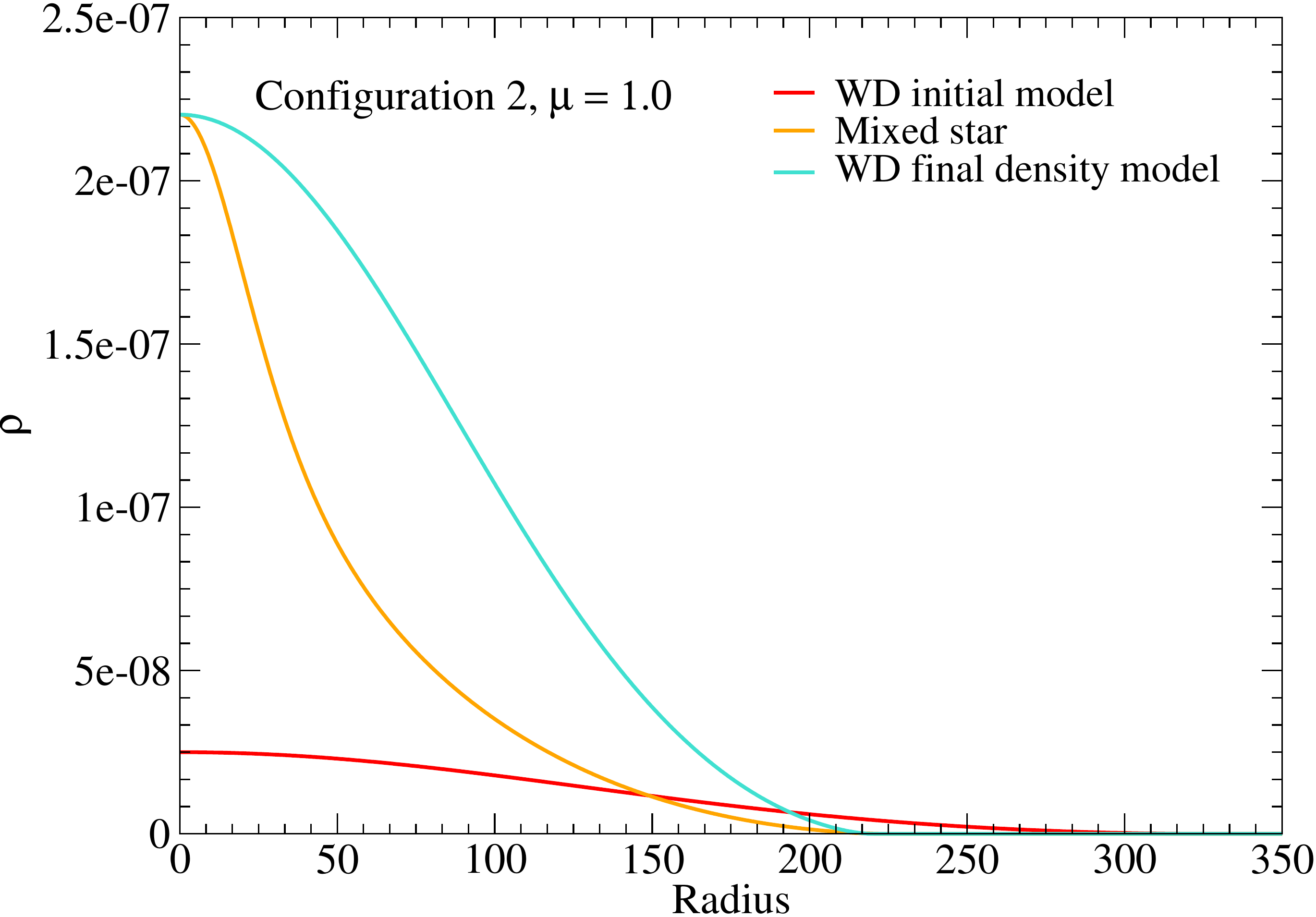}
\includegraphics[width=0.95\linewidth]{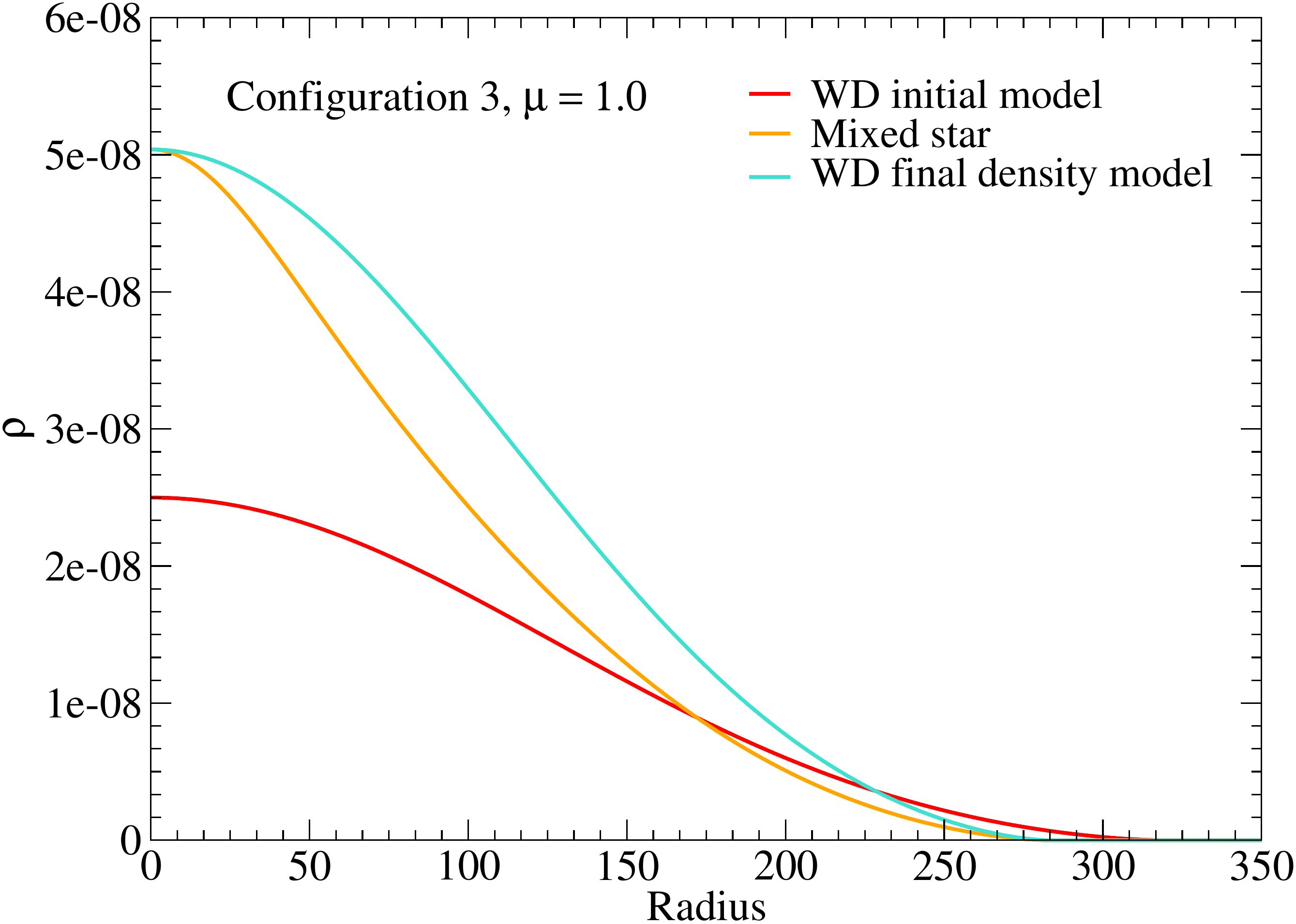}
\includegraphics[width=0.95\linewidth]{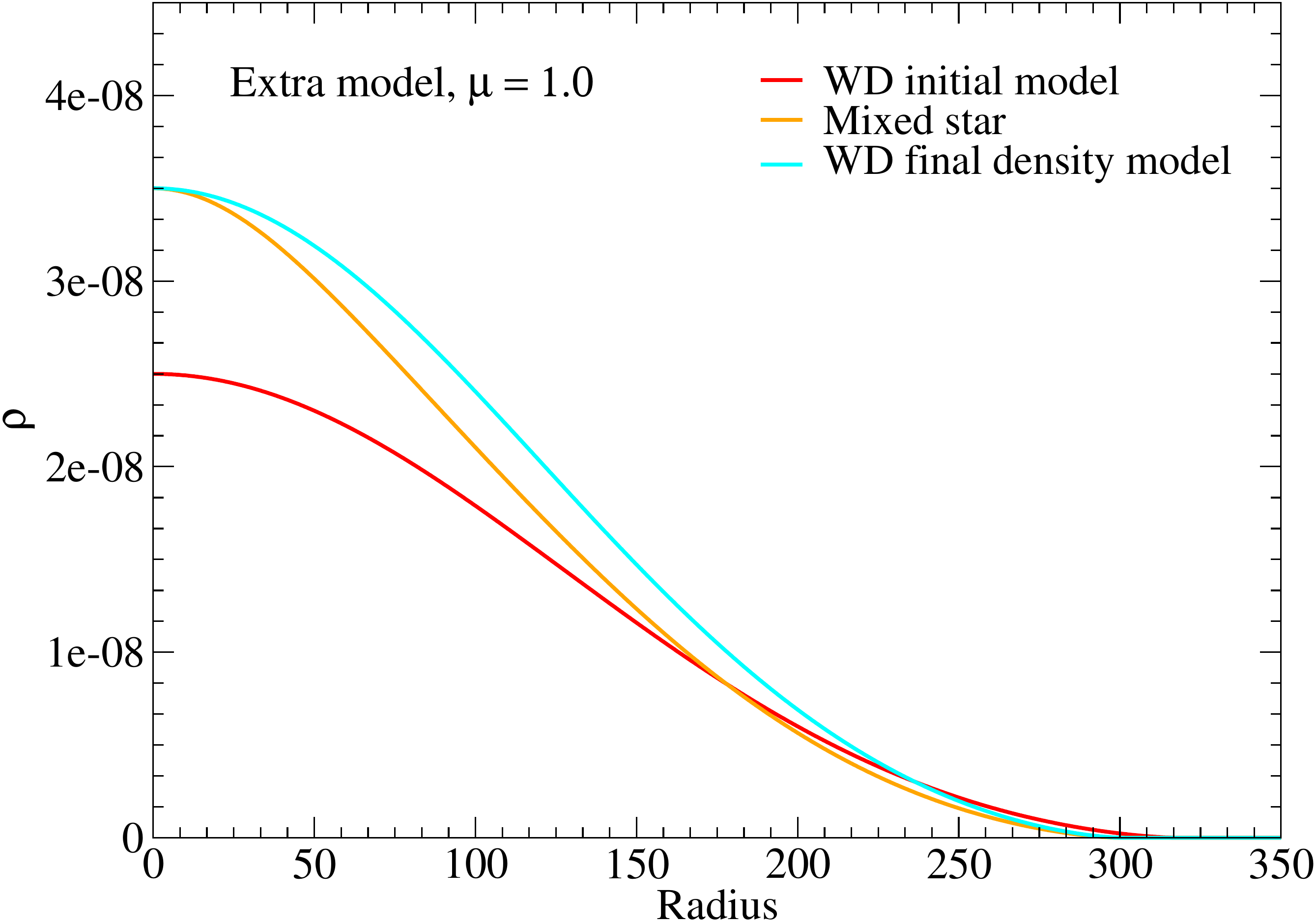}
\caption{Top panel: Radial profile of the fermion rest-mass density of different static models: the initial white dwarf (red line), the final mixed star from Configuration 2 and $\mu=1.0$ (orange line), and a white dwarf with the same central density than the fermion part of the mixed star (cyan line). Middle and bottom panels: same for Configuration 3 and an extra model.}
\label{fig6}
\end{figure}

For Configurations 2 and 3 (Figs.~\ref{fig2} and \ref{fig3}, respectively), the initial amplitude of the scalar cloud is largely decreased compared to Configuration 1, leading to initial masses of $M_{\rm{SF}}=0.183$ and $0.051$, respectively. Because of this, the endpoint of the scalar field evolution is a more diluted BS in both cases. %A mixed star is formed with most of the scalar field energy density inside the white dwarf.
Once more, the mixing process produces the white dwarf to collapse to a more compact star, but significantly less dense than in the previous Configuration 1 (see middle and bottom panels of Fig.~\ref{fig4}). In this way, the maximum value of the fermion rest-mass density increases by an order of magnitude in Configuration 2 and a few times its initial value in Configuration 3 for $\mu=1.0$ and $\mu=0.5$. Moreover, the white dwarf radius decreases from an initial value of $R_{\rm{WD}}=318.2$ to a final value of $\sim215$ and $\sim250$ for the Configurations 2 and 3, respectively.

In contrast, when $\mu=0.1$ (last column in Figs.~\ref{fig2} and~\ref{fig3}), the final mixed object is approximately the same white dwarf surrounded by a faint scalar field cloud. The blue line in the middle and bottom panels of Fig.~\ref{fig4} shows that the white dwarf is perturbed by the scalar field and oscillates around a slightly larger central value. 

The numerical evolutions confirm that mixed white-dwarf--boson stars can be formed from the collapse (or accretion) of a scalar cloud onto a white dwarf. Since the final objects are still perturbed and oscillating at the end of the simulations, for Configurations 2 and 3 (we leave out Configuration 1 since the white dwarf evolves into a neutron-star-like object) we fit $\rho_c$, $\phi_c$, and the BS and white dwarf masses, as shown in Fig.~\ref{fig5} for $\mu=1.0$, to build the corresponding static mixed star models. We make use of the code described in~\cite{di2020dynamical,di2021dynamical} to build the models. Then, we compute the gravitational redshift at the surface of the white dwarf and compare these mixed solutions with white dwarf models without scalar field. The total mass of the mixed fermion-boson stars, $M_{\rm{Total}}$, and the final BS masses, $M_{\rm{BS}}$, are given in Table~\ref{tab:table2}. Given that there is no loss of fermionic matter in any case, the white dwarf retains its initial mass for all Configurations, $M_{\rm{WD}}=0.562$.

\begin{table}
\caption{From left to 
right the columns indicate the initial configuration, the particle mass $\mu$, the total mass of the static mixed star model $M_{\rm{Total}}$, the mass of the BS $M_{\rm{BS}}$, the mass of the new isolated white dwarf with the same central rest-mass density $M'_{\rm{WD}}$, the redshift computed at the radius of the white dwarf $z_{g}$ for mixed stars, the redshift computed at the radius of the new white dwarf $z'_{g}$, and the fraction between the mixed star and new white dwarf redshift. The mass of the white dwarf in the mixed star is always $M_{\rm{WD}}=0.562$ and, for reference, the gravitational redshift at its surface is $z_g^{\rm{WD}}=0.0036$.
}\label{tab:table2}
\begin{ruledtabular}
\begin{tabular}{cccccccc}
Conf.&$\mu$&$M_{\rm{Total}}$&$M_{\rm{BS}}$&$M'_{\rm{WD}}$&$z_{g}$&$z'^{\rm{WD}}_{g}$&$z_{g}/z'^{\rm{WD}}_{g}$\\
\hline
%WD&-&0.562&0.00&0.562&0.0018&0.0018&1.000\\
%\hline
2&1.0&0.685&0.126&1.626&0.0062&0.0150&0.413\\
2&0.5&0.739&0.180&1.392&0.0072&0.0121&0.595\\
2&0.1&0.582&0.021&0.570&0.0037&0.0036&1.023\\
\hline
3&1.0&0.599&0.038&0.795&0.0043&0.0057&0.753\\
3&0.5&0.610&0.049&0.749&0.0044&0.0052&0.841\\
3&0.1&0.566&0.005&0.564&0.00358&0.00356&1.006\\
\hline
Extra&1.0&0.579&0.018&0.664&0.0039&0.0044&0.874
\end{tabular}
\end{ruledtabular}
\end{table}

\subsection{Gravitational redshift}

Highly compact BSs could be formed within white dwarfs, causing the fermion star to collapse to a remarkably denser configuration, deviating from its previous white-dwarf-like nature as seen in the previous section. %For small BS masses (less than 10\% of the initial white-dwarf mass $M_{\rm{WD}}$), 
On the other hand, the presence of a more diluted low-mass BS %( $M_{\rm{BS}}/M_{\rm{WD}}\leq0.1$) 
also produces an additional gravitational pull which leads to a growth of the central rest-mass density of the fermion star, but it remains within the bounds of standard white dwarfs. Since the white dwarf mass is conserved, the higher compactness of the mixed star results in an increase of the gravitational-field strength at the fermion star surface when compared to white dwarfs without scalar field.

In Fig.~\ref{fig6} we compare the rest-mass density radial profile of our initial white dwarf defined in Table~\ref{tab:table1} without scalar field (red lines) with the fermion rest-mass density of the static mixed stars (orange lines) obtained from Configurations 2 and 3 with $\mu=1.0$ (top and middle panels). We find that the fermion star resulting from the mixing process resembles an isolated white dwarf with the same central rest-mass density (cyan line) but with larger mass ($M'_{\rm{WD}}\geq M_{\rm{WD}}$). If white dwarfs are mainly defined by their central density (mass, radius, and also assuming that the intrinsic electromagnetic emission would be the same or sufficiently similar), our results may indicate that a particular white dwarf could be mimicked by a mixed star composed by a BS and a less massive white dwarf. From Fig.~\ref{fig6}, a trend is clear: the smaller the BS mass, the more the profile of the final fermion star approaches that of a more massive white dwarf, with both converging to the initial star when the scalar field vanishes. We include an extra model with even lower BS mass (bottom panel in Fig.~\ref{fig6}) to further illustrate the trend. From the top panel of Fig.~\ref{fig6} and Table~\ref{tab:table2} we see that the model corresponding to Configuration 2 with $\mu=1.0$ has a significantly massive and compact BS, leading to a mimicked white dwarf with a mass that would be too large, surpassing the Chandrasekhar limit mass ($M'_{\rm{WD}}>1.4$). In this case, our toy model, which approximates an isolated white dwarf to our fermion star, would no longer be valid.  %A range of 

As previously outlined, the outcome of the mixed system is a white dwarf embedded in a  self-gravitating BS that modifies its gravitational field. Therefore, the gravitational redshift at the white dwarf surface is modified. The gravitational redshift inside a BS is given by~\cite{schunck1997gravitational}
\begin{equation}\label{zgr}
1+z_{g}(r)=1/\alpha(r). 
\end{equation}
%\exp\left(-\frac{\alpha(r)}{2}\right).
%
where $z_{g}$ is the gravitational redshift and $\alpha$ is the lapse function at some radius $r$ (see Eq.~\ref{metric}), which is basically the $g_{tt}$ component given that the shift is zero. 

\begin{figure}[t!]
\begin{tabular}{ p{0.5\linewidth}  p{0.5\linewidth} }
\end{tabular}
\\
\includegraphics[width=0.95\linewidth]{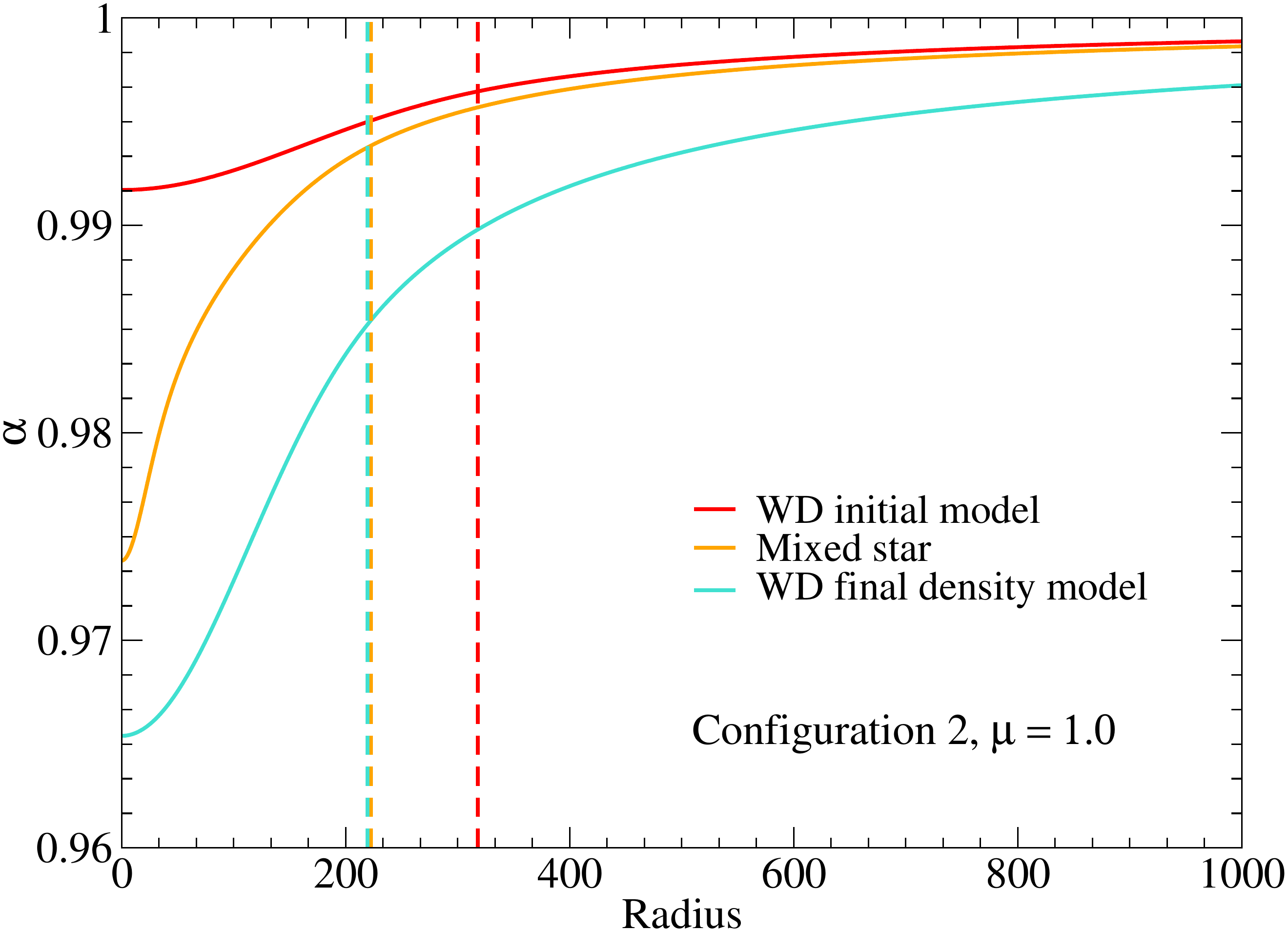}
\includegraphics[width=0.95\linewidth]{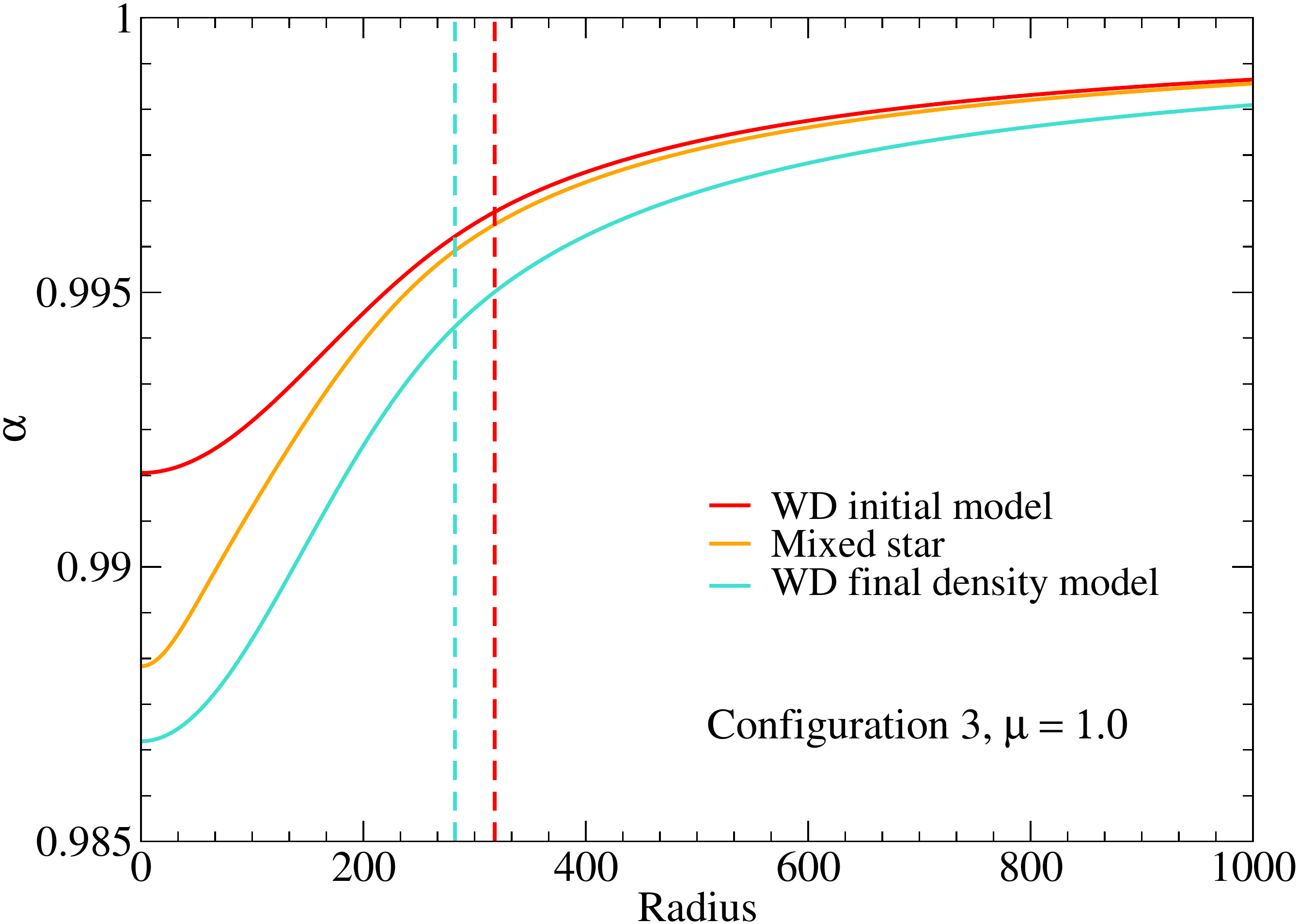}
\includegraphics[width=0.95\linewidth]{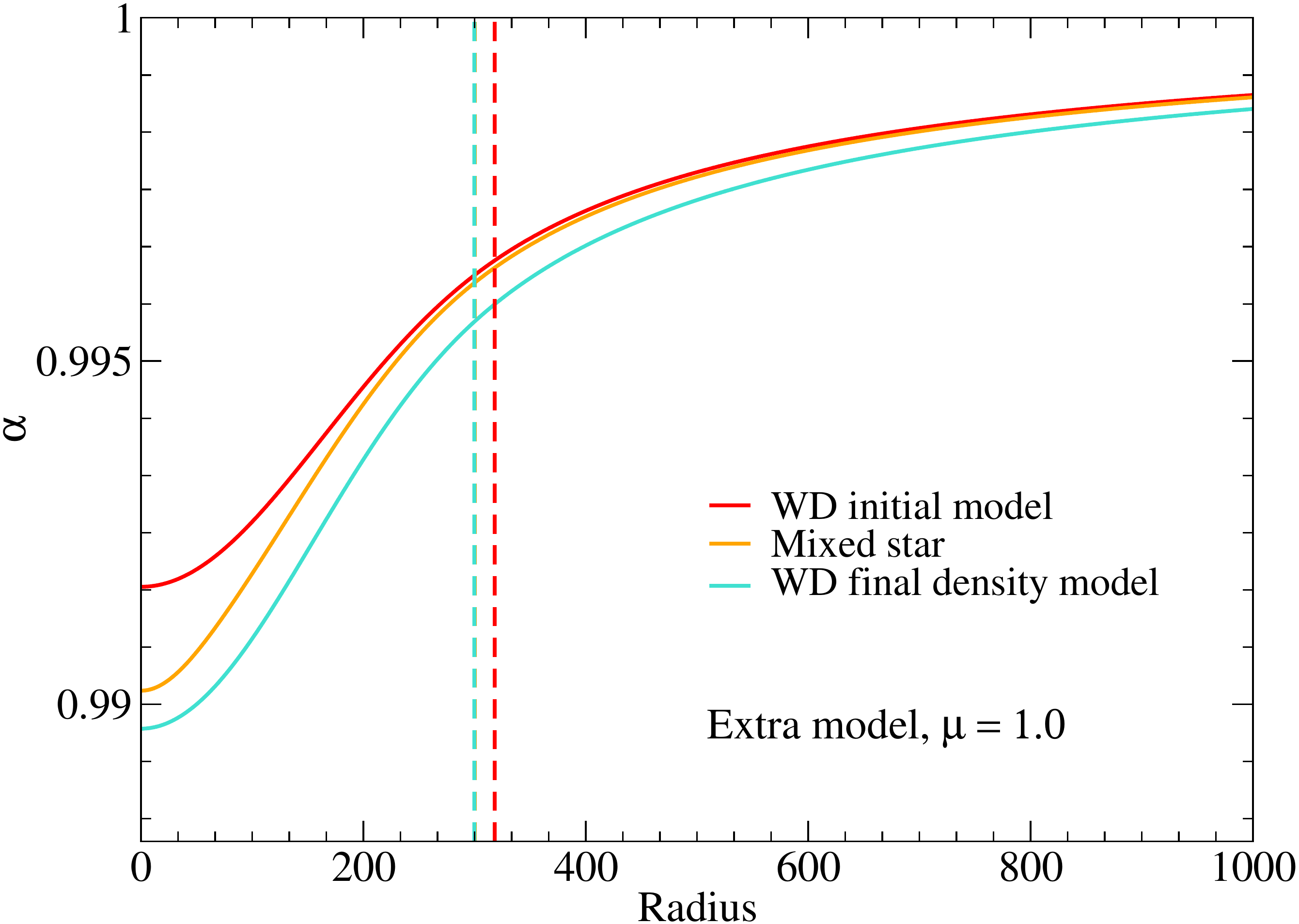}
\caption{Top panel: Radial profile of the lapse function $\alpha$ of different static models: the initial white dwarf (red line), the final mixed star from Configuration 2 and $\mu=1.0$ (orange line), and a white dwarf with the same central density than the mixed star (cyan line). Horizontal dashed lines indicate the white-dwarf radii of the different models. Middle and bottom panels: same for Configuration 3 and an extra model. The dashed orange and cyan lines are superimposed.}
\label{fig7}
\end{figure}

We compare in Fig.~\ref{fig7} the radial profiles of the lapse, $\alpha$, for the initial white dwarf model, three mixed star models with $\mu=1.0$ (Configurations 2 and 3, and the extra model), and the new more massive white dwarfs shown in Fig.~\ref{fig6}. The dashed horizontal lines correspond to the radius where the fermion star surface is located and where the value of the gravitational redshift is computed. Since the BS radius is smaller than the radius of the white dwarf, the lapse only deviates crucially at small radius.

%\pis{

Finally, in the last three columns of Table~\ref{tab:table2} we provide the gravitational redshift on the white dwarf surface as calculated from Eq.~\ref{zgr}. Depending on the BS and the boson particle masses, we observe that the gravitational redshift can differ from -60\% (Configuration 2, $mu=1.0$) to +2\% (Configuration 2, $\mu=0.1$). Such values decrease when we take lower BS masses. With a 3\% of BS mass with respect to the total mass (extra model, $\mu=1.0$), the difference is reduced to a 13\%. It is then possible to further reduce the bosonic dark matter within the white dwarf in order to obtain changes in the gravitational redshift of a few percent to match current observational estimated errors in the measurements~\cite{parsons2017testing,joyce2018gravitational,romero2019white,chandra2020gravitational}.

\subsection{Methods to estimate the mass of white dwarfs}

%In this section, we briefly describe the different techniques used to measure white dwarf's masses from astronomical observations.

There are two main techniques to derive the mass of white dwarfs: by direct measurement from observations or by means of theoretical mass-radius relations. %observational tested. 

Direct measurements require the white dwarf to be embedded in an astrometric binary system so to obtain the dynamical masses from orbital fits, where a reliable distance measurement is also needed. Then, from spectroscopic observations one can derive the gravitational redshift and couple it with the orbital fits to obtain a reliable mass estimation \citep{parsons2017testing}. Unfortunately, just a very few objects check all these boxes \citep[see][for a comprehensive list of such studies]{holberg12}.

The scarcity of such a complex mix of particularities in a system necessitates an alternative method and the undoubtedly most employed one relies on the use of theoretical mass-radius relations to measure the masses of white dwarfs. These are derived from evolutionary cooling sequences \citep[see e.g.][]{fontaine2001potential,bedard20}. To make use of mass-radius relations, one needs to characterise the photospheric parameters of the white dwarf, i.e. the effective temperature $T_{\rm{eff}}$, the surface gravity $\log{g}$, and the chemical composition. Then, for a given chemical composition, any pair of $T_{\rm{eff}}-\log{g}$ will lead to the expected mass and radius.

Although the mass-radius relations have been compared with observational measures for consistency proves, several of these tests had relied on mass-radius relations to a small extent %and are thus arguable proofs 
\citep[see e.g.][]{provencal98,holberg12}. This was not the case of \cite{parsons2017testing}, who carried out an exhaustive analysis of 16 eclipsing binaries. They obtained mass and radius measurements which were model-independent and found an excellent agreement between the measured and expected values.

Even though the analysis perpetuated by \cite{parsons2017testing} proves an important milestone into the confirmation of the theoretical mass-radius relation, it relies upon the gravitational redshift to derive the mass. As we have shown with this toy model, a mixed star (white dwarf + BS) would show a larger gravitational field and different EM emission than the same isolated white dwarf. When compared with another isolated white dwarf with same central rest-mass density, the different gravitational field of the mixed star system would lead to a potentially smaller derived (and miscalculated) white dwarf mass. Thus, one might wonder if the agreement found by \citeauthor{parsons2017testing}\,\cite{parsons2017testing} would be as strong if more systems with such particularities were available or if, in fact, further observations of model-independent measurements of white dwarf masses and radii would point towards the existence of mixed objects.

%%%%%%%%%%%%%%%%%%%%%%%%%%%%%%%%%%%%%%%%%%%%%
\section{Conclusions} 
\label{sec:conclusions}
%%%%%%%%%%%%%%%%%%%%%%%%%%%%%%%%%%%%%%%%%%%%%
We have performed numerical simulations of dynamical accretion of scalar field cloud onto a white dwarf described by a polytropic equation of state, varying the particle mass $\mu$ and the initial mass of the cloud. We found that the final object is a stable mixed white-dwarf--boson star. The initial white dwarf is affected by the accretion and collapse of the scalar field when forming the BS, which leads to an increase in its density. In particular for large values of $\mu$ and massive clouds, the white dwarf could grow into a low-mass neutron star. On the other hand, low-mass initial scalar fields form more dilute BSs and the fermion star evolves into a white dwarf with larger central density and whose energy density profile is akin to that of an isolated white dwarf with the same central density. However, the total mass of the mixed star can be smaller or larger, producing a different gravitational redshift effect at the surface of the white dwarf. Assuming that the electromagnetic emission of the fermion star is close to the emission of the equivalent isolated white dwarf with same central density, the effect of the dark matter core could induce slight differences in the mass estimates. 

In astronomical observations of white dwarfs, the gravitational redshift can be measured from the electromagnetic emission that occurs mainly in the photosphere and is used to derive the mass of the star. We have seen, as a proof of concept, that if BSs form inside white dwarfs, the gravitational redshift changes depending on the parameters of the final BS. If the estimated mass given by the gravitational redshift is not consistent with the mass given by other methods, this difference could suggest the presence of a massive dark object that only interacts through gravity. We have used BSs described by a complex scalar field minimally coupled to gravity, but in principle any dark matter object could produce similar variations in the gravitational redshift if it accumulates within fermion stars.

The mass estimates of white dwarfs using the gravitational redshift are in very good agreement with theoretical mass-radius relations~\cite{parsons2017testing,joyce2018gravitational,romero2019white,chandra2020gravitational}. However, such estimates are still subject to uncertainties~\cite{barstow2017sirius}. For instance, in~\cite{pasquini2019masses} the masses of the Hyades white dwarfs were measured using the gravitational redshift, finding that the error in the mass-radius is estimated to be about 5\% (2\% in Sirius B~\cite{barstow2017sirius,joyce2018gravitational}), obtaining systematically smaller masses. These results indicate that any variation of the gravitational redshift produce by dark matter should be small and probably difficult to decouple from uncertainties associated with observations. If an ultralight bosonic field could form a bosonic star inside a white dwarf, the star should be dilute enough and produce a modification in the gravitational redshift within those errors. In this sense, it would be interesting to study if the vector BSs recently proposed from the analysis of GW190521~\cite{bustillo2021gw190521}, made of an ultralight bosonic particle with $\mu_V\sim10^{-12}$ eV and maximum mass $M_{\rm{max}}\sim170\,M_{\odot}$, would have masses and compactness similar to white dwarfs, resulting in smaller variations in the gravitational field than those presented here.

This study could be extended to more realistic equation of states describing white dwarfs together with smaller BS masses that would give more accurate results, and to build systematic static models of mixed white-dwarf--boson stars  as in~\cite{valdez2013dynamical,valdez2020fermion,di2020dynamical,di2021dynamical,DiGiovanni:2021ejn} to explore the parameter space of these solutions. The ultralight bosonic particle mass of BSs or, in general, gravitational properties of dark stars could be derived from the measurement of this effect.

%%%%%%%%%%%%%%%%%%%%%%%%%%%%%%%%%%%%%%%%%%%%%%%
\section*{Acknowledgements}
%%%%%%%%%%%%%%%%%%%%%%%%%%%%%%%%%%%%%%%%%%%%%%%

We thank Pablo Rodriguez-Gil, Pedro Cunha, Fabrizio Di Giovanni, and Jos\'e A. Font for useful discussions and valuable comments. PI acknowledges financial support from the Spanish Ministry of Economy and Competitiveness (MINECO) under the 2015 Severo Ochoa Programme MINECO SEV–2015–0548. This work was supported by the Center for Research and Development in Mathematics and Applications (CIDMA) through the Portuguese Foundation for Science and Technology (FCT - Funda\c c\~ao para a Ci\^encia e a Tecnologia), references UIDB/04106/2020 and UIDP/04106/2020, by national funds (OE), through FCT, I.P., in the scope of the framework contract foreseen in the numbers 4, 5 and 6 of the article 23, of the Decree-Law 57/2016, of August 29, changed by Law 57/2017, of July 19 and by the projects PTDC/FIS-OUT/28407/2017,  CERN/FIS-PAR/0027/2019, PTDC/FIS-AST/3041/2020, and CERN/FIS-PAR/0024/2021. This work has further been supported by  the  European  Union's  Horizon  2020  research  and  innovation  (RISE) programme H2020-MSCA-RISE-2017 Grant No.~FunFiCO-777740. NSG was also supported by the Spanish Ministerio de Universidades, reference UP2021-044, within the European Union-Next Generation EU. We would like to acknowledge networking support by the COST Action GWverse CA16104 through the Short Term Scientific Mission CA16104-48211.

%%%%%%%%%%%%%%%%%%%%%%
%%%   REFERENCES   %%%
%%%%%%%%%%%%%%%%%%%%%%

\bibliography{num-rel}

\begin{thebibliography}{88}
\expandafter\ifx\csname natexlab\endcsname\relax\def\natexlab#1{#1}\fi
\expandafter\ifx\csname bibnamefont\endcsname\relax
  \def\bibnamefont#1{#1}\fi
\expandafter\ifx\csname bibfnamefont\endcsname\relax
  \def\bibfnamefont#1{#1}\fi
\expandafter\ifx\csname citenamefont\endcsname\relax
  \def\citenamefont#1{#1}\fi
\expandafter\ifx\csname url\endcsname\relax
  \def\url#1{\texttt{#1}}\fi
\expandafter\ifx\csname urlprefix\endcsname\relax\def\urlprefix{URL }\fi
\providecommand{\bibinfo}[2]{#2}
\providecommand{\eprint}[2][]{\url{#2}}

\bibitem[{\citenamefont{Abbott et~al.}(2016)\citenamefont{Abbott, Abbott,
  Abbott, Abernathy, Acernese, Ackley, Adams, Adams, Addesso, Adhikari
  et~al.}}]{abbott2016observation}
\bibinfo{author}{\bibfnamefont{B.~P.} \bibnamefont{Abbott}},
  \bibinfo{author}{\bibfnamefont{R.}~\bibnamefont{Abbott}},
  \bibinfo{author}{\bibfnamefont{T.}~\bibnamefont{Abbott}},
  \bibinfo{author}{\bibfnamefont{M.}~\bibnamefont{Abernathy}},
  \bibinfo{author}{\bibfnamefont{F.}~\bibnamefont{Acernese}},
  \bibinfo{author}{\bibfnamefont{K.}~\bibnamefont{Ackley}},
  \bibinfo{author}{\bibfnamefont{C.}~\bibnamefont{Adams}},
  \bibinfo{author}{\bibfnamefont{T.}~\bibnamefont{Adams}},
  \bibinfo{author}{\bibfnamefont{P.}~\bibnamefont{Addesso}},
  \bibinfo{author}{\bibfnamefont{R.}~\bibnamefont{Adhikari}},
  \bibnamefont{et~al.}, \bibinfo{journal}{Physical review letters}
  \textbf{\bibinfo{volume}{116}}, \bibinfo{pages}{061102}
  (\bibinfo{year}{2016}).

\bibitem[{\citenamefont{Abbott et~al.}(2017)\citenamefont{Abbott, Abbott,
  Abbott, Acernese, Ackley, Adams, Adams, Addesso, Adhikari, Adya
  et~al.}}]{abbott2017gw170817}
\bibinfo{author}{\bibfnamefont{B.~P.} \bibnamefont{Abbott}},
  \bibinfo{author}{\bibfnamefont{R.}~\bibnamefont{Abbott}},
  \bibinfo{author}{\bibfnamefont{T.}~\bibnamefont{Abbott}},
  \bibinfo{author}{\bibfnamefont{F.}~\bibnamefont{Acernese}},
  \bibinfo{author}{\bibfnamefont{K.}~\bibnamefont{Ackley}},
  \bibinfo{author}{\bibfnamefont{C.}~\bibnamefont{Adams}},
  \bibinfo{author}{\bibfnamefont{T.}~\bibnamefont{Adams}},
  \bibinfo{author}{\bibfnamefont{P.}~\bibnamefont{Addesso}},
  \bibinfo{author}{\bibfnamefont{R.}~\bibnamefont{Adhikari}},
  \bibinfo{author}{\bibfnamefont{V.}~\bibnamefont{Adya}}, \bibnamefont{et~al.},
  \bibinfo{journal}{Physical Review Letters} \textbf{\bibinfo{volume}{119}},
  \bibinfo{pages}{161101} (\bibinfo{year}{2017}).

\bibitem[{\citenamefont{Abbott et~al.}(2019)\citenamefont{Abbott, Abbott,
  Abbott, Abraham, Acernese, Ackley, Adams, Adhikari, Adya, Affeldt
  et~al.}}]{abbott2019gwtc}
\bibinfo{author}{\bibfnamefont{B.}~\bibnamefont{Abbott}},
  \bibinfo{author}{\bibfnamefont{R.}~\bibnamefont{Abbott}},
  \bibinfo{author}{\bibfnamefont{T.}~\bibnamefont{Abbott}},
  \bibinfo{author}{\bibfnamefont{S.}~\bibnamefont{Abraham}},
  \bibinfo{author}{\bibfnamefont{F.}~\bibnamefont{Acernese}},
  \bibinfo{author}{\bibfnamefont{K.}~\bibnamefont{Ackley}},
  \bibinfo{author}{\bibfnamefont{C.}~\bibnamefont{Adams}},
  \bibinfo{author}{\bibfnamefont{R.}~\bibnamefont{Adhikari}},
  \bibinfo{author}{\bibfnamefont{V.}~\bibnamefont{Adya}},
  \bibinfo{author}{\bibfnamefont{C.}~\bibnamefont{Affeldt}},
  \bibnamefont{et~al.}, \bibinfo{journal}{Physical Review X}
  \textbf{\bibinfo{volume}{9}}, \bibinfo{pages}{031040} (\bibinfo{year}{2019}).

\bibitem[{\citenamefont{Abbott et~al.}(2020{\natexlab{a}})\citenamefont{Abbott,
  Abbott, Abraham, Acernese, Ackley, Adams, Adams, Adhikari, Adya, Affeldt
  et~al.}}]{abbott2020gwtc}
\bibinfo{author}{\bibfnamefont{R.}~\bibnamefont{Abbott}},
  \bibinfo{author}{\bibfnamefont{T.}~\bibnamefont{Abbott}},
  \bibinfo{author}{\bibfnamefont{S.}~\bibnamefont{Abraham}},
  \bibinfo{author}{\bibfnamefont{F.}~\bibnamefont{Acernese}},
  \bibinfo{author}{\bibfnamefont{K.}~\bibnamefont{Ackley}},
  \bibinfo{author}{\bibfnamefont{A.}~\bibnamefont{Adams}},
  \bibinfo{author}{\bibfnamefont{C.}~\bibnamefont{Adams}},
  \bibinfo{author}{\bibfnamefont{R.}~\bibnamefont{Adhikari}},
  \bibinfo{author}{\bibfnamefont{V.}~\bibnamefont{Adya}},
  \bibinfo{author}{\bibfnamefont{C.}~\bibnamefont{Affeldt}},
  \bibnamefont{et~al.}, \bibinfo{journal}{arXiv preprint arXiv:2010.14527}
  (\bibinfo{year}{2020}{\natexlab{a}}).

\bibitem[{\citenamefont{Abbott et~al.}(2020{\natexlab{b}})\citenamefont{Abbott,
  Abbott, Abraham, Acernese, Ackley, Adams, Adhikari, Adya, Affeldt, Agathos
  et~al.}}]{abbott2020gw190814}
\bibinfo{author}{\bibfnamefont{R.}~\bibnamefont{Abbott}},
  \bibinfo{author}{\bibfnamefont{T.}~\bibnamefont{Abbott}},
  \bibinfo{author}{\bibfnamefont{S.}~\bibnamefont{Abraham}},
  \bibinfo{author}{\bibfnamefont{F.}~\bibnamefont{Acernese}},
  \bibinfo{author}{\bibfnamefont{K.}~\bibnamefont{Ackley}},
  \bibinfo{author}{\bibfnamefont{C.}~\bibnamefont{Adams}},
  \bibinfo{author}{\bibfnamefont{R.}~\bibnamefont{Adhikari}},
  \bibinfo{author}{\bibfnamefont{V.}~\bibnamefont{Adya}},
  \bibinfo{author}{\bibfnamefont{C.}~\bibnamefont{Affeldt}},
  \bibinfo{author}{\bibfnamefont{M.}~\bibnamefont{Agathos}},
  \bibnamefont{et~al.}, \bibinfo{journal}{The Astrophysical Journal Letters}
  \textbf{\bibinfo{volume}{896}}, \bibinfo{pages}{L44}
  (\bibinfo{year}{2020}{\natexlab{b}}).

\bibitem[{\citenamefont{Abbott et~al.}(2020{\natexlab{c}})\citenamefont{Abbott,
  Abbott, Abraham, Acernese, Ackley, Adams, Adhikari, Adya, Affeldt, Agathos
  et~al.}}]{abbott2020gw190521}
\bibinfo{author}{\bibfnamefont{R.}~\bibnamefont{Abbott}},
  \bibinfo{author}{\bibfnamefont{T.}~\bibnamefont{Abbott}},
  \bibinfo{author}{\bibfnamefont{S.}~\bibnamefont{Abraham}},
  \bibinfo{author}{\bibfnamefont{F.}~\bibnamefont{Acernese}},
  \bibinfo{author}{\bibfnamefont{K.}~\bibnamefont{Ackley}},
  \bibinfo{author}{\bibfnamefont{C.}~\bibnamefont{Adams}},
  \bibinfo{author}{\bibfnamefont{R.}~\bibnamefont{Adhikari}},
  \bibinfo{author}{\bibfnamefont{V.}~\bibnamefont{Adya}},
  \bibinfo{author}{\bibfnamefont{C.}~\bibnamefont{Affeldt}},
  \bibinfo{author}{\bibfnamefont{M.}~\bibnamefont{Agathos}},
  \bibnamefont{et~al.}, \bibinfo{journal}{Physical review letters}
  \textbf{\bibinfo{volume}{125}}, \bibinfo{pages}{101102}
  (\bibinfo{year}{2020}{\natexlab{c}}).

\bibitem[{\citenamefont{Collaboration et~al.}(2019)}]{event2019first}
\bibinfo{author}{\bibfnamefont{E.~H.~T.} \bibnamefont{Collaboration}}
  \bibnamefont{et~al.}, \bibinfo{journal}{arXiv preprint arXiv:1906.11238}
  (\bibinfo{year}{2019}).

\bibitem[{\citenamefont{Akiyama et~al.}(2019)\citenamefont{Akiyama, Alberdi,
  Alef, Asada, Azulay, Baczko, Ball, Balokovi{\'c}, Barrett, Bintley
  et~al.}}]{akiyama2019first}
\bibinfo{author}{\bibfnamefont{K.}~\bibnamefont{Akiyama}},
  \bibinfo{author}{\bibfnamefont{A.}~\bibnamefont{Alberdi}},
  \bibinfo{author}{\bibfnamefont{W.}~\bibnamefont{Alef}},
  \bibinfo{author}{\bibfnamefont{K.}~\bibnamefont{Asada}},
  \bibinfo{author}{\bibfnamefont{R.}~\bibnamefont{Azulay}},
  \bibinfo{author}{\bibfnamefont{A.-K.} \bibnamefont{Baczko}},
  \bibinfo{author}{\bibfnamefont{D.}~\bibnamefont{Ball}},
  \bibinfo{author}{\bibfnamefont{M.}~\bibnamefont{Balokovi{\'c}}},
  \bibinfo{author}{\bibfnamefont{J.}~\bibnamefont{Barrett}},
  \bibinfo{author}{\bibfnamefont{D.}~\bibnamefont{Bintley}},
  \bibnamefont{et~al.}, \bibinfo{journal}{The Astrophysical Journal Letters}
  \textbf{\bibinfo{volume}{875}}, \bibinfo{pages}{L5} (\bibinfo{year}{2019}).

\bibitem[{\citenamefont{Bustillo et~al.}(2021)\citenamefont{Bustillo,
  Sanchis-Gual, Torres-Forn{\'e}, Font, Vajpeyi, Smith, Herdeiro, Radu, and
  Leong}}]{bustillo2021gw190521}
\bibinfo{author}{\bibfnamefont{J.~C.} \bibnamefont{Bustillo}},
  \bibinfo{author}{\bibfnamefont{N.}~\bibnamefont{Sanchis-Gual}},
  \bibinfo{author}{\bibfnamefont{A.}~\bibnamefont{Torres-Forn{\'e}}},
  \bibinfo{author}{\bibfnamefont{J.~A.} \bibnamefont{Font}},
  \bibinfo{author}{\bibfnamefont{A.}~\bibnamefont{Vajpeyi}},
  \bibinfo{author}{\bibfnamefont{R.}~\bibnamefont{Smith}},
  \bibinfo{author}{\bibfnamefont{C.}~\bibnamefont{Herdeiro}},
  \bibinfo{author}{\bibfnamefont{E.}~\bibnamefont{Radu}}, \bibnamefont{and}
  \bibinfo{author}{\bibfnamefont{S.~H.} \bibnamefont{Leong}},
  \bibinfo{journal}{Physical Review Letters} \textbf{\bibinfo{volume}{126}},
  \bibinfo{pages}{081101} (\bibinfo{year}{2021}).

\bibitem[{\citenamefont{Olivares et~al.}(2020)\citenamefont{Olivares, Younsi,
  Fromm, De~Laurentis, Porth, Mizuno, Falcke, Kramer, and
  Rezzolla}}]{olivares2020tell}
\bibinfo{author}{\bibfnamefont{H.}~\bibnamefont{Olivares}},
  \bibinfo{author}{\bibfnamefont{Z.}~\bibnamefont{Younsi}},
  \bibinfo{author}{\bibfnamefont{C.~M.} \bibnamefont{Fromm}},
  \bibinfo{author}{\bibfnamefont{M.}~\bibnamefont{De~Laurentis}},
  \bibinfo{author}{\bibfnamefont{O.}~\bibnamefont{Porth}},
  \bibinfo{author}{\bibfnamefont{Y.}~\bibnamefont{Mizuno}},
  \bibinfo{author}{\bibfnamefont{H.}~\bibnamefont{Falcke}},
  \bibinfo{author}{\bibfnamefont{M.}~\bibnamefont{Kramer}}, \bibnamefont{and}
  \bibinfo{author}{\bibfnamefont{L.}~\bibnamefont{Rezzolla}},
  \bibinfo{journal}{Monthly Notices of the Royal Astronomical Society}
  \textbf{\bibinfo{volume}{497}}, \bibinfo{pages}{521} (\bibinfo{year}{2020}).

\bibitem[{\citenamefont{Herdeiro et~al.}(2021)\citenamefont{Herdeiro, Pombo,
  Radu, Cunha, and Sanchis-Gual}}]{herdeiro2021imitation}
\bibinfo{author}{\bibfnamefont{C.~A.} \bibnamefont{Herdeiro}},
  \bibinfo{author}{\bibfnamefont{A.~M.} \bibnamefont{Pombo}},
  \bibinfo{author}{\bibfnamefont{E.}~\bibnamefont{Radu}},
  \bibinfo{author}{\bibfnamefont{P.~V.} \bibnamefont{Cunha}}, \bibnamefont{and}
  \bibinfo{author}{\bibfnamefont{N.}~\bibnamefont{Sanchis-Gual}},
  \bibinfo{journal}{Journal of Cosmology and Astroparticle Physics}
  \textbf{\bibinfo{volume}{2021}}, \bibinfo{pages}{051} (\bibinfo{year}{2021}).

\bibitem[{\citenamefont{Herdeiro and Radu}(2014)}]{herdeiro2014kerr}
\bibinfo{author}{\bibfnamefont{C.~A.} \bibnamefont{Herdeiro}} \bibnamefont{and}
  \bibinfo{author}{\bibfnamefont{E.}~\bibnamefont{Radu}},
  \bibinfo{journal}{Physical review letters} \textbf{\bibinfo{volume}{112}},
  \bibinfo{pages}{221101} (\bibinfo{year}{2014}).

\bibitem[{\citenamefont{Kaup}(1968)}]{kaup1968klein}
\bibinfo{author}{\bibfnamefont{D.~J.} \bibnamefont{Kaup}},
  \bibinfo{journal}{Physical Review} \textbf{\bibinfo{volume}{172}},
  \bibinfo{pages}{1331} (\bibinfo{year}{1968}).

\bibitem[{\citenamefont{Ruffini and Bonazzola}(1969)}]{ruffini1969systems}
\bibinfo{author}{\bibfnamefont{R.}~\bibnamefont{Ruffini}} \bibnamefont{and}
  \bibinfo{author}{\bibfnamefont{S.}~\bibnamefont{Bonazzola}},
  \bibinfo{journal}{Physical Review} \textbf{\bibinfo{volume}{187}},
  \bibinfo{pages}{1767} (\bibinfo{year}{1969}).

\bibitem[{\citenamefont{Jetzer}(1992)}]{jetzer1992boson}
\bibinfo{author}{\bibfnamefont{P.}~\bibnamefont{Jetzer}},
  \bibinfo{journal}{Physics Reports} \textbf{\bibinfo{volume}{220}},
  \bibinfo{pages}{163} (\bibinfo{year}{1992}).

\bibitem[{\citenamefont{Schunck and Mielke}(2003)}]{schunck2003general}
\bibinfo{author}{\bibfnamefont{F.~E.} \bibnamefont{Schunck}} \bibnamefont{and}
  \bibinfo{author}{\bibfnamefont{E.~W.} \bibnamefont{Mielke}},
  \bibinfo{journal}{Classical and Quantum Gravity}
  \textbf{\bibinfo{volume}{20}}, \bibinfo{pages}{R301} (\bibinfo{year}{2003}).

\bibitem[{\citenamefont{Liebling and Palenzuela}(2017)}]{liebling2017dynamical}
\bibinfo{author}{\bibfnamefont{S.~L.} \bibnamefont{Liebling}} \bibnamefont{and}
  \bibinfo{author}{\bibfnamefont{C.}~\bibnamefont{Palenzuela}},
  \bibinfo{journal}{Living Reviews in Relativity}
  \textbf{\bibinfo{volume}{20}}, \bibinfo{pages}{5} (\bibinfo{year}{2017}).

\bibitem[{\citenamefont{Arvanitaki et~al.}(2010)\citenamefont{Arvanitaki,
  Dimopoulos, Dubovsky, Kaloper, and March-Russell}}]{arvanitaki2010string}
\bibinfo{author}{\bibfnamefont{A.}~\bibnamefont{Arvanitaki}},
  \bibinfo{author}{\bibfnamefont{S.}~\bibnamefont{Dimopoulos}},
  \bibinfo{author}{\bibfnamefont{S.}~\bibnamefont{Dubovsky}},
  \bibinfo{author}{\bibfnamefont{N.}~\bibnamefont{Kaloper}}, \bibnamefont{and}
  \bibinfo{author}{\bibfnamefont{J.}~\bibnamefont{March-Russell}},
  \bibinfo{journal}{Physical Review D} \textbf{\bibinfo{volume}{81}},
  \bibinfo{pages}{123530} (\bibinfo{year}{2010}).

\bibitem[{\citenamefont{Freitas et~al.}(2021)\citenamefont{Freitas, Herdeiro,
  Morais, Onofre, Pasechnik, Radu, Sanchis-Gual, and Santos}}]{Freitas:2021cfi}
\bibinfo{author}{\bibfnamefont{F.~F.} \bibnamefont{Freitas}},
  \bibinfo{author}{\bibfnamefont{C.~A.~R.} \bibnamefont{Herdeiro}},
  \bibinfo{author}{\bibfnamefont{A.~P.} \bibnamefont{Morais}},
  \bibinfo{author}{\bibfnamefont{A.}~\bibnamefont{Onofre}},
  \bibinfo{author}{\bibfnamefont{R.}~\bibnamefont{Pasechnik}},
  \bibinfo{author}{\bibfnamefont{E.}~\bibnamefont{Radu}},
  \bibinfo{author}{\bibfnamefont{N.}~\bibnamefont{Sanchis-Gual}},
  \bibnamefont{and} \bibinfo{author}{\bibfnamefont{R.}~\bibnamefont{Santos}}
  (\bibinfo{year}{2021}), \eprint{2107.09493}.

\bibitem[{\citenamefont{Matos et~al.}(2000)\citenamefont{Matos, Guzm{\'a}n, and
  Urena-L{\'o}pez}}]{matos2000scalar}
\bibinfo{author}{\bibfnamefont{T.}~\bibnamefont{Matos}},
  \bibinfo{author}{\bibfnamefont{F.~S.} \bibnamefont{Guzm{\'a}n}},
  \bibnamefont{and} \bibinfo{author}{\bibfnamefont{L.~A.}
  \bibnamefont{Urena-L{\'o}pez}}, \bibinfo{journal}{Classical and Quantum
  Gravity} \textbf{\bibinfo{volume}{17}}, \bibinfo{pages}{1707}
  (\bibinfo{year}{2000}).

\bibitem[{\citenamefont{Hu et~al.}(2000)\citenamefont{Hu, Barkana, and
  Gruzinov}}]{hu2000fuzzy}
\bibinfo{author}{\bibfnamefont{W.}~\bibnamefont{Hu}},
  \bibinfo{author}{\bibfnamefont{R.}~\bibnamefont{Barkana}}, \bibnamefont{and}
  \bibinfo{author}{\bibfnamefont{A.}~\bibnamefont{Gruzinov}},
  \bibinfo{journal}{Physical Review Letters} \textbf{\bibinfo{volume}{85}},
  \bibinfo{pages}{1158} (\bibinfo{year}{2000}).

\bibitem[{\citenamefont{Brito et~al.}(2016{\natexlab{a}})\citenamefont{Brito,
  Cardoso, Herdeiro, and Radu}}]{brito2016proca}
\bibinfo{author}{\bibfnamefont{R.}~\bibnamefont{Brito}},
  \bibinfo{author}{\bibfnamefont{V.}~\bibnamefont{Cardoso}},
  \bibinfo{author}{\bibfnamefont{C.~A.} \bibnamefont{Herdeiro}},
  \bibnamefont{and} \bibinfo{author}{\bibfnamefont{E.}~\bibnamefont{Radu}},
  \bibinfo{journal}{Physics Letters B} \textbf{\bibinfo{volume}{752}},
  \bibinfo{pages}{291} (\bibinfo{year}{2016}{\natexlab{a}}).

\bibitem[{\citenamefont{Seidel and Suen}(1994)}]{seidel1994formation}
\bibinfo{author}{\bibfnamefont{E.}~\bibnamefont{Seidel}} \bibnamefont{and}
  \bibinfo{author}{\bibfnamefont{W.-M.} \bibnamefont{Suen}},
  \bibinfo{journal}{Physical review letters} \textbf{\bibinfo{volume}{72}},
  \bibinfo{pages}{2516} (\bibinfo{year}{1994}).

\bibitem[{\citenamefont{Guzman and
  Urena-Lopez}(2006)}]{guzman2006gravitational}
\bibinfo{author}{\bibfnamefont{F.~S.} \bibnamefont{Guzman}} \bibnamefont{and}
  \bibinfo{author}{\bibfnamefont{L.~A.} \bibnamefont{Urena-Lopez}},
  \bibinfo{journal}{The Astrophysical Journal} \textbf{\bibinfo{volume}{645}},
  \bibinfo{pages}{814} (\bibinfo{year}{2006}).

\bibitem[{\citenamefont{Di~Giovanni et~al.}(2018)\citenamefont{Di~Giovanni,
  Sanchis-Gual, Herdeiro, and Font}}]{di2018dynamical}
\bibinfo{author}{\bibfnamefont{F.}~\bibnamefont{Di~Giovanni}},
  \bibinfo{author}{\bibfnamefont{N.}~\bibnamefont{Sanchis-Gual}},
  \bibinfo{author}{\bibfnamefont{C.~A.} \bibnamefont{Herdeiro}},
  \bibnamefont{and} \bibinfo{author}{\bibfnamefont{J.~A.} \bibnamefont{Font}},
  \bibinfo{journal}{Physical Review D} \textbf{\bibinfo{volume}{98}},
  \bibinfo{pages}{064044} (\bibinfo{year}{2018}).

\bibitem[{\citenamefont{Gleiser}(1988)}]{gleiser1988stability}
\bibinfo{author}{\bibfnamefont{M.}~\bibnamefont{Gleiser}},
  \bibinfo{journal}{Physical Review D} \textbf{\bibinfo{volume}{38}},
  \bibinfo{pages}{2376} (\bibinfo{year}{1988}).

\bibitem[{\citenamefont{Gleiser and Watkins}(1989)}]{gleiser1989gravitational}
\bibinfo{author}{\bibfnamefont{M.}~\bibnamefont{Gleiser}} \bibnamefont{and}
  \bibinfo{author}{\bibfnamefont{R.}~\bibnamefont{Watkins}},
  \bibinfo{journal}{Nuclear Physics B} \textbf{\bibinfo{volume}{319}},
  \bibinfo{pages}{733} (\bibinfo{year}{1989}).

\bibitem[{\citenamefont{Lee and Pang}(1989{\natexlab{a}})}]{lee1989stability}
\bibinfo{author}{\bibfnamefont{T.}~\bibnamefont{Lee}} \bibnamefont{and}
  \bibinfo{author}{\bibfnamefont{Y.}~\bibnamefont{Pang}},
  \bibinfo{journal}{Nuclear Physics B} \textbf{\bibinfo{volume}{315}},
  \bibinfo{pages}{477} (\bibinfo{year}{1989}{\natexlab{a}}).

\bibitem[{\citenamefont{Seidel and Suen}(1990)}]{seidel1990dynamical}
\bibinfo{author}{\bibfnamefont{E.}~\bibnamefont{Seidel}} \bibnamefont{and}
  \bibinfo{author}{\bibfnamefont{W.-M.} \bibnamefont{Suen}},
  \bibinfo{journal}{Physical Review D} \textbf{\bibinfo{volume}{42}},
  \bibinfo{pages}{384} (\bibinfo{year}{1990}).

\bibitem[{\citenamefont{Hawley and Choptuik}(2000)}]{hawley2000boson}
\bibinfo{author}{\bibfnamefont{S.~H.} \bibnamefont{Hawley}} \bibnamefont{and}
  \bibinfo{author}{\bibfnamefont{M.~W.} \bibnamefont{Choptuik}},
  \bibinfo{journal}{Physical Review D} \textbf{\bibinfo{volume}{62}},
  \bibinfo{pages}{104024} (\bibinfo{year}{2000}).

\bibitem[{\citenamefont{Guzm{\'a}n}(2009)}]{guzman2009three}
\bibinfo{author}{\bibfnamefont{F.}~\bibnamefont{Guzm{\'a}n}},
  \bibinfo{journal}{Revista mexicana de f{\'\i}sica}
  \textbf{\bibinfo{volume}{55}}, \bibinfo{pages}{321} (\bibinfo{year}{2009}).

\bibitem[{\citenamefont{Escorihuela-Tom\`as
  et~al.}(2017)\citenamefont{Escorihuela-Tom\`as, Sanchis-Gual, Degollado, and
  Font}}]{Escorihuela-Tomas:2017uac}
\bibinfo{author}{\bibfnamefont{A.}~\bibnamefont{Escorihuela-Tom\`as}},
  \bibinfo{author}{\bibfnamefont{N.}~\bibnamefont{Sanchis-Gual}},
  \bibinfo{author}{\bibfnamefont{J.~C.} \bibnamefont{Degollado}},
  \bibnamefont{and} \bibinfo{author}{\bibfnamefont{J.~A.} \bibnamefont{Font}},
  \bibinfo{journal}{Phys. Rev. D} \textbf{\bibinfo{volume}{96}},
  \bibinfo{pages}{024015} (\bibinfo{year}{2017}), \eprint{1704.08023}.

\bibitem[{\citenamefont{Sanchis-Gual et~al.}(2017)\citenamefont{Sanchis-Gual,
  Herdeiro, Radu, Degollado, and Font}}]{sanchis2017numerical}
\bibinfo{author}{\bibfnamefont{N.}~\bibnamefont{Sanchis-Gual}},
  \bibinfo{author}{\bibfnamefont{C.}~\bibnamefont{Herdeiro}},
  \bibinfo{author}{\bibfnamefont{E.}~\bibnamefont{Radu}},
  \bibinfo{author}{\bibfnamefont{J.~C.} \bibnamefont{Degollado}},
  \bibnamefont{and} \bibinfo{author}{\bibfnamefont{J.~A.} \bibnamefont{Font}},
  \bibinfo{journal}{Physical Review D} \textbf{\bibinfo{volume}{95}},
  \bibinfo{pages}{104028} (\bibinfo{year}{2017}).

\bibitem[{\citenamefont{Sanchis-Gual
  et~al.}(2019{\natexlab{a}})\citenamefont{Sanchis-Gual, Di~Giovanni,
  Zilh{\~a}o, Herdeiro, Cerd{\'a}-Dur{\'a}n, Font, and
  Radu}}]{sanchis2019nonlinear}
\bibinfo{author}{\bibfnamefont{N.}~\bibnamefont{Sanchis-Gual}},
  \bibinfo{author}{\bibfnamefont{F.}~\bibnamefont{Di~Giovanni}},
  \bibinfo{author}{\bibfnamefont{M.}~\bibnamefont{Zilh{\~a}o}},
  \bibinfo{author}{\bibfnamefont{C.}~\bibnamefont{Herdeiro}},
  \bibinfo{author}{\bibfnamefont{P.}~\bibnamefont{Cerd{\'a}-Dur{\'a}n}},
  \bibinfo{author}{\bibfnamefont{J.}~\bibnamefont{Font}}, \bibnamefont{and}
  \bibinfo{author}{\bibfnamefont{E.}~\bibnamefont{Radu}},
  \bibinfo{journal}{Physical review letters} \textbf{\bibinfo{volume}{123}},
  \bibinfo{pages}{221101} (\bibinfo{year}{2019}{\natexlab{a}}).

\bibitem[{\citenamefont{Di~Giovanni
  et~al.}(2020{\natexlab{a}})\citenamefont{Di~Giovanni, Sanchis-Gual,
  Cerd\'a-Dur\'an, Zilh\~ao, Herdeiro, Font, and Radu}}]{DiGiovanni:2020ror}
\bibinfo{author}{\bibfnamefont{F.}~\bibnamefont{Di~Giovanni}},
  \bibinfo{author}{\bibfnamefont{N.}~\bibnamefont{Sanchis-Gual}},
  \bibinfo{author}{\bibfnamefont{P.}~\bibnamefont{Cerd\'a-Dur\'an}},
  \bibinfo{author}{\bibfnamefont{M.}~\bibnamefont{Zilh\~ao}},
  \bibinfo{author}{\bibfnamefont{C.}~\bibnamefont{Herdeiro}},
  \bibinfo{author}{\bibfnamefont{J.~A.} \bibnamefont{Font}}, \bibnamefont{and}
  \bibinfo{author}{\bibfnamefont{E.}~\bibnamefont{Radu}},
  \bibinfo{journal}{Phys. Rev. D} \textbf{\bibinfo{volume}{102}},
  \bibinfo{pages}{124009} (\bibinfo{year}{2020}{\natexlab{a}}),
  \eprint{2010.05845}.

\bibitem[{\citenamefont{Siemonsen and East}(2021)}]{siemonsen2021stability}
\bibinfo{author}{\bibfnamefont{N.}~\bibnamefont{Siemonsen}} \bibnamefont{and}
  \bibinfo{author}{\bibfnamefont{W.~E.} \bibnamefont{East}},
  \bibinfo{journal}{Physical Review D} \textbf{\bibinfo{volume}{103}},
  \bibinfo{pages}{044022} (\bibinfo{year}{2021}).

\bibitem[{\citenamefont{Palenzuela et~al.}(2017)\citenamefont{Palenzuela, Pani,
  Bezares, Cardoso, Lehner, and Liebling}}]{palenzuela2017gravitational}
\bibinfo{author}{\bibfnamefont{C.}~\bibnamefont{Palenzuela}},
  \bibinfo{author}{\bibfnamefont{P.}~\bibnamefont{Pani}},
  \bibinfo{author}{\bibfnamefont{M.}~\bibnamefont{Bezares}},
  \bibinfo{author}{\bibfnamefont{V.}~\bibnamefont{Cardoso}},
  \bibinfo{author}{\bibfnamefont{L.}~\bibnamefont{Lehner}}, \bibnamefont{and}
  \bibinfo{author}{\bibfnamefont{S.}~\bibnamefont{Liebling}},
  \bibinfo{journal}{Physical Review D} \textbf{\bibinfo{volume}{96}},
  \bibinfo{pages}{104058} (\bibinfo{year}{2017}).

\bibitem[{\citenamefont{Bezares et~al.}(2017)\citenamefont{Bezares, Palenzuela,
  and Bona}}]{bezares2017final}
\bibinfo{author}{\bibfnamefont{M.}~\bibnamefont{Bezares}},
  \bibinfo{author}{\bibfnamefont{C.}~\bibnamefont{Palenzuela}},
  \bibnamefont{and} \bibinfo{author}{\bibfnamefont{C.}~\bibnamefont{Bona}},
  \bibinfo{journal}{Physical Review D} \textbf{\bibinfo{volume}{95}},
  \bibinfo{pages}{124005} (\bibinfo{year}{2017}).

\bibitem[{\citenamefont{Bezares and
  Palenzuela}(2018)}]{bezares2018gravitational}
\bibinfo{author}{\bibfnamefont{M.}~\bibnamefont{Bezares}} \bibnamefont{and}
  \bibinfo{author}{\bibfnamefont{C.}~\bibnamefont{Palenzuela}},
  \bibinfo{journal}{Classical and Quantum Gravity}
  \textbf{\bibinfo{volume}{35}}, \bibinfo{pages}{234002}
  (\bibinfo{year}{2018}).

\bibitem[{\citenamefont{Sanchis-Gual
  et~al.}(2019{\natexlab{b}})\citenamefont{Sanchis-Gual, Herdeiro, Font, Radu,
  and Di~Giovanni}}]{sanchis2019head}
\bibinfo{author}{\bibfnamefont{N.}~\bibnamefont{Sanchis-Gual}},
  \bibinfo{author}{\bibfnamefont{C.}~\bibnamefont{Herdeiro}},
  \bibinfo{author}{\bibfnamefont{J.~A.} \bibnamefont{Font}},
  \bibinfo{author}{\bibfnamefont{E.}~\bibnamefont{Radu}}, \bibnamefont{and}
  \bibinfo{author}{\bibfnamefont{F.}~\bibnamefont{Di~Giovanni}},
  \bibinfo{journal}{Physical Review D} \textbf{\bibinfo{volume}{99}},
  \bibinfo{pages}{024017} (\bibinfo{year}{2019}{\natexlab{b}}).

\bibitem[{\citenamefont{Bezares et~al.}(2022)\citenamefont{Bezares,
  Bo{\v{s}}kovi{\'c}, Liebling, Palenzuela, Pani, and
  Barausse}}]{bezares2022gravitational}
\bibinfo{author}{\bibfnamefont{M.}~\bibnamefont{Bezares}},
  \bibinfo{author}{\bibfnamefont{M.}~\bibnamefont{Bo{\v{s}}kovi{\'c}}},
  \bibinfo{author}{\bibfnamefont{S.}~\bibnamefont{Liebling}},
  \bibinfo{author}{\bibfnamefont{C.}~\bibnamefont{Palenzuela}},
  \bibinfo{author}{\bibfnamefont{P.}~\bibnamefont{Pani}}, \bibnamefont{and}
  \bibinfo{author}{\bibfnamefont{E.}~\bibnamefont{Barausse}},
  \bibinfo{journal}{arXiv preprint arXiv:2201.06113}  (\bibinfo{year}{2022}).

\bibitem[{\citenamefont{Cunha et~al.}(2015)\citenamefont{Cunha, Herdeiro, Radu,
  and R{\'u}narsson}}]{cunha2015shadows}
\bibinfo{author}{\bibfnamefont{P.~V.} \bibnamefont{Cunha}},
  \bibinfo{author}{\bibfnamefont{C.~A.} \bibnamefont{Herdeiro}},
  \bibinfo{author}{\bibfnamefont{E.}~\bibnamefont{Radu}}, \bibnamefont{and}
  \bibinfo{author}{\bibfnamefont{H.~F.} \bibnamefont{R{\'u}narsson}},
  \bibinfo{journal}{Physical review letters} \textbf{\bibinfo{volume}{115}},
  \bibinfo{pages}{211102} (\bibinfo{year}{2015}).

\bibitem[{\citenamefont{Cunha et~al.}(2017)\citenamefont{Cunha, Font, Herdeiro,
  Radu, Sanchis-Gual, and Zilhao}}]{cunha2017lensing}
\bibinfo{author}{\bibfnamefont{P.~V.} \bibnamefont{Cunha}},
  \bibinfo{author}{\bibfnamefont{J.~A.} \bibnamefont{Font}},
  \bibinfo{author}{\bibfnamefont{C.}~\bibnamefont{Herdeiro}},
  \bibinfo{author}{\bibfnamefont{E.}~\bibnamefont{Radu}},
  \bibinfo{author}{\bibfnamefont{N.}~\bibnamefont{Sanchis-Gual}},
  \bibnamefont{and} \bibinfo{author}{\bibfnamefont{M.}~\bibnamefont{Zilhao}},
  \bibinfo{journal}{Physical Review D} \textbf{\bibinfo{volume}{96}},
  \bibinfo{pages}{104040} (\bibinfo{year}{2017}).

\bibitem[{\citenamefont{Jetzer}(1990)}]{jetzer1990stability}
\bibinfo{author}{\bibfnamefont{P.}~\bibnamefont{Jetzer}},
  \bibinfo{journal}{Physics Letters B} \textbf{\bibinfo{volume}{243}},
  \bibinfo{pages}{36} (\bibinfo{year}{1990}).

\bibitem[{\citenamefont{Henriques et~al.}(1990)\citenamefont{Henriques, Liddle,
  and Moorhouse}}]{henriques1990stability}
\bibinfo{author}{\bibfnamefont{A.}~\bibnamefont{Henriques}},
  \bibinfo{author}{\bibfnamefont{A.~R.} \bibnamefont{Liddle}},
  \bibnamefont{and}
  \bibinfo{author}{\bibfnamefont{R.}~\bibnamefont{Moorhouse}},
  \bibinfo{journal}{Physics Letters B} \textbf{\bibinfo{volume}{251}},
  \bibinfo{pages}{511} (\bibinfo{year}{1990}).

\bibitem[{\citenamefont{Valdez-Alvarado
  et~al.}(2013)\citenamefont{Valdez-Alvarado, Palenzuela, Alic, and
  Urena-L{\'o}pez}}]{valdez2013dynamical}
\bibinfo{author}{\bibfnamefont{S.}~\bibnamefont{Valdez-Alvarado}},
  \bibinfo{author}{\bibfnamefont{C.}~\bibnamefont{Palenzuela}},
  \bibinfo{author}{\bibfnamefont{D.}~\bibnamefont{Alic}}, \bibnamefont{and}
  \bibinfo{author}{\bibfnamefont{L.~A.} \bibnamefont{Urena-L{\'o}pez}},
  \bibinfo{journal}{Physical Review D} \textbf{\bibinfo{volume}{87}},
  \bibinfo{pages}{084040} (\bibinfo{year}{2013}).

\bibitem[{\citenamefont{Valdez-Alvarado
  et~al.}(2020)\citenamefont{Valdez-Alvarado, Becerril, and
  Ure{\~n}a-L{\'o}pez}}]{valdez2020fermion}
\bibinfo{author}{\bibfnamefont{S.}~\bibnamefont{Valdez-Alvarado}},
  \bibinfo{author}{\bibfnamefont{R.}~\bibnamefont{Becerril}}, \bibnamefont{and}
  \bibinfo{author}{\bibfnamefont{L.~A.} \bibnamefont{Ure{\~n}a-L{\'o}pez}},
  \bibinfo{journal}{Physical Review D} \textbf{\bibinfo{volume}{102}},
  \bibinfo{pages}{064038} (\bibinfo{year}{2020}).

\bibitem[{\citenamefont{Di~Giovanni
  et~al.}(2020{\natexlab{b}})\citenamefont{Di~Giovanni, Fakhry, Sanchis-Gual,
  Degollado, and Font}}]{di2020dynamical}
\bibinfo{author}{\bibfnamefont{F.}~\bibnamefont{Di~Giovanni}},
  \bibinfo{author}{\bibfnamefont{S.}~\bibnamefont{Fakhry}},
  \bibinfo{author}{\bibfnamefont{N.}~\bibnamefont{Sanchis-Gual}},
  \bibinfo{author}{\bibfnamefont{J.~C.} \bibnamefont{Degollado}},
  \bibnamefont{and} \bibinfo{author}{\bibfnamefont{J.~A.} \bibnamefont{Font}},
  \bibinfo{journal}{Physical Review D} \textbf{\bibinfo{volume}{102}},
  \bibinfo{pages}{084063} (\bibinfo{year}{2020}{\natexlab{b}}).

\bibitem[{\citenamefont{Di~Giovanni
  et~al.}(2021{\natexlab{a}})\citenamefont{Di~Giovanni, Fakhry, Sanchis-Gual,
  Degollado, and Font}}]{di2021dynamical}
\bibinfo{author}{\bibfnamefont{F.}~\bibnamefont{Di~Giovanni}},
  \bibinfo{author}{\bibfnamefont{S.}~\bibnamefont{Fakhry}},
  \bibinfo{author}{\bibfnamefont{N.}~\bibnamefont{Sanchis-Gual}},
  \bibinfo{author}{\bibfnamefont{J.~C.} \bibnamefont{Degollado}},
  \bibnamefont{and} \bibinfo{author}{\bibfnamefont{J.~A.} \bibnamefont{Font}},
  \bibinfo{journal}{Class. Quant. Grav.} \textbf{\bibinfo{volume}{38}},
  \bibinfo{pages}{194001} (\bibinfo{year}{2021}{\natexlab{a}}),
  \eprint{2105.00530}.

\bibitem[{\citenamefont{Di~Giovanni
  et~al.}(2021{\natexlab{b}})\citenamefont{Di~Giovanni, Sanchis-Gual,
  Cerd\'a-Dur\'an, and Font}}]{DiGiovanni:2021ejn}
\bibinfo{author}{\bibfnamefont{F.}~\bibnamefont{Di~Giovanni}},
  \bibinfo{author}{\bibfnamefont{N.}~\bibnamefont{Sanchis-Gual}},
  \bibinfo{author}{\bibfnamefont{P.}~\bibnamefont{Cerd\'a-Dur\'an}},
  \bibnamefont{and} \bibinfo{author}{\bibfnamefont{J.~A.} \bibnamefont{Font}}
  (\bibinfo{year}{2021}{\natexlab{b}}), \eprint{2110.11997}.

\bibitem[{\citenamefont{Brito et~al.}(2015)\citenamefont{Brito, Cardoso, and
  Okawa}}]{brito2015accretion}
\bibinfo{author}{\bibfnamefont{R.}~\bibnamefont{Brito}},
  \bibinfo{author}{\bibfnamefont{V.}~\bibnamefont{Cardoso}}, \bibnamefont{and}
  \bibinfo{author}{\bibfnamefont{H.}~\bibnamefont{Okawa}},
  \bibinfo{journal}{Physical review letters} \textbf{\bibinfo{volume}{115}},
  \bibinfo{pages}{111301} (\bibinfo{year}{2015}).

\bibitem[{\citenamefont{Brito et~al.}(2016{\natexlab{b}})\citenamefont{Brito,
  Cardoso, Macedo, Okawa, and Palenzuela}}]{brito2016interaction}
\bibinfo{author}{\bibfnamefont{R.}~\bibnamefont{Brito}},
  \bibinfo{author}{\bibfnamefont{V.}~\bibnamefont{Cardoso}},
  \bibinfo{author}{\bibfnamefont{C.~F.} \bibnamefont{Macedo}},
  \bibinfo{author}{\bibfnamefont{H.}~\bibnamefont{Okawa}}, \bibnamefont{and}
  \bibinfo{author}{\bibfnamefont{C.}~\bibnamefont{Palenzuela}},
  \bibinfo{journal}{Physical Review D} \textbf{\bibinfo{volume}{93}},
  \bibinfo{pages}{044045} (\bibinfo{year}{2016}{\natexlab{b}}).

\bibitem[{\citenamefont{Bezares et~al.}(2019)\citenamefont{Bezares, Vigan{\`o},
  and Palenzuela}}]{bezares2019gravitational}
\bibinfo{author}{\bibfnamefont{M.}~\bibnamefont{Bezares}},
  \bibinfo{author}{\bibfnamefont{D.}~\bibnamefont{Vigan{\`o}}},
  \bibnamefont{and}
  \bibinfo{author}{\bibfnamefont{C.}~\bibnamefont{Palenzuela}},
  \bibinfo{journal}{Physical Review D} \textbf{\bibinfo{volume}{100}},
  \bibinfo{pages}{044049} (\bibinfo{year}{2019}).

\bibitem[{\citenamefont{Lee and Pang}(1989{\natexlab{b}})}]{Lee:1988av}
\bibinfo{author}{\bibfnamefont{T.~D.} \bibnamefont{Lee}} \bibnamefont{and}
  \bibinfo{author}{\bibfnamefont{Y.}~\bibnamefont{Pang}},
  \bibinfo{journal}{Nucl. Phys.} \textbf{\bibinfo{volume}{B315}},
  \bibinfo{pages}{477} (\bibinfo{year}{1989}{\natexlab{b}}),
  \bibinfo{note}{[,129(1988)]}.

\bibitem[{\citenamefont{Balakrishna et~al.}(1998)\citenamefont{Balakrishna,
  Seidel, and Suen}}]{balakrishna1998dynamical}
\bibinfo{author}{\bibfnamefont{J.}~\bibnamefont{Balakrishna}},
  \bibinfo{author}{\bibfnamefont{E.}~\bibnamefont{Seidel}}, \bibnamefont{and}
  \bibinfo{author}{\bibfnamefont{W.-M.} \bibnamefont{Suen}},
  \bibinfo{journal}{Physical Review D} \textbf{\bibinfo{volume}{58}},
  \bibinfo{pages}{104004} (\bibinfo{year}{1998}).

\bibitem[{\citenamefont{Sanchis-Gual et~al.}(2022)\citenamefont{Sanchis-Gual,
  Herdeiro, and Radu}}]{Sanchis-Gual:2021phr}
\bibinfo{author}{\bibfnamefont{N.}~\bibnamefont{Sanchis-Gual}},
  \bibinfo{author}{\bibfnamefont{C.~A.} \bibnamefont{Herdeiro}},
  \bibnamefont{and} \bibinfo{author}{\bibfnamefont{E.}~\bibnamefont{Radu}},
  \bibinfo{journal}{Classical and Quantum Gravity}  (\bibinfo{year}{2022}).

\bibitem[{\citenamefont{Bernal et~al.}(2010)\citenamefont{Bernal, Barranco,
  Alic, and Palenzuela}}]{bernal2010multistate}
\bibinfo{author}{\bibfnamefont{A.}~\bibnamefont{Bernal}},
  \bibinfo{author}{\bibfnamefont{J.}~\bibnamefont{Barranco}},
  \bibinfo{author}{\bibfnamefont{D.}~\bibnamefont{Alic}}, \bibnamefont{and}
  \bibinfo{author}{\bibfnamefont{C.}~\bibnamefont{Palenzuela}},
  \bibinfo{journal}{Physical Review D} \textbf{\bibinfo{volume}{81}},
  \bibinfo{pages}{044031} (\bibinfo{year}{2010}).

\bibitem[{\citenamefont{Sanchis-Gual et~al.}(2021)\citenamefont{Sanchis-Gual,
  Di~Giovanni, Herdeiro, Radu, and Font}}]{Sanchis-Gual:2021edp}
\bibinfo{author}{\bibfnamefont{N.}~\bibnamefont{Sanchis-Gual}},
  \bibinfo{author}{\bibfnamefont{F.}~\bibnamefont{Di~Giovanni}},
  \bibinfo{author}{\bibfnamefont{C.}~\bibnamefont{Herdeiro}},
  \bibinfo{author}{\bibfnamefont{E.}~\bibnamefont{Radu}}, \bibnamefont{and}
  \bibinfo{author}{\bibfnamefont{J.~A.} \bibnamefont{Font}},
  \bibinfo{journal}{Phys. Rev. Lett.} \textbf{\bibinfo{volume}{126}},
  \bibinfo{pages}{241105} (\bibinfo{year}{2021}), \eprint{2103.12136}.

\bibitem[{\citenamefont{Alcubierre et~al.}(2019)\citenamefont{Alcubierre,
  Barranco, Bernal, Degollado, Diez-Tejedor, Megevand, N{\'u}{\~n}ez, and
  Sarbach}}]{alcubierre2019dynamical}
\bibinfo{author}{\bibfnamefont{M.}~\bibnamefont{Alcubierre}},
  \bibinfo{author}{\bibfnamefont{J.}~\bibnamefont{Barranco}},
  \bibinfo{author}{\bibfnamefont{A.}~\bibnamefont{Bernal}},
  \bibinfo{author}{\bibfnamefont{J.~C.} \bibnamefont{Degollado}},
  \bibinfo{author}{\bibfnamefont{A.}~\bibnamefont{Diez-Tejedor}},
  \bibinfo{author}{\bibfnamefont{M.}~\bibnamefont{Megevand}},
  \bibinfo{author}{\bibfnamefont{D.}~\bibnamefont{N{\'u}{\~n}ez}},
  \bibnamefont{and} \bibinfo{author}{\bibfnamefont{O.}~\bibnamefont{Sarbach}},
  \bibinfo{journal}{Classical and Quantum Gravity}
  \textbf{\bibinfo{volume}{36}}, \bibinfo{pages}{215013}
  (\bibinfo{year}{2019}).

\bibitem[{\citenamefont{Guzm{\'a}n and
  Ure{\~n}a-L{\'o}pez}(2020)}]{guzman2020gravitational}
\bibinfo{author}{\bibfnamefont{F.}~\bibnamefont{Guzm{\'a}n}} \bibnamefont{and}
  \bibinfo{author}{\bibfnamefont{L.~A.} \bibnamefont{Ure{\~n}a-L{\'o}pez}},
  \bibinfo{journal}{Physical Review D} \textbf{\bibinfo{volume}{101}},
  \bibinfo{pages}{081302} (\bibinfo{year}{2020}).

\bibitem[{\citenamefont{Jaramillo et~al.}(2020)\citenamefont{Jaramillo,
  Sanchis-Gual, Barranco, Bernal, Degollado, Herdeiro, and
  N{\'u}{\~n}ez}}]{jaramillo2020dynamical}
\bibinfo{author}{\bibfnamefont{V.}~\bibnamefont{Jaramillo}},
  \bibinfo{author}{\bibfnamefont{N.}~\bibnamefont{Sanchis-Gual}},
  \bibinfo{author}{\bibfnamefont{J.}~\bibnamefont{Barranco}},
  \bibinfo{author}{\bibfnamefont{A.}~\bibnamefont{Bernal}},
  \bibinfo{author}{\bibfnamefont{J.~C.} \bibnamefont{Degollado}},
  \bibinfo{author}{\bibfnamefont{C.}~\bibnamefont{Herdeiro}}, \bibnamefont{and}
  \bibinfo{author}{\bibfnamefont{D.}~\bibnamefont{N{\'u}{\~n}ez}},
  \bibinfo{journal}{Physical Review D} \textbf{\bibinfo{volume}{101}},
  \bibinfo{pages}{124020} (\bibinfo{year}{2020}).

\bibitem[{\citenamefont{Fontaine et~al.}(2001)\citenamefont{Fontaine, Brassard,
  and Bergeron}}]{fontaine2001potential}
\bibinfo{author}{\bibfnamefont{G.}~\bibnamefont{Fontaine}},
  \bibinfo{author}{\bibfnamefont{P.}~\bibnamefont{Brassard}}, \bibnamefont{and}
  \bibinfo{author}{\bibfnamefont{P.}~\bibnamefont{Bergeron}},
  \bibinfo{journal}{Publications of the Astronomical Society of the Pacific}
  \textbf{\bibinfo{volume}{113}}, \bibinfo{pages}{409} (\bibinfo{year}{2001}).

\bibitem[{\citenamefont{Leung et~al.}(2013)\citenamefont{Leung, Chu, Lin, and
  Wong}}]{leung2013dark}
\bibinfo{author}{\bibfnamefont{S.-C.} \bibnamefont{Leung}},
  \bibinfo{author}{\bibfnamefont{M.-C.} \bibnamefont{Chu}},
  \bibinfo{author}{\bibfnamefont{L.-M.} \bibnamefont{Lin}}, \bibnamefont{and}
  \bibinfo{author}{\bibfnamefont{K.-W.} \bibnamefont{Wong}},
  \bibinfo{journal}{Physical Review D} \textbf{\bibinfo{volume}{87}},
  \bibinfo{pages}{123506} (\bibinfo{year}{2013}).

\bibitem[{\citenamefont{Leung et~al.}(2019)\citenamefont{Leung, Zha, Chu, Lin,
  and Nomoto}}]{leung2019accretion}
\bibinfo{author}{\bibfnamefont{S.-C.} \bibnamefont{Leung}},
  \bibinfo{author}{\bibfnamefont{S.}~\bibnamefont{Zha}},
  \bibinfo{author}{\bibfnamefont{M.-C.} \bibnamefont{Chu}},
  \bibinfo{author}{\bibfnamefont{L.-M.} \bibnamefont{Lin}}, \bibnamefont{and}
  \bibinfo{author}{\bibfnamefont{K.}~\bibnamefont{Nomoto}},
  \bibinfo{journal}{The Astrophysical Journal} \textbf{\bibinfo{volume}{884}},
  \bibinfo{pages}{9} (\bibinfo{year}{2019}).

\bibitem[{\citenamefont{Zha et~al.}(2019)\citenamefont{Zha, Chu, Leung, and
  Lin}}]{zha2019accretion}
\bibinfo{author}{\bibfnamefont{S.}~\bibnamefont{Zha}},
  \bibinfo{author}{\bibfnamefont{M.-C.} \bibnamefont{Chu}},
  \bibinfo{author}{\bibfnamefont{S.-C.} \bibnamefont{Leung}}, \bibnamefont{and}
  \bibinfo{author}{\bibfnamefont{L.-M.} \bibnamefont{Lin}},
  \bibinfo{journal}{The Astrophysical Journal} \textbf{\bibinfo{volume}{883}},
  \bibinfo{pages}{13} (\bibinfo{year}{2019}).

\bibitem[{\citenamefont{Parsons et~al.}(2017)\citenamefont{Parsons,
  G{\"a}nsicke, Marsh, Ashley, Bours, Breedt, Burleigh, Copperwheat, Dhillon,
  Green et~al.}}]{parsons2017testing}
\bibinfo{author}{\bibfnamefont{S.}~\bibnamefont{Parsons}},
  \bibinfo{author}{\bibfnamefont{B.~T.} \bibnamefont{G{\"a}nsicke}},
  \bibinfo{author}{\bibfnamefont{T.}~\bibnamefont{Marsh}},
  \bibinfo{author}{\bibfnamefont{R.}~\bibnamefont{Ashley}},
  \bibinfo{author}{\bibfnamefont{M.}~\bibnamefont{Bours}},
  \bibinfo{author}{\bibfnamefont{E.}~\bibnamefont{Breedt}},
  \bibinfo{author}{\bibfnamefont{M.}~\bibnamefont{Burleigh}},
  \bibinfo{author}{\bibfnamefont{C.}~\bibnamefont{Copperwheat}},
  \bibinfo{author}{\bibfnamefont{V.}~\bibnamefont{Dhillon}},
  \bibinfo{author}{\bibfnamefont{M.}~\bibnamefont{Green}},
  \bibnamefont{et~al.}, \bibinfo{journal}{Monthly Notices of the Royal
  Astronomical Society} \textbf{\bibinfo{volume}{470}}, \bibinfo{pages}{4473}
  (\bibinfo{year}{2017}).

\bibitem[{\citenamefont{Joyce et~al.}(2018)\citenamefont{Joyce, Barstow,
  Holberg, Bond, Casewell, and Burleigh}}]{joyce2018gravitational}
\bibinfo{author}{\bibfnamefont{S.~R.} \bibnamefont{Joyce}},
  \bibinfo{author}{\bibfnamefont{M.~A.} \bibnamefont{Barstow}},
  \bibinfo{author}{\bibfnamefont{J.~B.} \bibnamefont{Holberg}},
  \bibinfo{author}{\bibfnamefont{H.~E.} \bibnamefont{Bond}},
  \bibinfo{author}{\bibfnamefont{S.~L.} \bibnamefont{Casewell}},
  \bibnamefont{and} \bibinfo{author}{\bibfnamefont{M.~R.}
  \bibnamefont{Burleigh}}, \bibinfo{journal}{Monthly Notices of the Royal
  Astronomical Society} \textbf{\bibinfo{volume}{481}}, \bibinfo{pages}{2361}
  (\bibinfo{year}{2018}).

\bibitem[{\citenamefont{Romero et~al.}(2019)\citenamefont{Romero, Kepler,
  Joyce, Lauffer, and C{\'o}rsico}}]{romero2019white}
\bibinfo{author}{\bibfnamefont{A.~D.} \bibnamefont{Romero}},
  \bibinfo{author}{\bibfnamefont{S.~O.} \bibnamefont{Kepler}},
  \bibinfo{author}{\bibfnamefont{S.~R.} \bibnamefont{Joyce}},
  \bibinfo{author}{\bibfnamefont{G.}~\bibnamefont{Lauffer}}, \bibnamefont{and}
  \bibinfo{author}{\bibfnamefont{A.~H.} \bibnamefont{C{\'o}rsico}},
  \bibinfo{journal}{Monthly Notices of the Royal Astronomical Society}
  \textbf{\bibinfo{volume}{484}}, \bibinfo{pages}{2711} (\bibinfo{year}{2019}).

\bibitem[{\citenamefont{Chandra et~al.}(2020)\citenamefont{Chandra, Hwang,
  Zakamska, and Cheng}}]{chandra2020gravitational}
\bibinfo{author}{\bibfnamefont{V.}~\bibnamefont{Chandra}},
  \bibinfo{author}{\bibfnamefont{H.-C.} \bibnamefont{Hwang}},
  \bibinfo{author}{\bibfnamefont{N.~L.} \bibnamefont{Zakamska}},
  \bibnamefont{and} \bibinfo{author}{\bibfnamefont{S.}~\bibnamefont{Cheng}},
  \bibinfo{journal}{The Astrophysical Journal} \textbf{\bibinfo{volume}{899}},
  \bibinfo{pages}{146} (\bibinfo{year}{2020}).

\bibitem[{\citenamefont{Bergeron et~al.}(1991)\citenamefont{Bergeron, Saffer,
  and Liebert}}]{bergeron1991spectroscopic}
\bibinfo{author}{\bibfnamefont{P.}~\bibnamefont{Bergeron}},
  \bibinfo{author}{\bibfnamefont{R.~A.} \bibnamefont{Saffer}},
  \bibnamefont{and} \bibinfo{author}{\bibfnamefont{J.}~\bibnamefont{Liebert}},
  in \emph{\bibinfo{booktitle}{White Dwarfs}} (\bibinfo{publisher}{Springer},
  \bibinfo{year}{1991}), pp. \bibinfo{pages}{75--87}.

\bibitem[{\citenamefont{Bergeron et~al.}(2011)\citenamefont{Bergeron, Wesemael,
  Dufour, Beauchamp, Hunter, Saffer, Gianninas, Ruiz, Limoges, Dufour
  et~al.}}]{bergeron2011comprehensive}
\bibinfo{author}{\bibfnamefont{P.}~\bibnamefont{Bergeron}},
  \bibinfo{author}{\bibfnamefont{F.}~\bibnamefont{Wesemael}},
  \bibinfo{author}{\bibfnamefont{P.}~\bibnamefont{Dufour}},
  \bibinfo{author}{\bibfnamefont{A.}~\bibnamefont{Beauchamp}},
  \bibinfo{author}{\bibfnamefont{C.}~\bibnamefont{Hunter}},
  \bibinfo{author}{\bibfnamefont{R.~A.} \bibnamefont{Saffer}},
  \bibinfo{author}{\bibfnamefont{A.}~\bibnamefont{Gianninas}},
  \bibinfo{author}{\bibfnamefont{M.}~\bibnamefont{Ruiz}},
  \bibinfo{author}{\bibfnamefont{M.-M.} \bibnamefont{Limoges}},
  \bibinfo{author}{\bibfnamefont{P.}~\bibnamefont{Dufour}},
  \bibnamefont{et~al.}, \bibinfo{journal}{The Astrophysical Journal}
  \textbf{\bibinfo{volume}{737}}, \bibinfo{pages}{28} (\bibinfo{year}{2011}).

\bibitem[{\citenamefont{Carrasco et~al.}(2014)\citenamefont{Carrasco,
  Catal{\'a}n, Jordi, Tremblay, Napiwotzki, Luri, Robin, and
  Kowalski}}]{carrasco2014gaia}
\bibinfo{author}{\bibfnamefont{J.}~\bibnamefont{Carrasco}},
  \bibinfo{author}{\bibfnamefont{S.}~\bibnamefont{Catal{\'a}n}},
  \bibinfo{author}{\bibfnamefont{C.}~\bibnamefont{Jordi}},
  \bibinfo{author}{\bibfnamefont{P.-E.} \bibnamefont{Tremblay}},
  \bibinfo{author}{\bibfnamefont{R.}~\bibnamefont{Napiwotzki}},
  \bibinfo{author}{\bibfnamefont{X.}~\bibnamefont{Luri}},
  \bibinfo{author}{\bibfnamefont{A.}~\bibnamefont{Robin}}, \bibnamefont{and}
  \bibinfo{author}{\bibfnamefont{P.}~\bibnamefont{Kowalski}},
  \bibinfo{journal}{Astronomy \& Astrophysics} \textbf{\bibinfo{volume}{565}},
  \bibinfo{pages}{A11} (\bibinfo{year}{2014}).

\bibitem[{\citenamefont{Barstow et~al.}(2017)\citenamefont{Barstow, Joyce,
  Casewell, Holberg, Bond, and Burleigh}}]{barstow2017sirius}
\bibinfo{author}{\bibfnamefont{M.}~\bibnamefont{Barstow}},
  \bibinfo{author}{\bibfnamefont{S.}~\bibnamefont{Joyce}},
  \bibinfo{author}{\bibfnamefont{S.}~\bibnamefont{Casewell}},
  \bibinfo{author}{\bibfnamefont{J.}~\bibnamefont{Holberg}},
  \bibinfo{author}{\bibfnamefont{H.}~\bibnamefont{Bond}}, \bibnamefont{and}
  \bibinfo{author}{\bibfnamefont{M.}~\bibnamefont{Burleigh}}, in
  \emph{\bibinfo{booktitle}{20th European White Dwarf Workshop}}
  (\bibinfo{year}{2017}), vol. \bibinfo{volume}{509}, p. \bibinfo{pages}{383}.

\bibitem[{\citenamefont{Pasquini et~al.}(2019)\citenamefont{Pasquini, Pala,
  Ludwig, L\~eao, de~Medeiros, and Weiss}}]{pasquini2019masses}
\bibinfo{author}{\bibfnamefont{L.}~\bibnamefont{Pasquini}},
  \bibinfo{author}{\bibfnamefont{A.}~\bibnamefont{Pala}},
  \bibinfo{author}{\bibfnamefont{H.-G.} \bibnamefont{Ludwig}},
  \bibinfo{author}{\bibfnamefont{I.}~\bibnamefont{L\~eao}},
  \bibinfo{author}{\bibfnamefont{J.}~\bibnamefont{de~Medeiros}},
  \bibnamefont{and} \bibinfo{author}{\bibfnamefont{A.}~\bibnamefont{Weiss}},
  \bibinfo{journal}{Astronomy \& Astrophysics} \textbf{\bibinfo{volume}{627}},
  \bibinfo{pages}{L8} (\bibinfo{year}{2019}).

\bibitem[{\citenamefont{Montero and Cordero-Carrion}(2012)}]{Montero:2012yr}
\bibinfo{author}{\bibfnamefont{P.~J.} \bibnamefont{Montero}} \bibnamefont{and}
  \bibinfo{author}{\bibfnamefont{I.}~\bibnamefont{Cordero-Carrion}},
  \bibinfo{journal}{Phys.Rev.} \textbf{\bibinfo{volume}{D85}},
  \bibinfo{pages}{124037} (\bibinfo{year}{2012}), \eprint{1204.5377}.

\bibitem[{\citenamefont{Sanchis-Gual
  et~al.}(2015{\natexlab{a}})\citenamefont{Sanchis-Gual, Degollado, Montero,
  and Font}}]{Sanchis-Gual:2015}
\bibinfo{author}{\bibfnamefont{N.}~\bibnamefont{Sanchis-Gual}},
  \bibinfo{author}{\bibfnamefont{J.~C.} \bibnamefont{Degollado}},
  \bibinfo{author}{\bibfnamefont{P.~J.} \bibnamefont{Montero}},
  \bibnamefont{and} \bibinfo{author}{\bibfnamefont{J.~A.} \bibnamefont{Font}},
  \bibinfo{journal}{Phys. Rev. D} \textbf{\bibinfo{volume}{91}},
  \bibinfo{pages}{043005} (\bibinfo{year}{2015}{\natexlab{a}}),
  \urlprefix\url{http://link.aps.org/doi/10.1103/PhysRevD.91.043005}.

\bibitem[{\citenamefont{Sanchis-Gual
  et~al.}(2015{\natexlab{b}})\citenamefont{Sanchis-Gual, Degollado, Montero,
  Font, and Mewes}}]{sanchis2015quasistationary}
\bibinfo{author}{\bibfnamefont{N.}~\bibnamefont{Sanchis-Gual}},
  \bibinfo{author}{\bibfnamefont{J.~C.} \bibnamefont{Degollado}},
  \bibinfo{author}{\bibfnamefont{P.~J.} \bibnamefont{Montero}},
  \bibinfo{author}{\bibfnamefont{J.~A.} \bibnamefont{Font}}, \bibnamefont{and}
  \bibinfo{author}{\bibfnamefont{V.}~\bibnamefont{Mewes}},
  \bibinfo{journal}{Physical Review D} \textbf{\bibinfo{volume}{92}},
  \bibinfo{pages}{083001} (\bibinfo{year}{2015}{\natexlab{b}}).

\bibitem[{\citenamefont{Baumgarte and Shapiro}(1998)}]{Baumgarte98}
\bibinfo{author}{\bibfnamefont{T.~W.} \bibnamefont{Baumgarte}}
  \bibnamefont{and} \bibinfo{author}{\bibfnamefont{S.~L.}
  \bibnamefont{Shapiro}}, \bibinfo{journal}{Phys. Rev. D}
  \textbf{\bibinfo{volume}{59}}, \bibinfo{pages}{024007}
  (\bibinfo{year}{1998}).

\bibitem[{\citenamefont{Shibata and Nakamura}(1995)}]{Shibata95}
\bibinfo{author}{\bibfnamefont{M.}~\bibnamefont{Shibata}} \bibnamefont{and}
  \bibinfo{author}{\bibfnamefont{T.}~\bibnamefont{Nakamura}},
  \bibinfo{journal}{Phys. Rev. D} \textbf{\bibinfo{volume}{52}},
  \bibinfo{pages}{5428} (\bibinfo{year}{1995}).

\bibitem[{\citenamefont{Alcubierre and Mendez}(2011)}]{Alcubierre:2010is}
\bibinfo{author}{\bibfnamefont{M.}~\bibnamefont{Alcubierre}} \bibnamefont{and}
  \bibinfo{author}{\bibfnamefont{M.~D.} \bibnamefont{Mendez}},
  \bibinfo{journal}{Gen.Rel.Grav.} \textbf{\bibinfo{volume}{43}},
  \bibinfo{pages}{2769} (\bibinfo{year}{2011}), \eprint{1010.4013}.

\bibitem[{\citenamefont{Bona et~al.}(1997)\citenamefont{Bona, Mass\'o, Seidel,
  and Stela}}]{Bona:1997prd}
\bibinfo{author}{\bibfnamefont{C.}~\bibnamefont{Bona}},
  \bibinfo{author}{\bibfnamefont{J.}~\bibnamefont{Mass\'o}},
  \bibinfo{author}{\bibfnamefont{E.}~\bibnamefont{Seidel}}, \bibnamefont{and}
  \bibinfo{author}{\bibfnamefont{J.}~\bibnamefont{Stela}},
  \bibinfo{journal}{Phys. Rev. D} \textbf{\bibinfo{volume}{56}},
  \bibinfo{pages}{3405} (\bibinfo{year}{1997}),
  \urlprefix\url{http://link.aps.org/doi/10.1103/PhysRevD.56.3405}.

\bibitem[{\citenamefont{Alcubierre et~al.}(2003)\citenamefont{Alcubierre,
  Br\"ugmann, Diener, Koppitz, Pollney, Seidel, and
  Takahashi}}]{Alcubierre:2003ab}
\bibinfo{author}{\bibfnamefont{M.}~\bibnamefont{Alcubierre}},
  \bibinfo{author}{\bibfnamefont{B.}~\bibnamefont{Br\"ugmann}},
  \bibinfo{author}{\bibfnamefont{P.}~\bibnamefont{Diener}},
  \bibinfo{author}{\bibfnamefont{M.}~\bibnamefont{Koppitz}},
  \bibinfo{author}{\bibfnamefont{D.}~\bibnamefont{Pollney}},
  \bibinfo{author}{\bibfnamefont{E.}~\bibnamefont{Seidel}}, \bibnamefont{and}
  \bibinfo{author}{\bibfnamefont{R.}~\bibnamefont{Takahashi}},
  \bibinfo{journal}{Phys. Rev. D} \textbf{\bibinfo{volume}{67}},
  \bibinfo{pages}{084023} (\bibinfo{year}{2003}),
  \urlprefix\url{http://link.aps.org/doi/10.1103/PhysRevD.67.084023}.

\bibitem[{\citenamefont{{Cordero-Carri{\'o}n} and
  {Cerd{\'a}-Dur{\'a}n}}(2012)}]{Isabel:2012arx}
\bibinfo{author}{\bibfnamefont{I.}~\bibnamefont{{Cordero-Carri{\'o}n}}}
  \bibnamefont{and}
  \bibinfo{author}{\bibfnamefont{P.}~\bibnamefont{{Cerd{\'a}-Dur{\'a}n}}},
  \bibinfo{journal}{ArXiv e-prints}  (\bibinfo{year}{2012}),
  \eprint{1211.5930}.

\bibitem[{\citenamefont{Cordero-Carri{\'o}n and
  Cerd{\'a}-Dur{\'a}n}(2014)}]{cordero2014partially}
\bibinfo{author}{\bibfnamefont{I.}~\bibnamefont{Cordero-Carri{\'o}n}}
  \bibnamefont{and}
  \bibinfo{author}{\bibfnamefont{P.}~\bibnamefont{Cerd{\'a}-Dur{\'a}n}}, in
  \emph{\bibinfo{booktitle}{Advances in Differential Equations and
  Applications}} (\bibinfo{publisher}{Springer}, \bibinfo{year}{2014}), pp.
  \bibinfo{pages}{267--278}.

\bibitem[{\citenamefont{Schunck and Liddle}(1997)}]{schunck1997gravitational}
\bibinfo{author}{\bibfnamefont{F.~E.} \bibnamefont{Schunck}} \bibnamefont{and}
  \bibinfo{author}{\bibfnamefont{A.~R.} \bibnamefont{Liddle}},
  \bibinfo{journal}{Physics Letters B} \textbf{\bibinfo{volume}{404}},
  \bibinfo{pages}{25} (\bibinfo{year}{1997}).

\bibitem[{\citenamefont{Holberg et~al.}(2012)\citenamefont{Holberg, Oswalt, and
  Barstow}}]{holberg12}
\bibinfo{author}{\bibfnamefont{J.~B.} \bibnamefont{Holberg}},
  \bibinfo{author}{\bibfnamefont{T.~D.} \bibnamefont{Oswalt}},
  \bibnamefont{and} \bibinfo{author}{\bibfnamefont{M.~A.}
  \bibnamefont{Barstow}}, \bibinfo{journal}{The Astronomical Journal}
  \textbf{\bibinfo{volume}{143}}, \bibinfo{pages}{68} (\bibinfo{year}{2012}).

\bibitem[{\citenamefont{{B{\'e}dard} et~al.}(2020)\citenamefont{{B{\'e}dard},
  {Bergeron}, {Brassard}, and {Fontaine}}}]{bedard20}
\bibinfo{author}{\bibfnamefont{A.}~\bibnamefont{{B{\'e}dard}}},
  \bibinfo{author}{\bibfnamefont{P.}~\bibnamefont{{Bergeron}}},
  \bibinfo{author}{\bibfnamefont{P.}~\bibnamefont{{Brassard}}},
  \bibnamefont{and}
  \bibinfo{author}{\bibfnamefont{G.}~\bibnamefont{{Fontaine}}},
  \bibinfo{journal}{\apj} \textbf{\bibinfo{volume}{901}}, \bibinfo{eid}{93}
  (\bibinfo{year}{2020}), \eprint{2008.07469}.

\bibitem[{\citenamefont{{Provencal} et~al.}(1998)\citenamefont{{Provencal},
  {Shipman}, {H{\o}g}, and {Thejll}}}]{provencal98}
\bibinfo{author}{\bibfnamefont{J.~L.} \bibnamefont{{Provencal}}},
  \bibinfo{author}{\bibfnamefont{H.~L.} \bibnamefont{{Shipman}}},
  \bibinfo{author}{\bibfnamefont{E.}~\bibnamefont{{H{\o}g}}}, \bibnamefont{and}
  \bibinfo{author}{\bibfnamefont{P.}~\bibnamefont{{Thejll}}},
  \bibinfo{journal}{\apj} \textbf{\bibinfo{volume}{494}}, \bibinfo{pages}{759}
  (\bibinfo{year}{1998}).

\end{thebibliography}

%%%%%%%%%%%%%%%
%%%   END   %%%
%%%%%%%%%%%%%%%
\end{document}